**Resonant Far-Infrared Spectroscopy of Flat-Band Fermions in Magic Angle Graphene**

Ayelet J. Uzan-Narovlansky[1,#,*], Ipsita Das[1,#,*], Juan F. Mendez Valderrama[1,#], Yue Tang[1,#], Jonah Herzog-Arbeitman[1,2], Haoyu Hu[1], Pengjie Wang[1,3,4], Zhaoyi Joy Zheng[1,5], Haosen Guan[1], Kenji Watanabe[6], Takashi Taniguchi[7], B. Andrei Bernevig[1,8,9], Sanfeng Wu[1*]

[1] Department of Physics, Princeton University, Princeton, New Jersey 08544, USA
[2] Center for Computational Quantum Physics, Flatiron Institute, New York, NY, 10010, USA
[3] Department of Physics, The Grainger College of Engineering, University of Illinois Urbana-Champaign, Urbana, Illinois 61801, United States
[4] Materials Research Laboratory, The Grainger College of Engineering, University of Illinois Urbana-Champaign, Urbana, Illinois 61801, United States
[5] Department of Electrical and Computer Engineering, Princeton University, Princeton, New Jersey 08544, USA
[6] Research Center for Electronic and Optical Materials, National Institute for Materials Science, 1-1 Namiki, Tsukuba 305-0044, Japan
[7] Research Center for Materials Nanoarchitectonics, National Institute for Materials Science, 1-1 Namiki, Tsukuba 305-0044, Japan
[8] Donostia International Physics Center, P. Manuel de Lardizabal 4, 20018 Donostia-San Sebastian, Spain
[9] IKERBASQUE, Basque Foundation for Science, Bilbao, Spain
[#] These authors contributed equally to this work
[*]Email: sanfengw@princeton.edu; auzan@princeton.edu; id8791@princeton.edu

**Moiré engineering in twisted two-dimensional (2D) materials radically alters low-energy bands, interactions and topological quantum states. Despite extensive studies[1–6], optical spectroscopy of interacting moiré bands in the characteristic far-infrared (FIR) regime has remained largely unexplored due to extreme experimental challenges. Using a newly developed millikelvin FIR platform[7], we report the observation of the long-sought-after characteristic FIR resonances of flat-band electrons in magic-angle twisted bilayer graphene (MATBG)[8–21]. We observe highly tunable spectroscopic signatures of interacting light and heavy fermions that constitute the flat bands in MATBG. Using the topological heavy-fermion model (THF)[22–25], we show that itinerant topological electrons act as an "antenna" that couples strongly to the optical field, with resonant frequencies renormalized by the hybridization with localized heavy electrons. We establish optical selection rules of MATBG which uncovers the key symmetry governing light-heavy fermion hybridization. At charge neutrality, we observe pronounced resonances at energies below the on-site Coulomb energy, implying the emergence of new many-body modes. Our experiments and modeling provide a fundamental understanding of light-matter interactions in MATBG and enable resonant optical spectroscopy of moiré bands down to millikelvin temperatures.**

## Main

Despite extensive studies of moiré materials[1–6], optical spectroscopy of highly interacting low energy bands in the characteristic far-infrared (FIR) or THz regime has remained largely unexplored due to outstanding experimental challenges. For instance, optical spectroscopic characteristics and FIR resonances of magic angle twisted bilayer graphene (MATBG)[8–21] remain elusive both experimentally and theoretically to date, despite eight years of worldwide investigation of this system since its experimental discovery. The same is true for other extensively investigated moiré systems [3–5] including twisted transition metal dichalcogenides. In principle, light-matter interactions in strongly correlated quantum materials featuring flat electronic bands are of interest for investigating interacting phases and searching for novel optical phenomena in new regimes**.** In conventional 2D electron gas systems featuring Gallilean translational symmetry and parabolic energy dispersion, Kohn's theorem implies that optical excitations under Landau quantization are insensitive to many body interactions[26]. Indeed, excitations in the fractional quantum Hall regime, such as the magneto-roton[27,28], are optically dark and its optical detection is limited to nonlinear responses[29,30]. In single layer graphene, the Dirac, instead of parabolic, dispersion violates the condition of Kohn's theorem, but many body interactions lead to small corrections to its cyclotron and optical resonances[31–35]. Intriguingly, twisted moiré materials offer remarkable flat-band systems in which the assumptions of Kohn's theorem, both the translational invariance and the parabolic band conditions, are maximally broken. This, in principle, allows for a plethora of strong (i.e., first order) low-energy optical excitations that are sensitive to interactions in moiré flat bands[36–39], providing exciting opportunities for investigating light-matter interactions in new correlated regimes.

Experimentally the spot size of an FIR beam (~ hundreds of μm in diameter) due to the diffraction limit is significantly larger than the typical size of moiré devices (a few μm), preventing the use of typical reflection or transmission spectroscopy. This challenge can be overcome by employing photocurrent-based measurements[40–45], including the successful implementation of photocurrent Fourier transform infrared (FTIR) spectroscopy for bilayer and ABC trilayer graphene systems[43,45]. Another prominent challenge is that strongly correlated phases and exotic states of many moiré systems appear at low temperatures, often in the sub-kelvin regime, which is rarely accessible in optical experiments[46–49]. To overcome these challenges, some of the authors have developed a state-of-the-art millikelvin FIR detection platform[7] (**Fig. 1A**), enabling for the first time the implementation of FIR spectroscopy of small 2D materials at ultralow temperatures. The platform is capable of performing spectroscopy at a base temperature below 50 mK with an record electron temperature down to ~ 450 mK at the center of radiation and in strong magnetic fields[7]. In this work, by employing this instrument we successfully resolve characteristic FIR resonances of interacting fermions of MATBG and establish foundational understanding of light-matter interactions of the flat-band electrons using the topological heavy-fermion framework[22–25].

## Millikelvin FIR Spectroscopy of MATBG

We employ photocurrent-based FTIR detection scheme[42,45], which extracts photo-induced electric signals from the 2D samples (**Fig. 1b**) by detecting the a.c. current ($I_{ph}$) or voltage ($V_{ph}$) modulations in the presence of a d.c. bias ($I_{dc}$) applied to the source contact, at the frequency of a chopper inserted in the incident beam path. The signals, recorded as a function of the delay time ($\tau$) between the two broadband FIR beams traveling respectively along the interferometer paths, are Fourier transformed to reveal spectroscopic information of the sample responses in the frequency domain. To achieve this, we fabricate high quality MATBG Hall bar devices (see **Methods** & **Extended Data Figs. 1** & **2**) using the standard cut and stack techniques, placed on a bottom gate which controls the carrier density and thus the filling factor ($\nu$), defined as the number of electrons per moiré unit cell[50]. **Figs. 1c** & **d** show transport characteristics of a typical device (D1). The transport fan diagram (**Fig. 1c**) of magnetoresistance ($R_{xx}$) taken at base temperature (with light off) develops features of a MATBG device with an extracted twist angle $\theta$ of 1.12° (**Methods**), including the resistive peak at charge neutrality point (CNP, $\nu = 0$), the strong band insulator (BI) states at full filling $\nu = \pm 4$, the correlated insulator peaks at half filling $\nu = \pm 2$ and at $\nu = +3$, and the superconductivity near $\nu = -2$ (**Fig. 1d**). Similar transport characteristics are observed in D2 ($\theta \sim 1.06°$) and D3 ($\theta \sim 1.05°$). All transport observations are consistent with previous studies[9,10,12,51].

Upon FIR radiation, the conductivity of the sample changes due to the absorption of light, which can be monitored by $I_{ph}$ or $V_{ph}$. **Figs. 1e** & **f** compare the gate tuned $V_{ph}$ (chopped light on) *v.s.* $R_{xx}$ (light off), revealing strong photo-induced signals near the insulating states at integer fillings. The photo signals are highly sensitive to temperatures, as shown in **Fig. 1g**, from which one finds that $V_{ph}$ are substantially enhanced below ~ 1.5 K. At CNP and half fillings, $V_{ph}$ is nearly negligible at 4 K, while being large at 0.5 K (**Fig. 1f**). The strong sub-kelvin signals not only reflect the fact that the correlated states are pronounced at such low temperatures but also enable the possibility of using the insulating states at integer fillings as a sensitive detector for spectroscopic measurement of MATBG. It is thus essential to reach such low temperatures in the presence of FIR radiation, which is the key capability of our instrument[7].

To obtain the FIR excitation spectra, the interferogram of $V_{ph}$ (or $I_{ph}$) *v.s.* $\tau$ is recorded at a finite $I_{dc}$ for each chosen integer filling (see **Figs. 2a** & **b** for a representative data taken at $\nu = +4$). The Fourier transformation of the signal yields the spectrum (**Fig. 2c**). The incident light source is broadband and can be filtered to select the detection window in each experiment. In this work we present data taken using a long pass filter[7] (LP10) combined with either a Mylar or KBr beam splitter that covers an energy window from sub-10 meV to ~ 160 meV (**Fig. 2c**). The features of MATBG manifest as peaks whereas the absorption due to optical elements in the path induces dips in the spectra. The entire optical setup is placed in sealed environment either filled with nitrogen (outside the fridge) or in vacuum (inside the fridge) to minimize air-molecules absorptions in the path.

**Filling-Dependent Resonances of Interacting Electrons**

We first resolve the FIR excitation spectra of MATBG observed at integer fillings of $\nu = 0$, $\pm 2$, and $\pm 4$, as shown in **Figs. 2d** & **e** for device D1 and **Figs. 2f** & **g** for device D2 at $B = 1$ T. Otherwise mentioned, all data are taken at the base temperature and normalized by the black body radiation spectrum of the light source (**Extended Data Fig. 3)**. Both devices show qualitatively consistent results. At $\nu = \pm 4$ (BI), we observe pronounced resonances above certain energies (~ 60 meV for D1 and ~ 40 meV for D2), below which the spectral weight is completely suppressed. The suppression at low energy, confirmed by measuring it using different optics configurations (**Fig. 2c**), clearly reveals an optical gap that is consistent with the development of a large band gap at $\nu = \pm 4$. The quantitative variation of the exact gap value and resonance energies between the two devices may be understood due to device dependent twist angle, strain and interaction effects[51–57]. Our data directly probes the gap structure of the state. The spectra show remarkable dependence on the filling factor. In sharp contrast to the BI states, when the doping is into the flat bands, substantial spectral weight develops at low energies. This is true for $\nu = 0$ and $\pm 2$, observed in both devices (**Figs. 2e** & **g**). At CNP ($\nu = 0$), a new low energy resonance around 26 meV (~ 20 meV) for D1 (D2) becomes most pronounced, distinguishing itself from all other doping.

In MATBG, the interacting low-energy flat-band physics is known to exhibit mixed characters of Dirac light fermions (**c**) that are itinerant and heavy fermions (**f**) localized at the AA stacking sites [11,22,25,39,58–66]. Theoretically, the framework based on the THF model[22–25] naturally describes the strongly correlated physics in which the **f**-electrons are subject to Hubbard interaction, the **c**-electrons carry the topology, and their hybridization gives rise to the rich phenomena of MATBG. Recent quantum twisting microscope (QTM) studies have indeed observed the distribution of the **c** and **f** sectors in momentum space, where the Hubbard-like bands of **f-**electrons have been resolved[64]. Whereas the **c**-electron spectrum appears to be broad as well in the data, the latest theoretical predictions[67–69] suggest that the light-sector excitations have much lower scattering rate and hence should exhibit sharp spectral features compared to the heavy sector. This situation raises a central question: can and how the interacting flat-band fermions develop bright optical resonances? Naively, one may argue that the **f** sector, featuring a high density of state, would favor participating optical transitions. We show that this is not the case and reveal critical roles of the topological **c** sectors and the **c**-**f** hybridization in the optical responses. In the next section and supplementary materials, we examine carefully the observed FIR resonances at full filling ($\nu$ = +4), by examining their magnetic field dependence. We establish fundamental understanding of the FIR resonances consistent with calculations enabled by the THF model.

## Signatures of Light and Heavy-Fermions

The magnetic field modifies the energy structure of MATBG via quantizing electron orbitals that strongly impact the FIR resonances. We focus on the electron side BI state at $\nu$ = +4. We measure the photovoltage ($V_{\mathrm{ph}}$) as a function of filling factor in the presence of magnetic field $B$ applied perpendicular to the 2D plane (see **Extended Data Fig. 4** for fan diagrams). **Figs. 3a** & **b** plot the excitation spectra of D1 & D2 taken under $B$. At finite fields, distinct resonant modes clearly develop. Two modes ($m_1$ & $m_3$) are visible in D1 while three modes ($m_1$-$m_3$) appear in D2. At high field, a fourth mode ($m_4$) develops in company with $m_3$. Modes $m_2$ & $m_3$ (**Fig. 3c**) increase their energies with increasing $B$ with a slope of approximately 3.4 meV/T, the same for both devices. In contrast, $m_1$ bends towards lower energy (i.e., a negative slope) upon increasing $B$. The negative slope is much larger in D2 compared to D1. We verified that the measurements yield consistent results using either photovoltage ($V_{\mathrm{ph}}$) and photocurrent ($I_{\mathrm{ph}}$) data **(Extended Data Fig. 5)**. Similar features are also observed for the hole side BI state at $\nu$ = -4 (**Extended Data Fig. 6**).

To understand the observed modes, we performed comprehensive calculations of optical conductivity (see details in **Supplementary Materials**), based on the THF model. The simulation (**Extended Data Figs. 7** & **8**), which includes lattice relaxation, shows remarkable agreement with the experimental data ($m_1$-$m_3$ modes), allowing us to attribute the resonances to bright transitions between empty and occupied Landau levels (LLs) of MATBG (**Figs. 3d-g**). Mode $m_3$ is the strongest, consistent with experiments, reflecting the dominant **c**-character of the electrons involved, while $m_1$ and $m_2$ are weaker ($m_2$ is invisible in D1 data). This agreement between simulations and experiments allows us to extract interacting parameters in the devices, summarized in **Extended Data Fig. 8 & Extended Data Table 1**, in general consistent with recent experiments[62–64]. We also theoretically investigated the impact of heterostrain, which e.g., plays a role in determining the brightness of $m_2$ mode (**Extended Data Fig. 9**).

Our combined experimental and theoretical results reveal the distinct roles of the **c**- and **f**-electrons, and their complex interplay, in determining the FIR response. First, because the velocity operator associated with the localized **f**-electrons is negligible (**Supplementary Materials**), the **c**-electrons couple much more strongly to the external optical field. The THF model makes it transparent that bright optical transitions across the insulator gap favor the **c**-sector (**c** → **c**) around the $\Gamma_M$ point, despite the high density of states associated with the **f**-electrons. The exact energies of the observed resonances ($m_1$-$m_3$) are, however, strongly renormalized by the hybridization with the strongly interacting **f**-electrons. The resonances are highly sensitive to interacting parameters that determine e.g., the bandwidth of the occupied bands (which are no longer flat with interactions) and the gap to the remote bands. In particular, the $m_1$ mode uniquely disperses *v.s.* $B$ with a negative slope (**Figs. 3a-c**). The energy difference between the zero-field and the high-field resonances of this mode measures the bandwidth lower bound of the interacting flat bands (**Fig. 3g**). Similarly, the extrapolated zero-field value of the $m_3$ mode measures the gap between the flat and remote bands at the $\Gamma_M$ point that is sensitive to the **c**-**f** hybridization (see **Supplementary Materials**).

To understand the qualitative distinction between the data observed in D1 and D2, we examine interacting parameters within the THF model. We note, beyond $U_1$, the Coulomb repulsion between **f**-electrons and the $\Gamma_3$ and $\Gamma_1$ **c**-electrons ($W_1$ and $W_3$ respectively), strongly impact the resonances. Increasing $W_3$ leads to a reduction of the overall width of the flat bands (see **Supplementary Materials**), hence influencing the negative slope of $m_1$ mode. The distinct value of $W_3/U_1$ between D1 (1.36) and D2 (1.15) can be naturally attributed to their difference in the twist angle (1.12$^\circ$ and 1.06$^\circ$ respectively). This is consistent with calculations[23] that have estimated that for a fixed dielectric environment, $W_3$ scales approximately quadratically with twist angle near the magic value, whereas $U_1$ exhibits only linear scaling. To directly illustrate the impact of **c**–**f** repulsion, **Figs. 3d** & **e** plot simulated optical conductivity of MATBG that contrast between two set of values of $W_1$ and $W_3$, differed by 17%, with all other parameters being fixed. The maps closely resemble the observations in D1 and D2, implying that the **c**–**f** interaction is essential in determining the optical fingerprints of MATBG. The conclusion is further supported by a systematic extraction of interacting parameters by best fitting to experimental spectra (**Extended Data Table 1**). We conclude that **c**-electrons effectively act as an "antenna" that couples to the external field while the interactions with **f**-electrons sensitively tune the resonant frequencies. This clear picture is obscured in the continuum model, which offers no straightforward distinction between different degrees of freedom impacting optical spectra.

## Optical Selection Rules and Hidden Organizing Symmetry in Fields

The excellent agreement between the measured FIR spectra and our theoretical results enables a direct determination of the optical selection rules in MATBG within the THF framework (see **Supplementary Materials**). While MATBG is, microscopically, a strongly moiré-hybridized system that generically develops Hofstadter bands at large magnetic flux, the THF construction connects it to a low-energy continuum theory augmented by **f**-electrons that can be canonically quantized[25] without invoking the Hofstadter problem. This description is controlled in the low $B$ limit and accommodates both interactions and magnetic field effects on equal footing, providing a practical route to compute the optical response. As in other graphene-based systems, most notably Bernal bilayer graphene, where in the rotationally invariant limit LLs are labeled by an integer index (N) and obey the well-known selection rule $\Delta|N| = \pm 1$, the THF model admits a natural rotationally invariant limit. The underlying structure reflects an emergent SO(2) rotational symmetry, explicitly broken by lattice-scale effects and strain, but robust at integer fillings where strain is negligible. In this limit, one may assign definite angular momenta, $m = 0, 1, 2, \ldots$ to the fermionic modes so that the total angular momentum operator that includes a pseudospin contribution generates the symmetry and enforces the optical selection rule $\Delta m = \pm 1$ (see **Supplementary Materials**).

In **Fig. 3g**, we explicitly label $m$, revealing that the LLs of **c**-electrons in the occupied flat bands ($\nu = 4$) follow a series of $m = 0, 1, 3, 2, 4 \ldots.$ at finite B and that the conduction band across the hybridization gap exhibits an unusual sequence of $m = 2, 3, 1, \ldots.$, implying Landau level crossings. This crossing in the conduction band, originated from the Rashba-like dispersion, has been previously revealed in transport measurement[70]. In **Extended Data Fig. 10**, we compare this LL sequence with that of a conventional spin Rashba-split band and indeed find close similarities. However, distinct from the spin-Rashba band, the LL index $m$ of MATBG, in which spin plays no role, is offset by 1. This angular momentum offset occurs in MATBG because the conduction band states at the $\Gamma_M$ point form a $\Gamma_3$ irreducible representation (irrep). In the THF model, this symmetry constraint is key to the **c**-**f** hybridization that inherently ties together the formation of the Rashba-like band without spin splitting, the usual offset of angular momentum and the localized **f-**electrons featuring p-wave orbitals. Our calculation shows that the angular momentum offset is contributed by both the **c-** and **f**-electrons carrying the $\Gamma_3$ irrep (see **Supplementary Materials**).

We emphasize that the angular momentum offset due to the $\Gamma_3$ irrep is essential for the optical selection rule (**Fig. 3h**), as if this were not present, the bright resonances would be a completely different set of transitions that cannot explain the experimental data. For instance, in the absence of the angular momentum offset, the $m_1$ mode would be dark and thus one can no longer explain the joint appearance of this mode that negatively disperses with $B$ in tandem with other modes ($m_2$ & $m_3$). Our data hence obeys an optical selection rule that reveals this intriguing additional angular momentum and the hidden organizing symmetry (the $\Gamma_3$ irrep) of the THF physics. Such information is uniquely captured by optical spectroscopy while being hidden from other measurement approaches.

**Many-body Optical Resonances at CNP**

We further discuss the low energy resonances observed at the CNP, as shown in **Figs. 4a & b**. Note that the bottom gate of D2 is single layer graphene, whose FIR resonances have been well characterized and not present in **Fig 4a**. We focus on resonances below the onsite Coulomb energy, ($U_1$ ~ 45 meV, see **Extended Data Table 1**), and hence transitions can be potentially attributed to empty and occupied flat bands of MATBG. At low fields (below 2 T), the excitations are centered around 20 meV and can be decomposed into two modes, one at ~ 12 meV and the other at ~ 24 meV (see **Figs. 4c** & **d** for fitting the spectra with Gaussian functions). Above 3 T, the lower energy mode is largely suppressed, leaving behind one pronounced mode that shifts its energy in $B$ with a slope of ~ 2 meV/T. With increasing $B$, this mode merges with a new mode that develops only at high fields (above 6 T), appearing at ~ 40 meV. Qualitatively similar data are observed in another graphene gated device (D3, **Extended Data Fig. 11**).

Strong low energy resonances corresponded to transitions associated with **c**-electrons in flat bands are expected for CNP state being (strain[71–73] or Mott[60,74–76]) semimetal. This is experimentally supported by the weakly resistive peak observed at the CNP in our devices (**Fig. 1d** & **Extended Data Figure**

**2**) as well as the recent QTM measurements[64]. The QTM experiments revealed an anomalous low energy mode[64] which may further complicate the interpretation of optical resonances in this regime. However, we note that our observation of strong low energy optical resonances is consistent with the recent predictions that the **c** sector excitations remain sharp[67–69]. The resonance near 40 meV at high *B* is novel as it is close to energy separation between the **f**-electrons, but we currently do not understand its exact origin. The potential formation of collective or composite modes, such as flat band excitons, can alter the response significantly[77,78]. Comprehensive simulations of transitions within flat bands, in the presence of strain and lattice relaxations, are necessary for understanding the nature of these excitations, which we leave for separate studies.

**Outlook**

Many questions are now open regarding the many-body optical resonances of interacting electrons at the CNP and partial filling of the flat bands, enabled by our experiments. Polarization-resolved spectroscopy could offer a critical degree of freedom to further resolve the transitions and their underlying quantum states. Beyond MATBG, our experimental approach is readily adaptable to other moiré systems, such as twisted bilayer $MoTe_2$, inviting experimental and theoretical efforts to uncover light-matter interactions in such strongly correlated systems.

**References**

1. Geim, A. K. & Grigorieva, I. V. Van der Waals heterostructures. *Nature* **499**, 419–425 (2013).
2. Ajayan, P., Kim, P. & Banerjee, K. Two-dimensional van der Waals materials. *Phys. Today* **69**, 38–44 (2016).
3. Andrei, E. Y. *et al.* The marvels of moiré materials. *Nat. Rev. Mater.* **6**, 201–206 (2021).
4. Kennes, D. M. *et al.* Moiré heterostructures as a condensed-matter quantum simulator. *Nat. Phys.* **17**, 155–163 (2021).
5. Nuckolls, K. P. & Yazdani, A. A microscopic perspective on moiré materials. *Nat. Rev. Mater.* **9**, 460–480 (2024).
6. Cao, T., Fu, L., Ju, L., Xiao, D. & Xu, X. Fractional Quantum Anomalous Hall Effect. (2025).
7. Onyszczak, M. *et al.* A platform for far-infrared spectroscopy of quantum materials at millikelvin temperatures. *Rev. Sci. Instrum.* **94**, 103903 (2023).
8. Bistritzer, R. & MacDonald, A. H. Moiré bands in twisted double-layer graphene. *Proc. Natl. Acad. Sci. U. S. A.* **108**, 12233–12237 (2011).
9. Cao, Y. *et al.* Correlated insulator behaviour at half-filling in magic-angle graphene superlattices. *Nature* **556**, 80–84 (2018).
10. Cao, Y. *et al.* Unconventional superconductivity in magic-angle graphene superlattices. *Nature* **556**, 43–50 (2018).
11. Xie, Y. *et al.* Spectroscopic signatures of many-body correlations in magic-angle twisted bilayer graphene. *Nature* **572**, 101–105 (2019).
12. Lu, X. *et al.* Superconductors, orbital magnets and correlated states in magic-angle bilayer graphene. *Nature* **574**, 653–657 (2019).
13. Yankowitz, M. *et al.* Tuning Superconductivity in Twisted Bilayer Graphene. *Science* **363**, 1058-1064 (2019).

14. Serlin, M. *et al.* Intrinsic Quantized Anomalous Hall Effect in a Moiré Heterostructure. *Science* **367**, 900-903 (2020).
15. Sharpe, A. L. *et al.* Emergent Ferromagnetism near Three-Quarters Filling in Twisted Bilayer Graphene. *Science* **365**, 605-608 (2019).
16. Wong, D. *et al.* Cascade of electronic transitions in magic-angle twisted bilayer graphene. *Nature* **582**, 198–202 (2020).
17. Nuckolls, K. P. *et al.* Strongly correlated Chern insulators in magic-angle twisted bilayer graphene. *Nature* **588**, 610–615 (2020).
18. Choi, Y. *et al.* Interaction-driven band flattening and correlated phases in twisted bilayer graphene. *Nat. Phys.* **17**, 1375–1381 (2021).
19. Xie, Y. *et al.* Fractional Chern insulators in magic-angle twisted bilayer graphene. *Nature* **600**, 439–443 (2021).
20. Nuckolls, K. P. *et al.* Quantum textures of the many-body wavefunctions in magic-angle graphene. *Nature* **620**, 525–532 (2023).
21. Chen, C. *et al.* Strong electron–phonon coupling in magic-angle twisted bilayer graphene. *Nature* **636**, 342–347 (2024).
22. Song, Z. D. & Bernevig, B. A. Magic-Angle Twisted Bilayer Graphene as a Topological Heavy Fermion Problem. *Phys. Rev. Lett.* **129**, (2022).
23. Călugăru, D. *et al.* Twisted bilayer graphene as topological heavy fermion: II. Analytical approximations of the model parameters. *Low Temp Phys.* **49**, 640-654 (2023).
24. Herzog-Arbeitman, J. *et al.* Topological heavy fermion model as an efficient representation of atomistic strain and relaxation in twisted bilayer graphene. *Phys. Rev. B* **112**, 125128 (2025).
25. Singh, K., Chew, A., Herzog-Arbeitman, J., Bernevig, B. A. & Vafek, O. Topological heavy fermions in magnetic field. *Nat. Commun.* **15**, 5257 (2024).
26. Kohn, W. Cyclotron Resonance and de Haas-van Alphen Oscillations of an Interacting Electron Gas. *Phys. Rev.* **123**, 1242–1244 (1961).
27. Girvin, S. M., Macdonald, A. H. & Platzman, P. M. Magneto-Roton Theory of Collective Excitations in the Fractional Quantum Hall Effect. *Phys. Rev. B* **33**, 2481-2494 (1986).
28. Yang, B., Hu, Z.-X., Papić, Z. & Haldane, F. D. M. Model Wave Functions for the Collective Modes and the Magnetoroton Theory of the Fractional Quantum Hall Effect. *Phys. Rev. Lett.* **108**, 256807 (2012).
29. Liang, J. *et al.* Evidence for chiral graviton modes in fractional quantum Hall liquids. *Nature* **628**, 78–83 (2024).
30. Pinczuk, A., Dennis, B. S., Pfeiffer, L. N. & West, K. Observation of collective excitations in the fractional quantum Hall effect. *Phys. Rev. Lett.* **70**, 3983–3986 (1993).
31. Jiang, Z. *et al.* Infrared spectroscopy of landau levels of graphene. *Phys. Rev. Lett.* **98**, (2007).
32. Henriksen, E. A. *et al.* Interaction-induced shift of the cyclotron resonance of graphene using infrared spectroscopy. *Phys. Rev. Lett.* **104**, (2010).
33. Shizuya, K. Many-body corrections to cyclotron resonance in monolayer and bilayer graphene. *Phys. Rev. B - Condens. Matter Mater. Phys.* **81**, (2010).

34. Faugeras, C. *et al.* Landau level spectroscopy of electron-electron interactions in graphene. *Phys. Rev. Lett.* **114**, (2015).
35. Pack, J. *et al.* Broken Symmetries and Kohn's Theorem in Graphene Cyclotron Resonance. *Phys. Rev. X* **10**, (2020).
36. Kumar, A., Xie, M. & MacDonald, A. H. Lattice collective modes from a continuum model of magic-angle twisted bilayer graphene. *Phys. Rev. B* **104**, 035119 (2021).
37. Kousa, B. M., Morales-Durán, N., Wolf, T. M. R., Khalaf, E. & MacDonald, A. H. Theory of Magnetoroton Bands in Moiré Materials. *Phys. Rev. Lett.* **135**, 246604 (2025).
38. Mao, D., Mendez-Valderrama, J. F. & Chowdhury, D. Low-energy optical absorption in correlated insulators: Projected sum rules and the role of quantum geometry. *Phys. Rev. B* **112**, 075116 (2025).
39. Calderón, M. J., Camjayi, A., Datta, A. & Bascones, E. Cascades in transport and optical conductivity of twisted bilayer graphene. *Phys. Rev. B* **112**, 1–7 (2025).
40. Woessner, A. *et al.* Near-field photocurrent nanoscopy on bare and encapsulated graphene. *Nat. Commun.* **7**, 10783 (2016).
41. Sunku, S. S. *et al.* Nano-photocurrent Mapping of Local Electronic Structure in Twisted Bilayer Graphene. *Nano Lett.* **20**, 2958–2964 (2020).
42. Ju, L. *et al.* Unconventional valley-dependent optical selection rules and landau level mixing in bilayer graphene. *Nat. Commun.* **11**, 2941 (2020).
43. Yang, J. *et al.* Spectroscopy signatures of electron correlations in a trilayer graphene/hBN moiré superlattice. *Science* **375**, 1295–1299 (2022).
44. Ma, Q., Krishna Kumar, R., Xu, S.-Y., Koppens, F. H. L. & Song, J. C. W. Photocurrent as a multiphysics diagnostic of quantum materials. *Nat. Rev. Phys.* **5**, 170–184 (2023).
45. Ju, L. *et al.* Tunable excitons in bilayer graphene. *Science* **358**, 907–910 (2017).
46. Deng, B. *et al.* Strong mid-infrared photoresponse in small-twist-angle bilayer graphene. *Nat. Photonics* **14**, 549–553 (2020).
47. Li, G. *et al.* Infrared Spectroscopy for Diagnosing Superlattice Minibands in Twisted Bilayer Graphene near the Magic Angle. *Nano Lett.* **24**, 15956–15963 (2024).
48. Krishna Kumar, R. *et al.* Terahertz photocurrent probe of quantum geometry and interactions in magic-angle twisted bilayer graphene. *Nat. Mater.* **24**, 1034–1041 (2025).
49. Persky, E. *et al.* Optical control of orbital magnetism in magic angle twisted bilayer graphene. arXiv.2503.21750 (2025).
50. Díez-Mérida, J. *et al.* High-yield fabrication of bubble-free magic-angle twisted bilayer graphene devices with high twist-angle homogeneity. *Newton* **1**, 100007 (2025).
51. Stepanov, P. *et al.* Untying the insulating and superconducting orders in magic-angle graphene. *Nature* **583**, 375–378 (2020).
52. Guinea, F. & Walet, N. R. Electrostatic effects, band distortions, and superconductivity in twisted graphene bilayers. *Proc. Natl. Acad. Sci. U. S. A.* **115**, 13174–13179 (2018).
53. Balents, L., Dean, C. R., Efetov, D. K. & Young, A. F. Superconductivity and strong correlations in moiré flat bands. *Nat. Phys.* **16**, 725–733 (2020).
54. Uri, A. *et al.* Mapping the twist-angle disorder and Landau levels in magic-angle graphene. *Nature* **581**, 47–52 (2020).

55. Parker, D. E., Soejima, T., Hauschild, J., Zaletel, M. P. & Bultinck, N. Strain-Induced Quantum Phase Transitions in Magic-Angle Graphene. *Phys. Rev. Lett.* **127**, (2021).
56. Saito, Y., Ge, J., Watanabe, K., Taniguchi, T. & Young, A. F. Independent superconductors and correlated insulators in twisted bilayer graphene. *Nat. Phys.* **16**, 926–930 (2020).
57. Zhang, L. *et al.* Correlated States in Strained Twisted Bilayer Graphene Away from the Magic Angle. *Nano Lett.* **22**, 3204–3211 (2022).
58. Xie, M. & MacDonald, A. H. Nature of the Correlated Insulator States in Twisted Bilayer Graphene. *Phys. Rev. Lett.* **124**, 097601 (2020).
59. Bultinck, N. *et al.* Ground State and Hidden Symmetry of Magic-Angle Graphene at Even Integer Filling. *Phys. Rev. X* **10**, 031034 (2020).
60. Hofmann, J. S., Khalaf, E., Vishwanath, A., Berg, E. & Lee, J. Y. Fermionic Monte Carlo Study of a Realistic Model of Twisted Bilayer Graphene. *Phys. Rev. X* **12**, 011061 (2022).
61. Vafek, O. & Kang, J. Renormalization Group Study of Hidden Symmetry in Twisted Bilayer Graphene with Coulomb Interactions. *Phys. Rev. Lett.* **125**, 257602 (2020).
62. Merino, R. L. *et al.* Interplay between light and heavy electron bands in magic-angle twisted bilayer graphene. *Nat. Phys.* **21**, 1078–1084 (2025).
63. Zhang, Z. *et al.* Heavy fermions, mass renormalization and local moments in magic-angle twisted bilayer graphene via planar tunneling spectroscopy. arXiv.2503.17875.
64. Xiao, J. *et al.* The Interacting Energy Bands of Magic Angle Twisted Bilayer Graphene Revealed by the Quantum Twisting Microscope. *Nature* **653**, 68–75 (2026).
65. Ledwith, P. J., Vishwanath, A. & Khalaf, E. Exotic Carriers from Concentrated Topology: Dirac Trions as the Origin of the Missing Spectral Weight in Twisted Bilayer Graphene. arXiv.2505.08779 (2025).
66. Huang, C. *et al.* Angle-tuned Gross-Neveu quantum criticality in twisted bilayer graphene. *Nat. Commun.* **16**, 7176 (2025).
67. Vituri, Y. & Berg, E. Controlled Loop Expansion for the Topological Heavy Fermion Model. arXiv.2604.14278 (2026).
68. Wei, N., von Oppen, F. & Glazman, L. I. Lifetime and spectral function of topological heavy fermions. arXiv.2604.14369 (2026).
69. Hu, H. *et al.* Twisted Bilayer Graphene Lifetimes at Integer Fillings: An Analytic Result. arXiv.2604.14303 (2026).
70. Das, I. *et al.* Symmetry-broken Chern insulators and Rashba-like Landau-level crossings in magic-angle bilayer graphene. *Nat. Phys.* **17**, 710–714 (2021).
71. Wagner, G., Kwan, Y. H., Bultinck, N., Simon, S. H. & Parameswaran, S. A. Global Phase Diagram of the Normal State of Twisted Bilayer Graphene. *Phys. Rev. Lett.* **128**, 156401 (2022).
72. Kwan, Y. H. *et al.* Kekulé Spiral Order at All Nonzero Integer Fillings in Twisted Bilayer Graphene. *Phys. Rev. X* **11**, 041063 (2021).
73. Crippa, L. *et al.* Dynamical correlation effects in twisted bilayer graphene under strain and lattice relaxation. arXiv.2509.19436 (2025).
74. Wang, Y.-J., Zhou, G.-D., Lian, B. & Song, Z.-D. Electron-phonon coupling in the topological heavy fermion model of twisted bilayer graphene. *Phys. Rev. B* **111**, 035110 (2025).

75. Ledwith, P. J., Dong, J., Vishwanath, A. & Khalaf, E. Nonlocal Moments and Mott Semimetal in the Chern Bands of Twisted Bilayer Graphene. *Phys. Rev. X* **15**, 021087 (2025).
76. Hu, H., Song, Z.-D. & Bernevig, B. A. Projected and Solvable Topological Heavy Fermion Model of Twisted Bilayer Graphene. arXiv.2502.14039 (2025).
77. Xie, H.-Y., Ghaemi, P., Mitrano, M. & Uchoa, B. Theory of topological exciton insulators and condensates in flat Chern bands. *Proc. Natl. Acad. Sci.* **121**, e2401644121 (2024).
78. Kwan, Y. H., Hu, Y., Simon, S. H. & Parameswaran, S. A. Exciton Band Topology in Spontaneous Quantum Anomalous Hall Insulators: Applications to Twisted Bilayer Graphene. *Phys. Rev. Lett.* **126**, 137601 (2021).

**Acknowledgement**

This work is mainly supported by the Gordon and Betty Moore Foundation through Grant GBMF11946 to S.W. and the Materials Research Science and Engineering Center (MRSEC) program of the National Science Foundation (DMR-2011750) awarded to S.W. S.W. acknowledges support from AFOSR (on device fabrications and transport) through awards FA9550-23-1-0140 and FA9550-25-1-0354, and the support from Princeton Catalysis Initiative (on FIR spectroscopy). BAB was supported by DOE Grant No. DE-SC001623. JFMV, HH and JHA were supported by Gordon and Betty Moore Foundation through Grant No. GBMF8685 towards the Princeton theory program, the Gordon and Betty Moore Foundation's EPiQS Initiative (Grant No. GBMF11070), the Global Collaborative Network Grant at Princeton University, the Simons Investigator Grant No. 404513, the NSF-MERSEC (Grant No. MERSEC DMR 2011750), Simons Collaboration on New Frontiers in Superconductivity (SFI-MPS- NFS-00006741-01), and the Schmidt Foundation at the Princeton University, and the Princeton Catalyst Initiative. H.H. was also supported partially by a European Research Council (ERC) under the European Union's Horizon 2020 research and innovation program (Grant Agreement No. 101020833). K.W. and T.T. acknowledge support from the JSPS KAKENHI (Grant Numbers 21H05233 and 23H02052), the CREST (JPMJCR24A5), JST and World Premier International Research Center Initiative (WPI), MEXT, Japan.

**Author Contributions**

A. J. U-N., I. D., Y. T. fabricated the devices, performed measurements and analyzed the data, assisted by P. W., Z. J. Z., and H. G., and supervised by S. W. J. F. M. V. performed theoretical calculations, assisted by J. H-A. and H. H., supervised by B. A. B. K. W. and T. T. provided hBN crystals. S. W., A. J. U-N., J. F. M. V., I. D., J. H-A. and B. A. B interpreted the results and wrote the paper with input from all authors.

**Competing Interests**

The authors declare no competing interests.

**Data Availability**

The data that support the findings of this study are available from the corresponding author upon reasonable request.

**Code Availability**

N.A.

**Methods**

**Device fabrication**

The MATBG samples were fabricated using a "cut-and-stack" method. Graphene was cut into two pieces using a tungsten tip. A poly(bisphenol A carbonate)/polydimethylsiloxane (PC/PDMS) stamp mounted on a glass slide was then used to pick up an hexagonal boron nitride (hBN) flake (10 ~ 15 nm), which then sequentially picked up the two graphene layers from a $Si{+}{+}/SiO_2$ substrate with a target twist angle of $\theta$ near 1.1°, followed by encapsulation with another hBN layer. In this study, either graphite (4 ~ 5 nm thick, D1) or single-layer graphene (D2 and D3) was used as gates to tune the carrier density. The bottom hBN flakes thicknesses were ~15 nm (D1), ~30 nm (D2) and ~10 nm (D3). Finally, the poly(bisphenol A carbonate) was melted at 180°C, and the complete stack was released onto an $O_2$-plasma-cleaned Si++/$SiO_2$ chip. Electrical connections to the twisted bilayer graphene were formed by $CHF_3/O_2$ plasma etching followed by deposition of Cr/Au (6 nm/70 nm) to create metallic edge contacts.

**Millikelvin FIR spectroscopy**

The FIR spectrum is resolved via photocurrent or photovoltage spectroscopy using a Fourier transform infrared (FTIR) spectrometer. The technical details of the instrument are reported in ref [7]. A Bruker FTIR spectrometer (Vertex v80) provides two collinear time-delayed beams, generated via internal infrared light source (globar bulb, unpolarized light), with a controlled time delay via a motorized delay stage in the interferometer. A long pass filter and the beam splitter can be alternated to support measurements with different spectral ranges. The infrared beams freely propagate and are focused onto a sample that cooled down inside a dilution refrigerator (Bluefors), at roughly the center of the equipped superconducting magnet. In the external optical set up, an optical chopper is used to switch on-off the light, at ~ 200 Hz frequency. The entire optical system is placed inside closed boxes that are being pumped or $N_2$ purged to minimized air absorption of the light.

For photocurrent and photovoltage ($I_{ph}$, $V_{ph}$) measurements a d.c. current ($I_{dc}$) is applied to the source contact via a Keithley 2400 source meter. The photo-induced signals are measured at the frequency of optical chopper, using standard low-frequency lock-in technique (Stanford Research SR860) and a current or voltage amplifiers. All the measurements are performed with an optical filter LP10 for a spectral range up to ~ 160 meV. Two beamsplitters were alternately used, to support interferogram at two different spectral ranges. The low energy beam splitter, Mylar, allows for interference measurements in the range of 5 - 80 meV while the KBr beam splitter allows better measurements above 50 meV. The interferograms (transformed later into spectra) were resolved at the fridge base temperature ~ 55 - 60 mK.

**Transport Measurements**

All measurements were performed in a Bluefors dilution refrigerator equipped with a superconducting magnet using a standard low-frequency lock-in technique (Stanford Research SR860 or SR830) with an excitation frequency of 23 Hz. To minimize electron heating, a low excitation current of 10 nA was used. Back gate voltages were controlled using a Keithley 2400 source meter.

**Twist Angle Extraction**

The Landau fan diagrams are used to estimate the twist angle $\theta$ in the measured devices. We use the relation $n_S = 8\theta^2/\sqrt{3}\,a^2$, where $a = 0.246$ nm is the lattice constant of graphene and $n_S$ is the charge carrier density corresponding to a fully filled superlattice unit cell. Quantum oscillations propagating outside the fully filled flat band were used to define $n_S$. The carrier density was calibrated with the trajectory of LLs and low-field Hall effect. The extracted twist angles are as follows: D1 ~ 1.12°, D2 ~ 1.06°, D3 ~ 1.05° with an error of ±0.02°

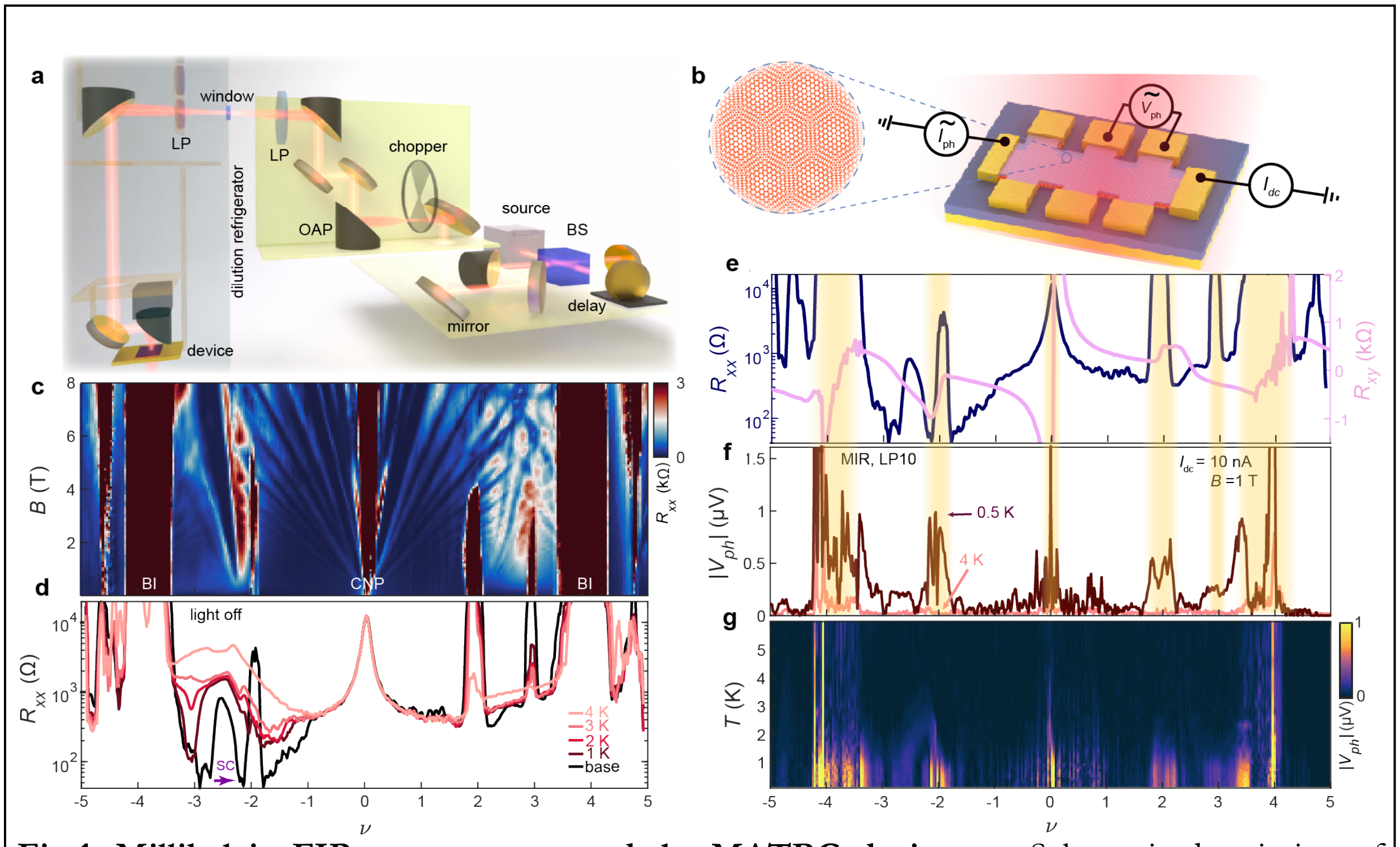

**Fig.1. Millikelvin FIR spectroscopy and the MATBG devices. a**, Schematic description of millikelvin FIR optical detection platform based on Fourier transform infrared spectroscopy. **b,** FIR excitation spectrum is measured via photocurrent ($I_{\mathrm{ph}}$) and photovoltage ($V_{\mathrm{ph}}$) in the presence of d.c. current ($I_{\mathrm{dc}}$) applied to the source contact. **c**, $R_{\mathrm{xx}}$ of D1 as a function of $B$ and $\nu$ taken at base temperature (with light off). **d**, Temperature ($T$) dependence of the $R_{\mathrm{xx}}$ traces, showing the insulating and superconducting (SC) states. **e,** A selected longitudinal and transverse resistivity ($R_{xx}$ & $R_{xy}$) traces taken with light off and at $B = 1$ T. **f,** Photovoltage ($V_{\mathrm{ph}}$) as a function of filling factor $\nu$, taken at $T = 4$ K (coral) and at $T = 0.5$ K (dark brown). The signals are measured with 10 nA d.c. current applied. The MATBG insulating states are marked by yellow shades at $\nu = 0, \pm 2, 3, \pm 4$. **g**, Temperature dependence of the photovoltage signal.

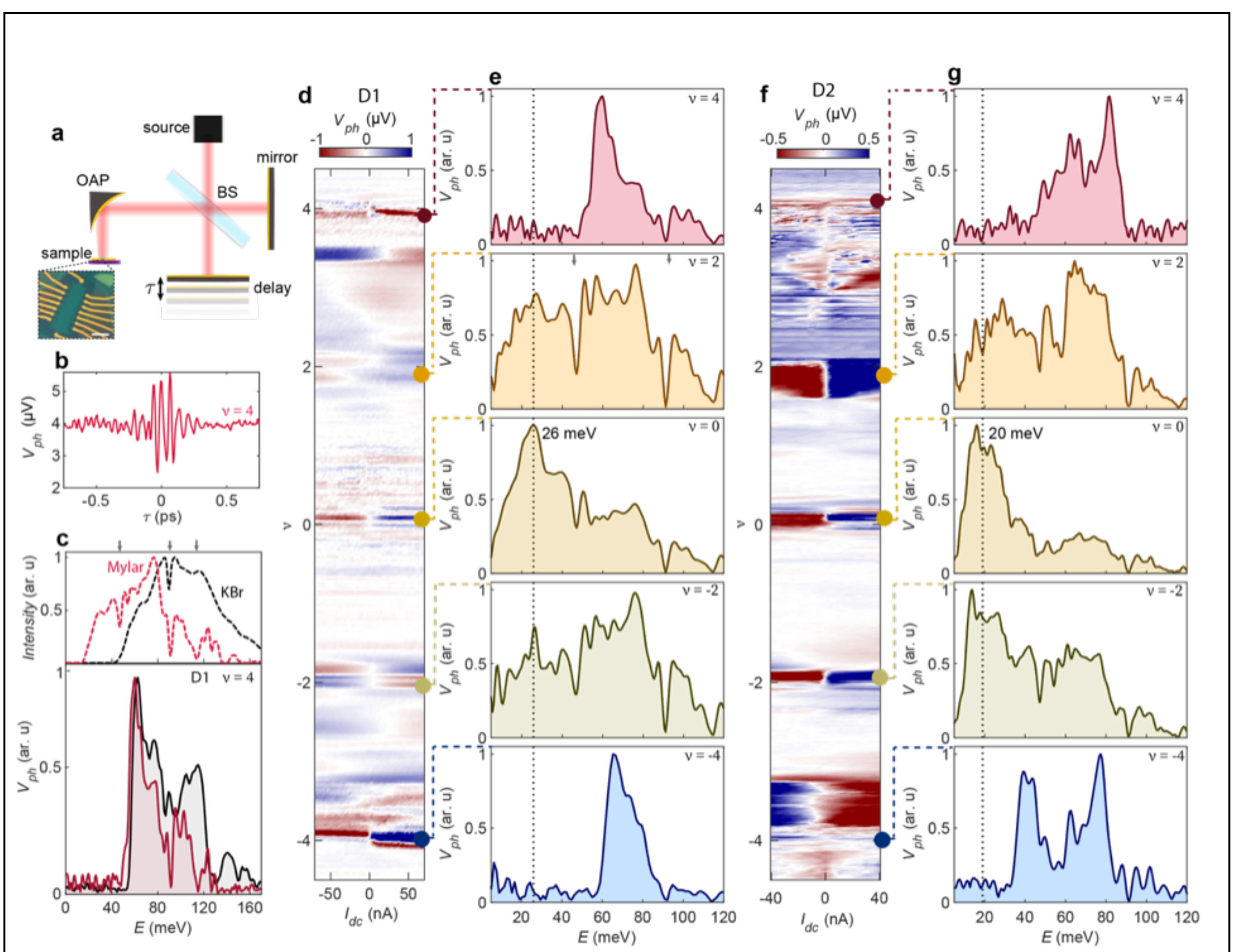

**Fig. 2. Filling dependence FIR spectra of MATBG**. **a & b,** Schematic description of FTIR spectroscopy (**a**) and a representative oscillating photovoltage interferogram (red curve) as a function of the time delay ($\tau$) between the two optical beams (**b**). **c**, Top: The reference input source spectrum, detected via external detector, highlight the best working range of the two beamsplitters (BS) of KBr (black) and Mylar (red). The features due to beamsplitter absorption (dips) are marked with gray arrows. Bottom: The excitation spectrum of D1 at filling $\nu = +4$, detected using the KBr (black) or Mylar (red) beamsplitter at B = 1 T and at the base $T$. **d** & **f**, Photovoltage map (D1 and D2, respectively) as a function of the bias current $I_{dc}$ and the filling factor $\nu$. **e** & **g**, Photovoltage FIR excitation spectra (D1 and D2 respectively), taken at different $\nu$. The data at **d-g** is taking using Mylar BS, at $B = 1$ T and at the base $T$.

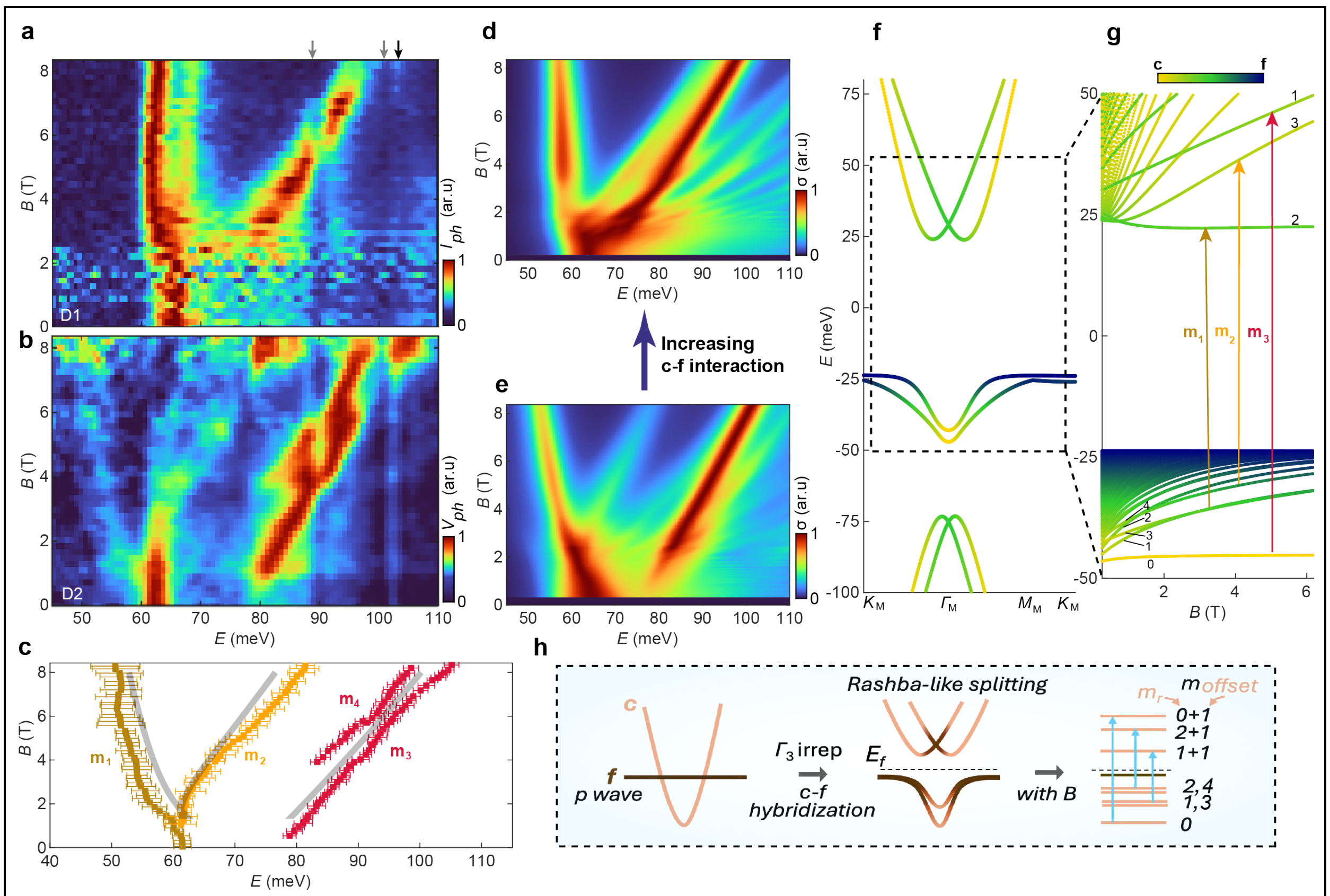

**Fig. 3. Interaction-driven resonances at $\nu$ = +4 and the optical selection rule. a & b**, The FIR spectra taken at $\nu$ = +4 as a function of $B$, of device D1 and D2, respectively. Data is taken at the base $T$ with KBr beamsplitter. Each spectrum is normalized by its maximum value. BS absorption lines are marked by gray arrows and the hBN phonon excitation at 102 meV is marked by the black arrow. **c**, Extraction of the resonance energy for each mode observed in D2 (**b**) as a function of $B$. The theoretical prediction for these excitations ($m_1$-$m_3$, extracted from **e**) is presented by the gray curves. **d** & **e**, Calculated infrared excitation spectra based on the THF model. **e** is a simulation with parameters that match best to data in **b** (D2) where **d** is obtained by increasing c-f interaction terms, $W_1$ & $W_3$, by 17%. **f**, The interacting band structure corresponding to calculations in **e**, at $\nu$ = +4 and $B$ = 0, in which the color encodes contributions of the light and heavy fermions (yellow to blue). **g**, Calculated LLs as a function of $B$, corresponding to the interacting bands shown in **f**. The experimentally observed excitation modes ($m_1$-$m_3$) and the LL index are indicated. **h**, Schematic description of the optical selection rules, visualized in the THF model. The angular momentum offset of the Rashba-like dispersive band is explicitly shown.

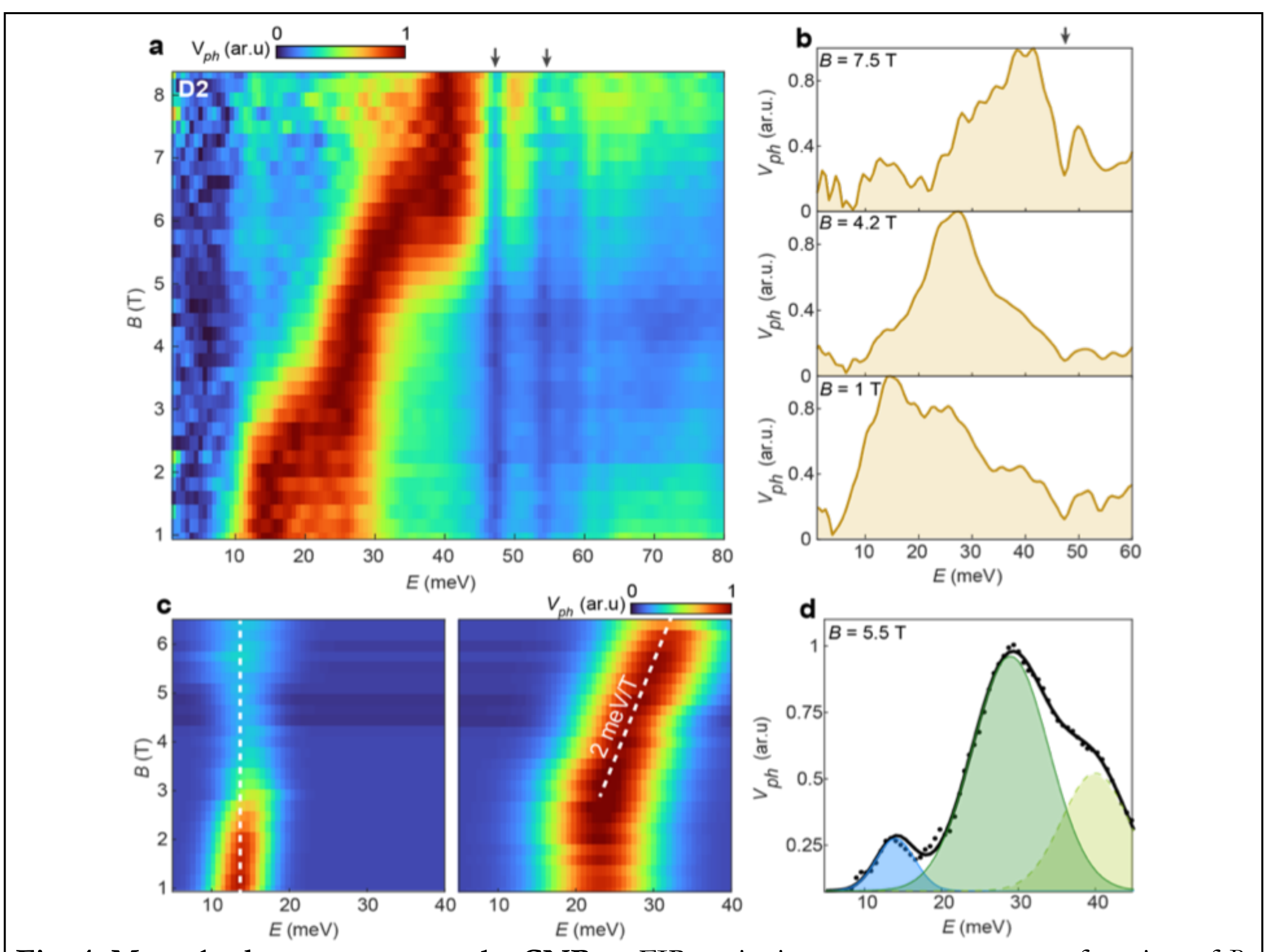

**Fig. 4. Many-body resonances at the CNP. a**, FIR excitation spectrum map, as a function of $B$, resolved at base $T$ using Mylar BS. Each spectrum at a given $B$ is normalized by its maximum value. The BS absorption lines are marked with gray arrows. **b**, Linecuts of the excitation spectra, at $B$ = 1 T, 4.2 T and 7.5 T, respectively. **c**, Left and right maps: Gaussian decomposition of the spectra, identifying two low-energy modes and their magnetic-field evolution (each mode is plotted in a separated map). The white dash lines indicate the shift (right map) or the absence (left map) of shift in its resonance energy. **d**, Demonstration of the Gaussian decomposition for the excitation spectrum resolved at 5.5 T. The lower and higher energy excitations plotted in **c** correspond to the blue/green curves.

## Extended Data Figures & Tables

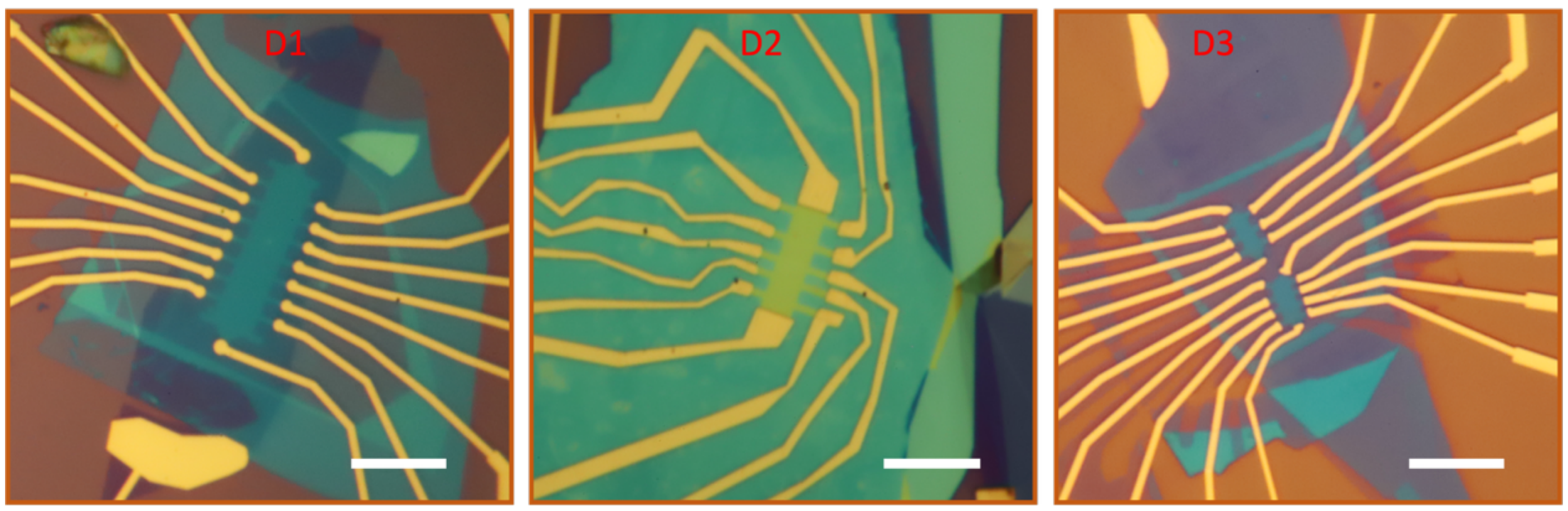

**Extended Data Figure 1. Optical images of three devices,** D1( $\theta = 1.12°$), D2 ($\theta = 1.06°$) and D3 ($\theta = 1.05°$), demonstrating bubble-free and clean channels. The scale bar in all the images is 10 μm.

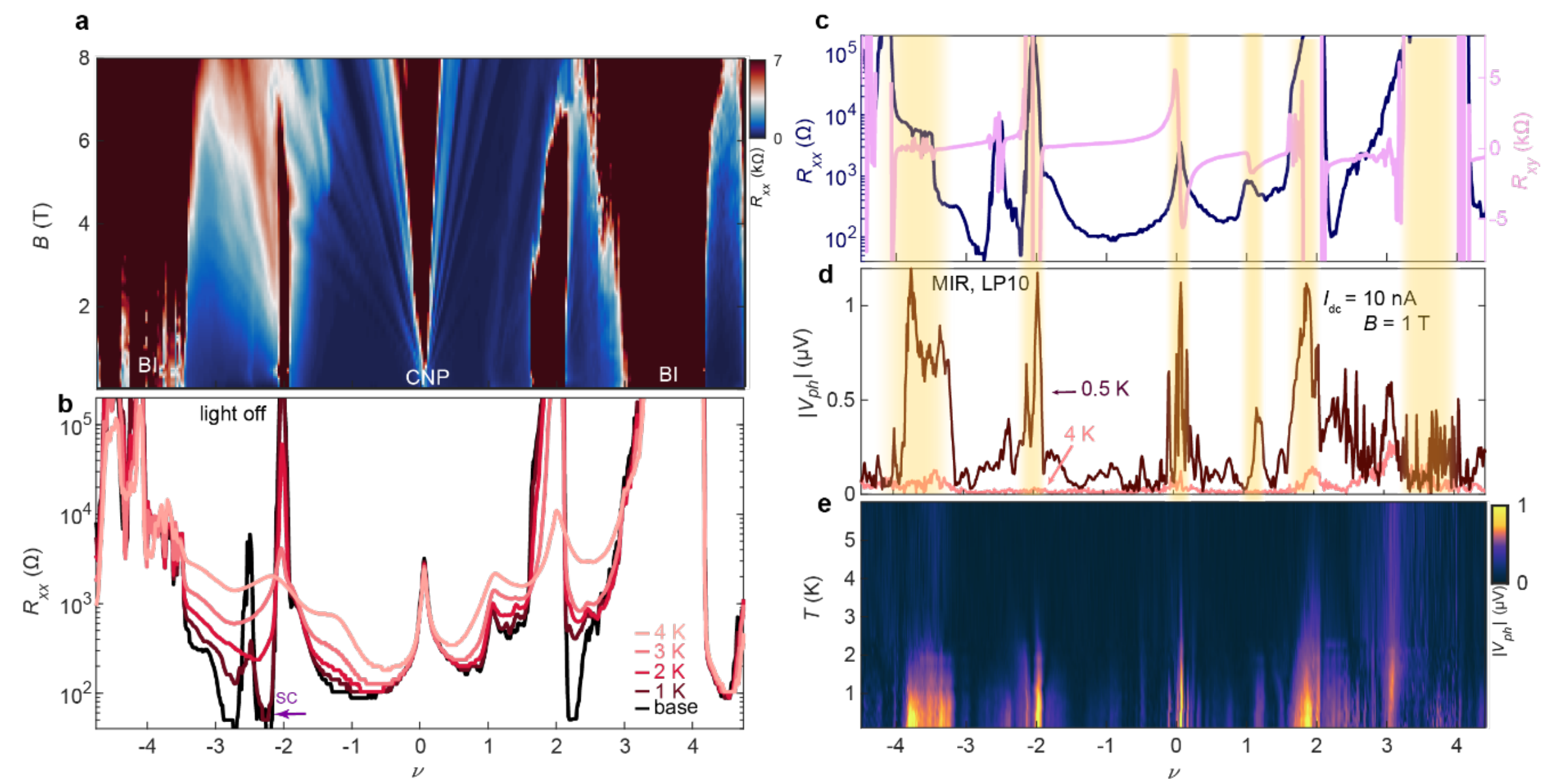

**Extended Data Figure 2. Electronic and optical responses of MATBG device D2. a** & **b**, Magnetic field and temperature dependent transport measurements (respectively), presenting the longitudinal resistance ($R_{xx}$) as a function of filling factor ($\nu$) with the light off. The magnetoresistance map in **a** is taken at the base $T$. **c,** $R_{xx}$ and $R_{xy}$ curves (dark blue and violet respectively), taken with the light off, at the base $T$ and $B$ = 1 T. **d,** Photovoltage signal as a function of filling $\nu$, taken at the $T$ = 4 K (coral) and $T$ = 0.5 K (dark brown). The signals are measured at $B$ = 1 T and with 10 nA d.c. current applied. The MATBG insulating states are marked by yellow shades at $\nu$ = 0, 1, ±2, ±4. **e**, Temperature dependence of the photovoltage signal at different fillings.

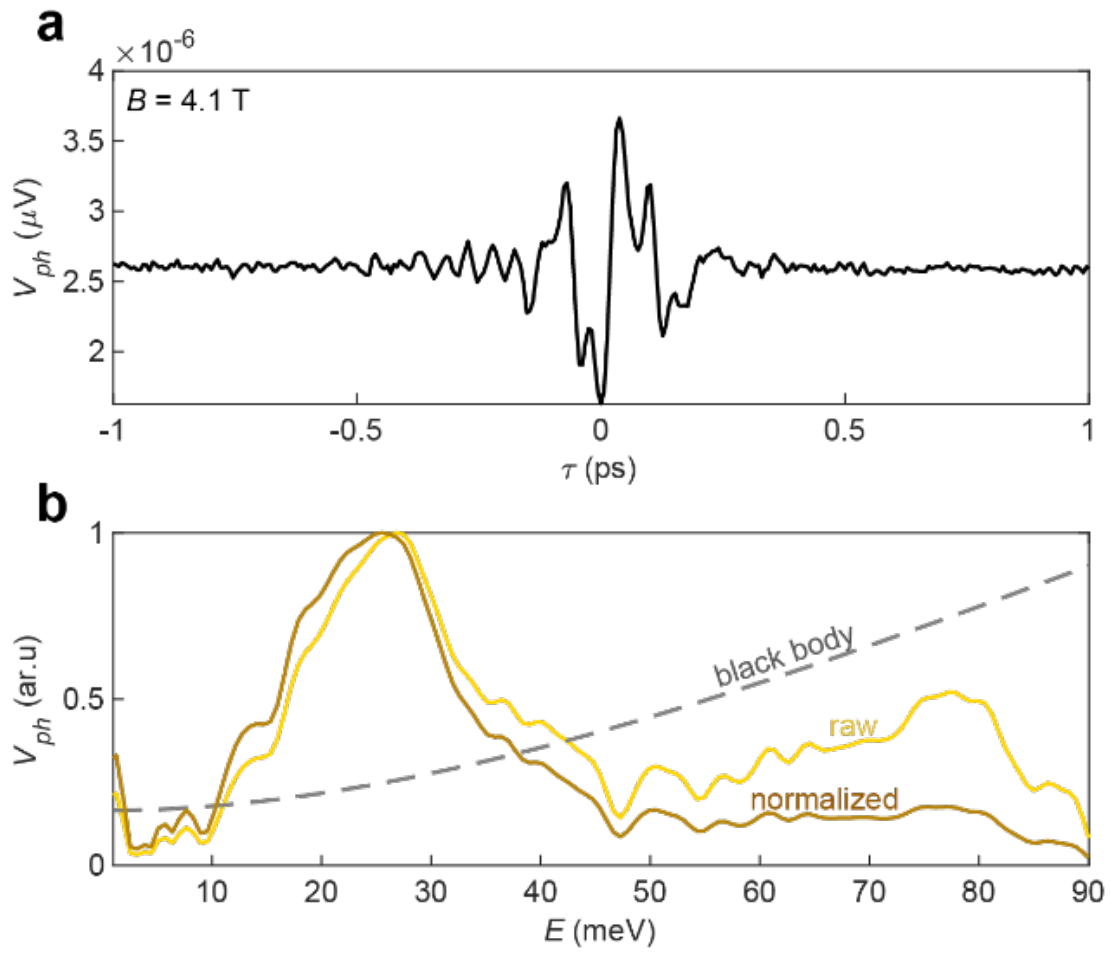

**Extended Data Figure 3. Spectrum normalization by the blackbody radiation a,** A representative CNP photovoltage interferogram (device D2), measured at base $T$ and $B$ = 4.1 T. **b,** Normalized (brown) and unnormalized (gold) FIR spectra, obtained via the Fourier transformation of the interferogram plotted in **a**. The normalization is performed according to the light source black body radiation ($T$ = 1200 K), plotted as the dash gray line.

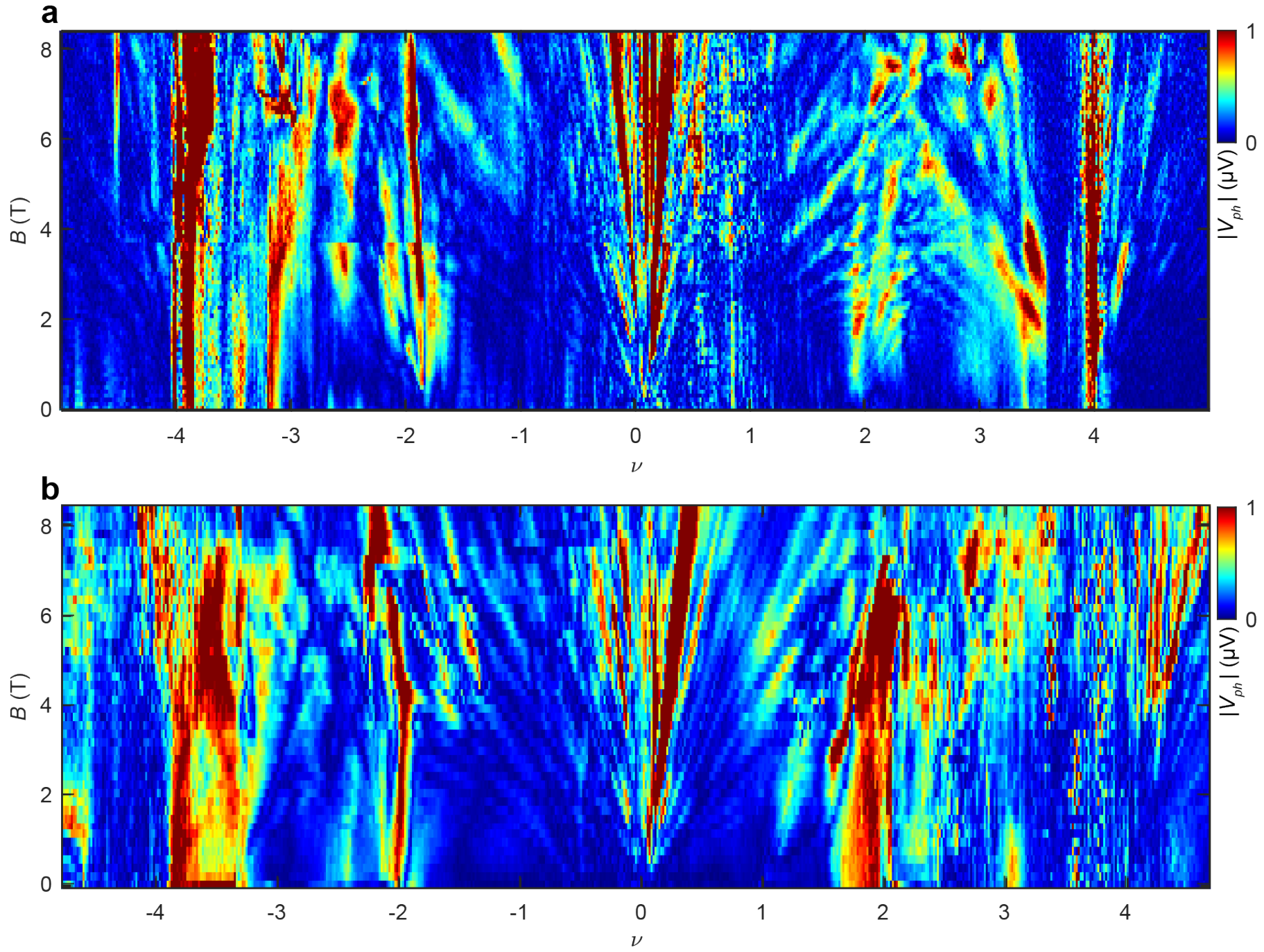

**Extended Data Figure 4. Photovoltage magnetic field maps of device D1/D2 (a/b).** Photovoltage as a function of $B$ and $\nu$, measured at the base $T$ with a 15 nA applied d.c. current. The signals are recorded using a LP10 filter and Mylar beamsplitter.

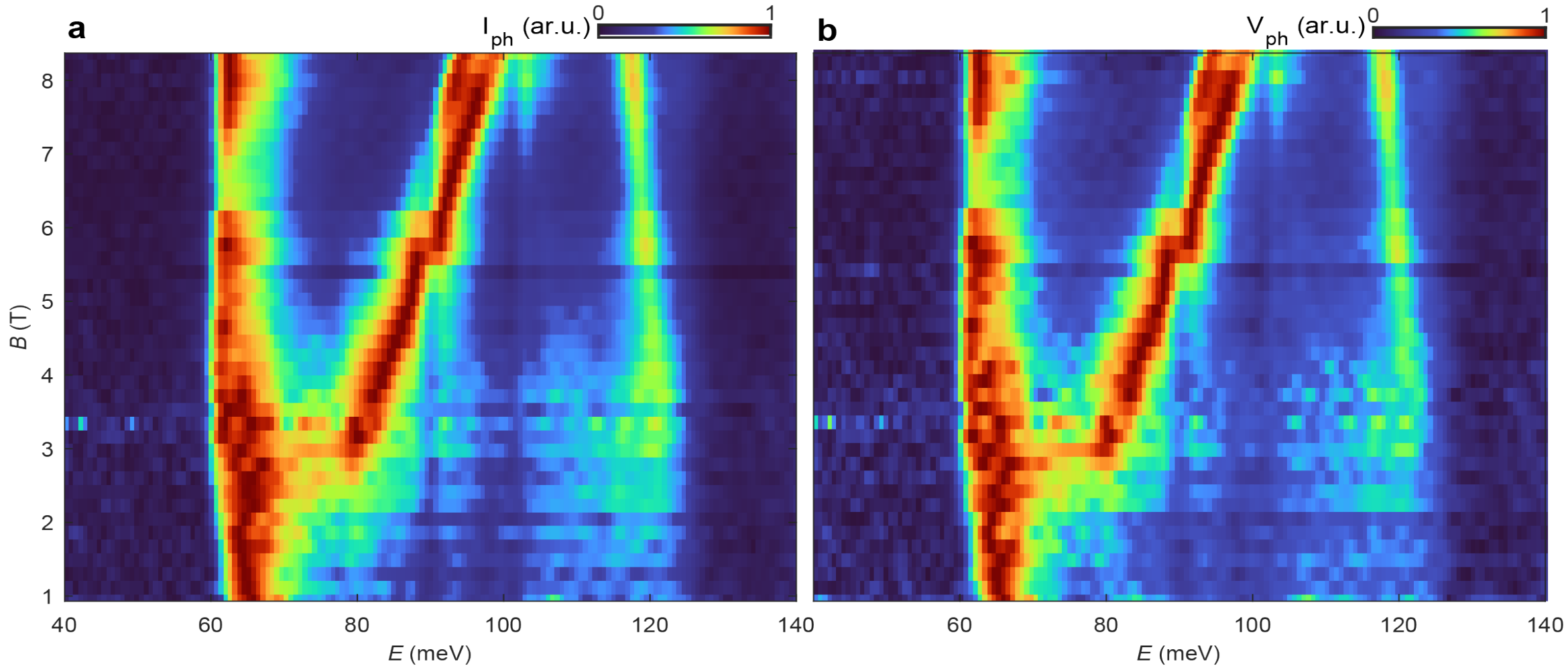

**Extended Data Figure 5. Consistent spectra taken via photocurrent (a) and photovoltage (b) signals.** Data were taken in device D1 at $\nu = +4$, during the second cooldown. The photo-induced signals were measured at the base $T$ with a 25 nA applied d.c. current, using LP10 filter (placed externally to the fridge) and KBr beamsplitter. The FIR spectrum is normalized by its maximum value for each magnetic field value. Compared to the first cooldown (see **Fig. 3a** and **Extended Data Fig. 6.**), the splitting of the $m_3$ mode (described as an additional $m_4$ mode in the main text) is more obvious while the splitting of $m_1$ mode is less clear here.

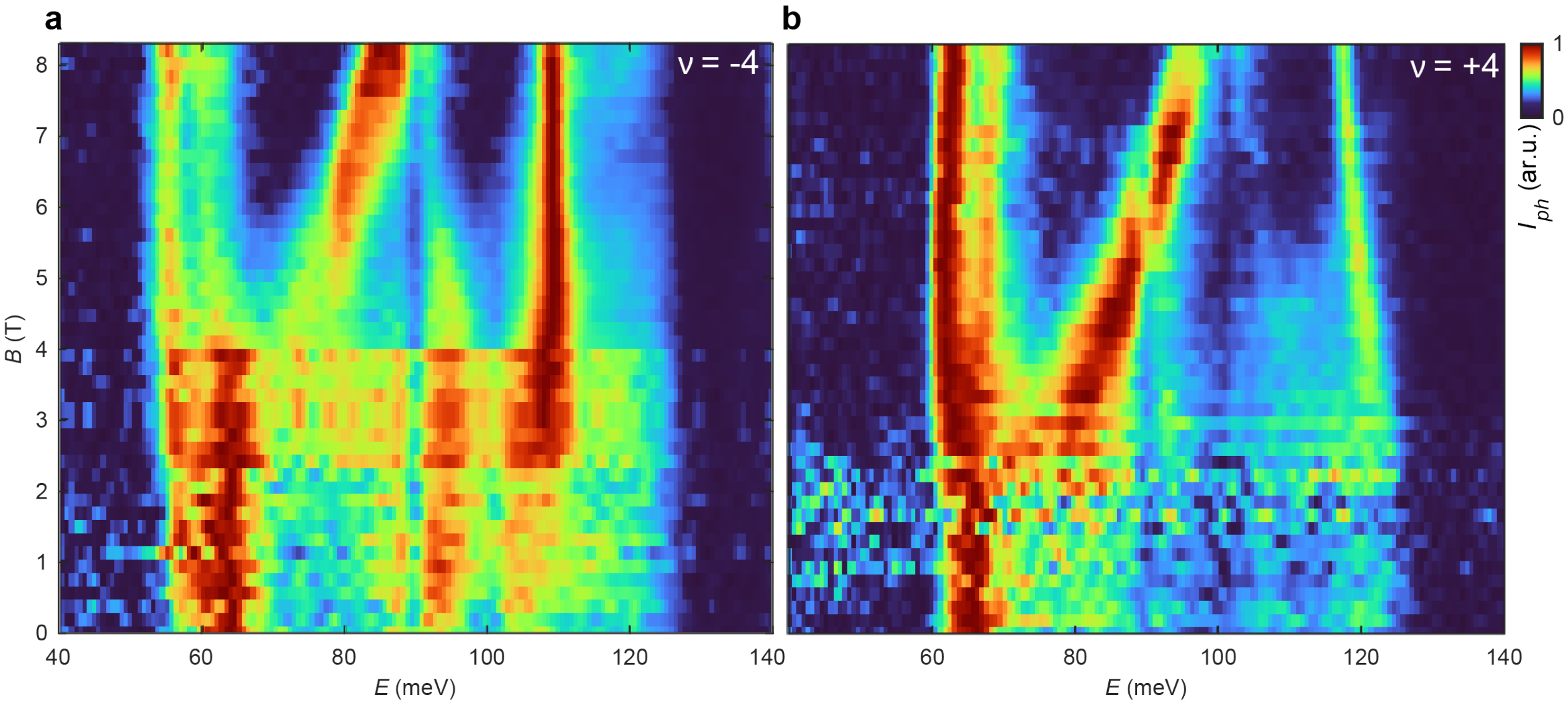

**Extended Data Figure 6. Hole- *v.s.* electron-BI spectra (a & b).** FIR spectrum (device D1) as a function of $B$, measured via photocurrent, at $\nu = \pm 4$ respectively. The spectra were taken at the base $T$ with a 25 nA applied d.c. current, using LP10 optical filter (placed inside the fridge) and KBr beamsplitter, during first cooldown. The FIR spectrum is normalized by its maximum value at each $B$. The bright high energy mode above 100 meV is also resonance of MATBG, whose origins are not the focus of this current paper. We also observe a splitting of $m_1$ mode in this cooldown, which is consistent with (but less clear in) the data taken in the second cooldown (see **Extended Data Fig. 5**). Data presented in main figures are from this cooldown.

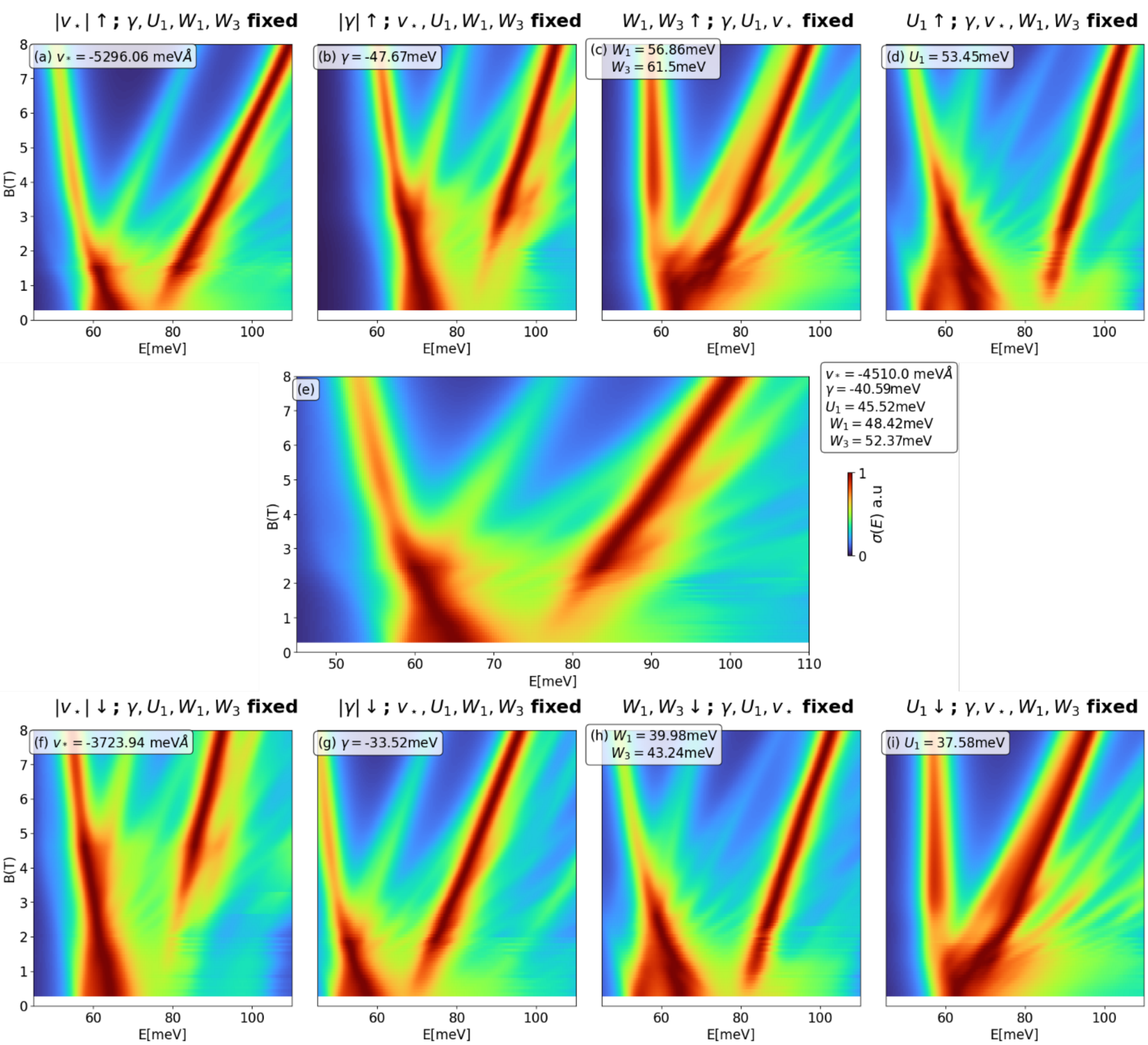

**Extended Data Figure 7. Impact of THF parameters on optical resonances.** Parameter sensitivity of the optical conductivity calculated using the Topological Heavy Fermion (THF) model at filling factor $\nu = +4$. The middle row, **e,** corresponds to the optimal parameter set determined for device D2. The top **a-d** and bottom **i-l** rows correspond to decreasing and increasing these parameters by 17%, respectively. Each column isolates the effect of a specific model parameter while keeping the rest constant: **a & f,** The **c**-electrons velocity, $\nu_*$, which primarily controls the slope of the $m_3$ mode. **b & g,** The hybridization strength $\gamma$, which governs the threshold for the onset of visible transitions. **c & h,** The repulsion between **f-** and **c**-electrons, $W_1$ and $W_3$, which modify the slope of the $m_1$ mode and the energy intercept of the $m_2$ mode. **d & i,** The Hubbard interaction on f-electrons, $U_1$, which controls the zero-field splitting between the $m_1$ and $m_3$ modes as well as the slope of the $m_1$ mode. (See **Supplementary Materials** for full parameter details).

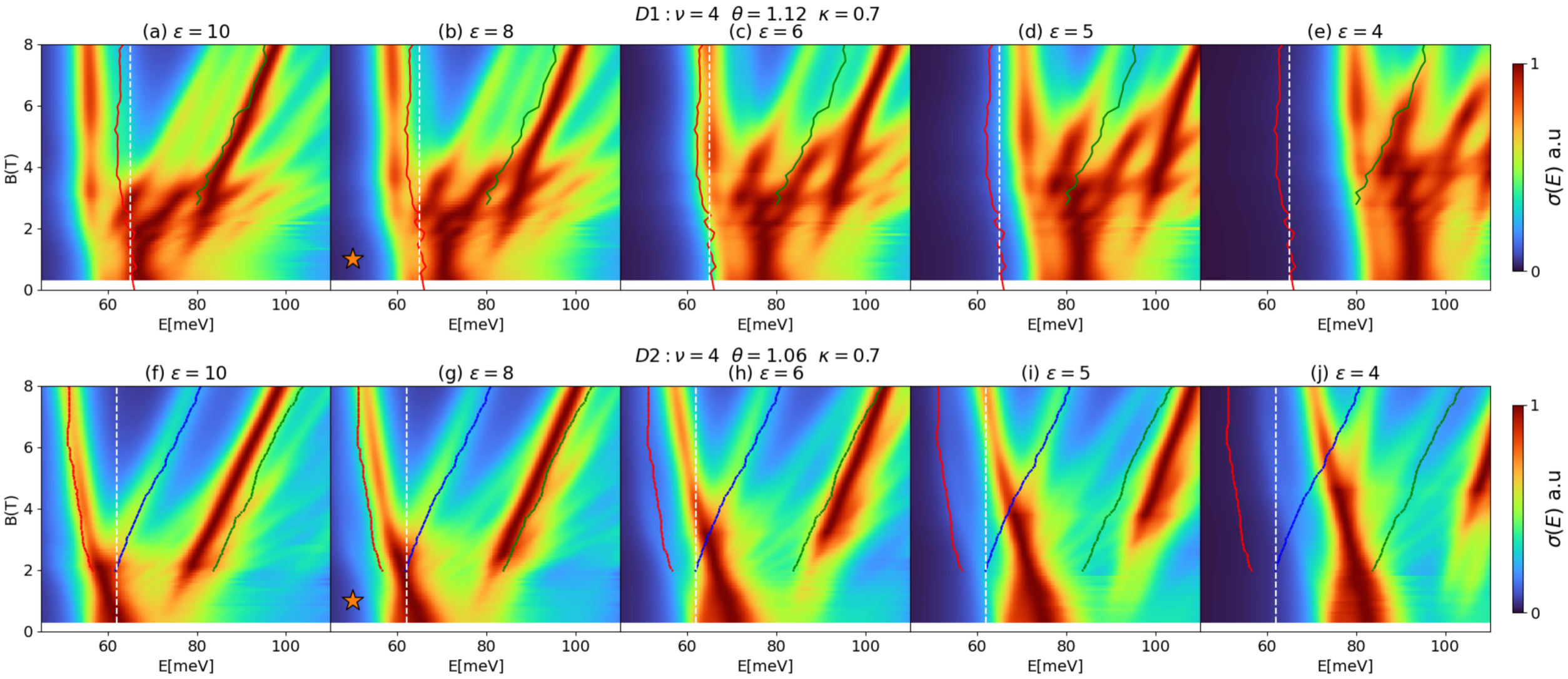

**Extended Data Figure 8. Calculating optical conductivity based on the THF model.** Shown are the results for device D1 (top row, **a**–**e**) and device D2 (bottom row, **f**–**j**) as the dielectric constant ε varies from 4 to 10, while the inter sublattice hopping ratio is fixed at $\varkappa = 0.7$. The orange star in panels **b** and **g** indicates the optimal value of ε that minimizes the discrepancy between the theoretical transition energies and the experimental peak positions. For details regarding the additional relaxation parameters used in this optimization, see **Supplementary Materials.**

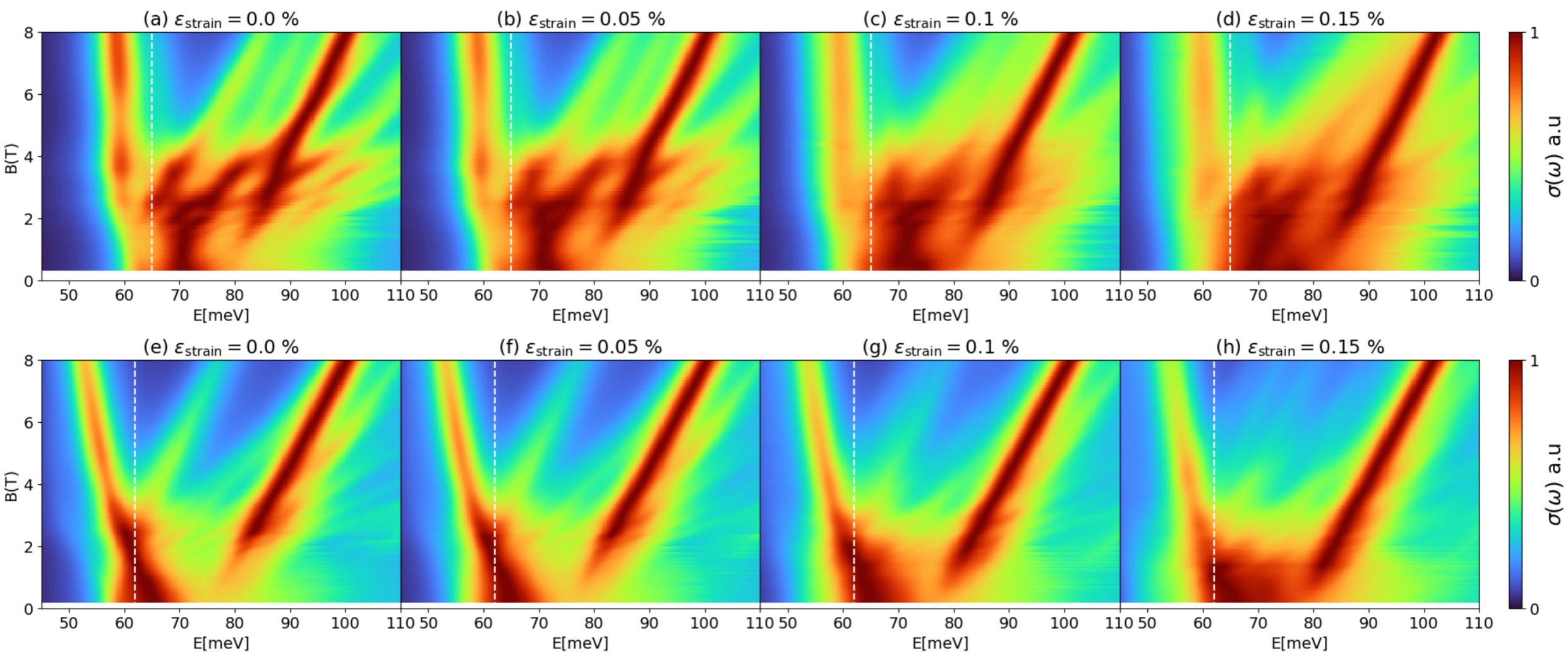

**Extended Data Figure 9. Effect of heterostrain on the optical conductivity.** Optical conductivity σ (ω) at $\nu = +4$, as a function of heterostrain, $\varepsilon_{strain}$, using the Topological Heavy Fermion (THF) model. For details on the calculation see **Supplementary Materials. a-d,** Show the evolution of σ (ω) with strain using the THF parameters consistent with D1 and **e-h** show the same for D2. In both **a-d** and **e-h** the transitions with energies in between those of $m_1$ and $m_3$ are smeared by strain. It strongly breaks the rotational symmetries and enhances LL mixing in the occupied flat bands reducing the dipole matrix elements to the lowest energy $m = 3$ LL in the empty remote bands (see **Supplementary Materials**).

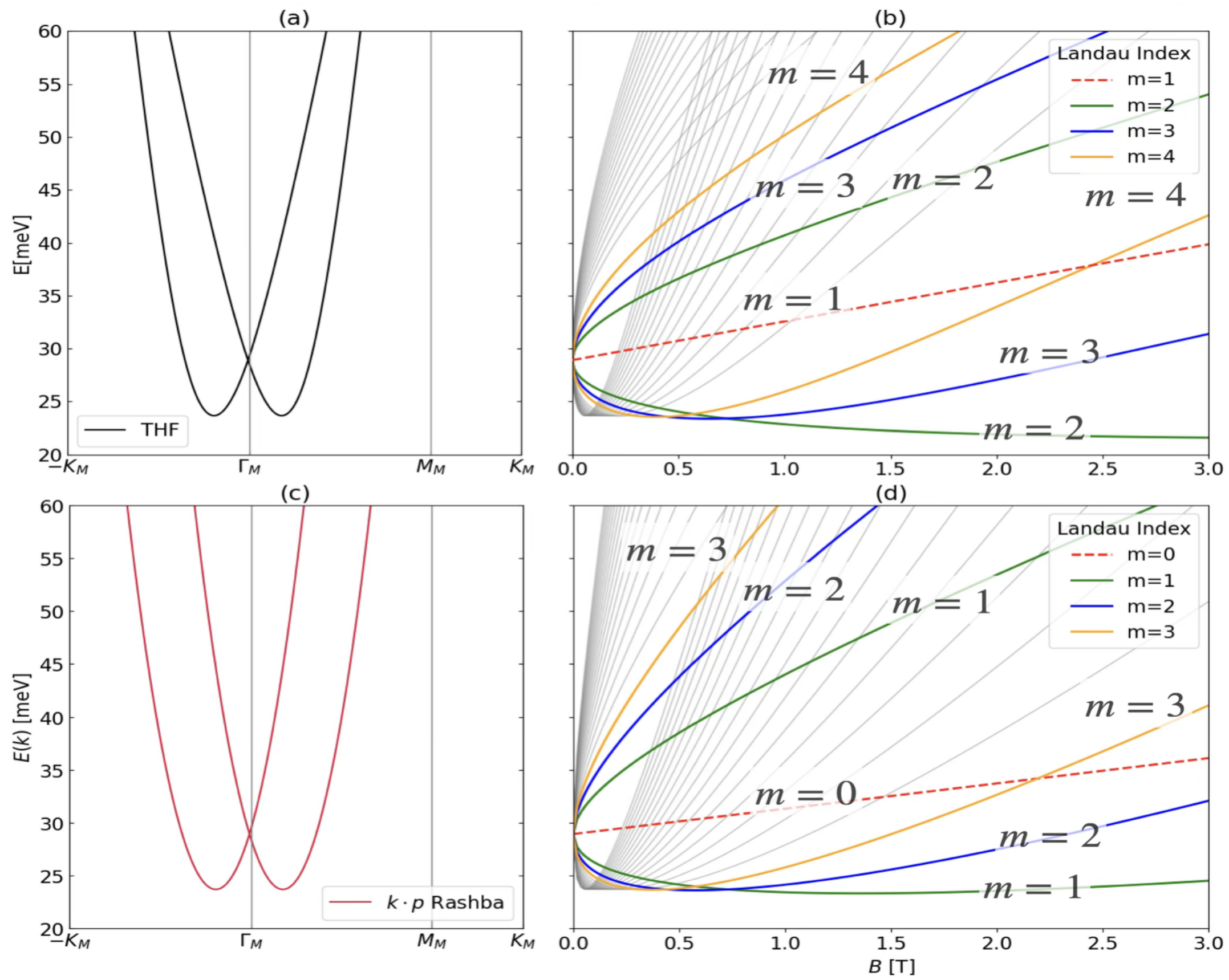

**Extended Data Figure 10. Comparison of LL sequence in MATBG *v.s.* pure Rashba bands,** Comparison of the interacting SO(2) symmetric THF model at filling $\nu = +4$ with a conventional spin Rashba model (see details in **Supplementary Materials**). **a,** Zero-field band conduction-band dispersion of the THF model around the $\Gamma_{\mathrm{M}}$ point. **b,** Landau level spectrum of the THF model as a function of magnetic field $B$. The first few levels are highlighted, showing an indexing starting at $m = 1$. **c,** Band dispersion of a naïve $k{\cdot}p$ Rashba expansion around $\Gamma_{\mathrm{M}}$, which reproduces the low-energy dispersion features of **a**. **d,** Landau level spectrum of the naïve $k{\cdot}p$ model. Note that while the energy dispersions are similar, the Landau level indexing differs by 1 (starting at $m = 0$ for the naïve expansion *v.s.* $m = 1$ for the THF model), reflecting the difference in angular momentum between the THF basis and the trivial expansion.

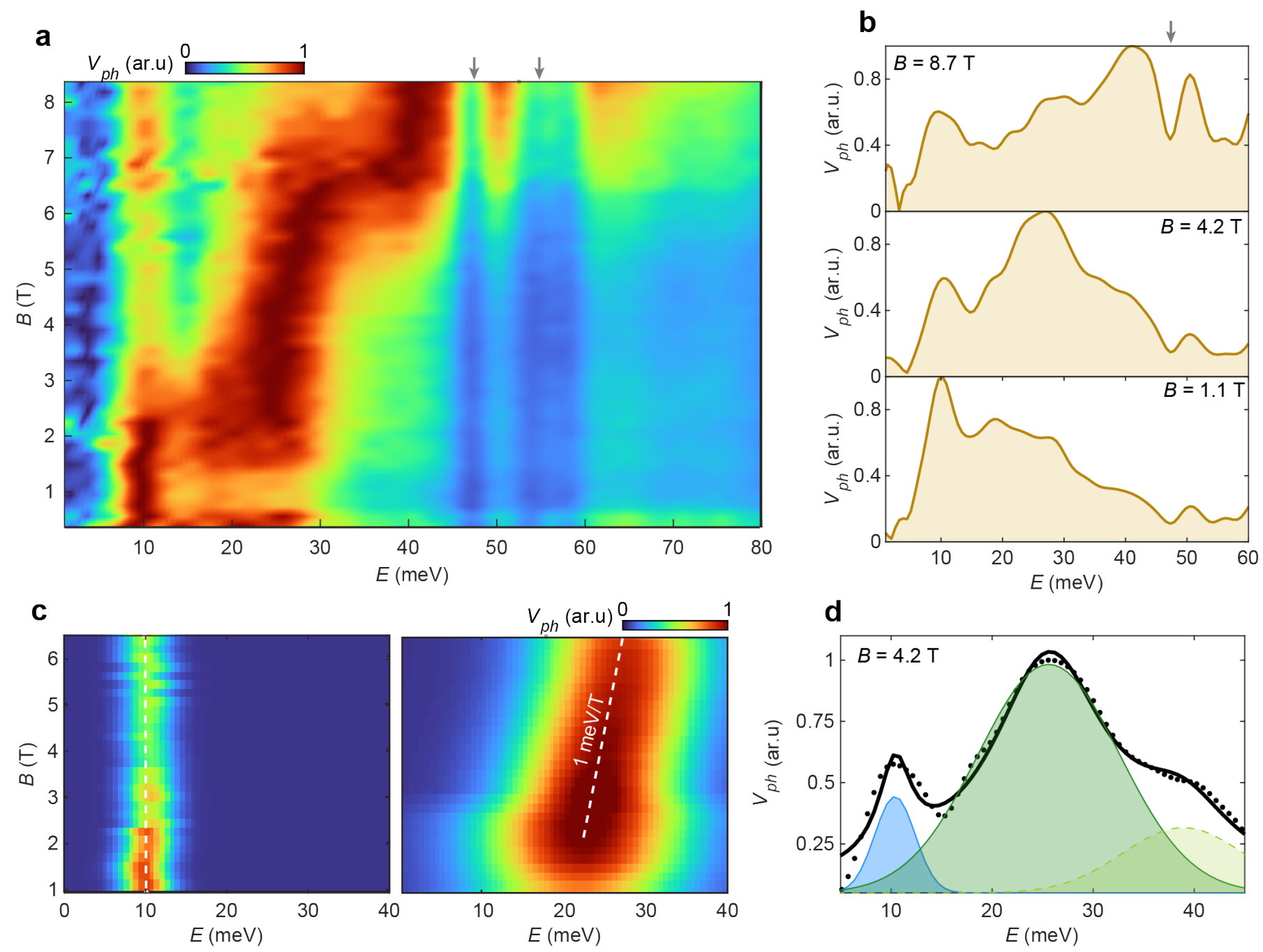

**Extended Data Figure 11. The magnetic field dependence of the CNP spectrum (device D3).** **a**, FIR excitation spectra as a function of $B$, resolved via photovoltage at base $T$. The spectrum is measured with Mylar BS and normalized by its maximum value for each $B$. The BS absorption dips are marked by gray arrows. **b**, Representative spectra, at $B$ = 1 T, 4.2 T and 8.7 T respectively. **c**, Gaussian decomposition of the excitation spectra, identifying two low energy excitations (at 10 meV and 25 meV) and their magnetic-field evolution (left and right maps). The white dash lines indicate the shift (right) or absence of shift (left) of the resonance energy with $B$. **d**, Demonstration of the Gaussian decomposition for the excitation spectrum resolved at 4.2 T. The resonances shown in the left/right maps in **c** correspond to the blue/green modes.

| Device | $\lambda$ | $\nu_*$ | $\nu'_*$ | $\gamma$ | $M$ | $U_1$ | $U_2$ | $W_1$ | $W_3$ | $J$ | $\mu_1$ | $\mu_2$ |
|---|---|---|---|---|---|---|---|---|---|---|---|---|
| D1 | 0.359 | -4658.0 | 1671.0 | -49.66 | 3.25 | 48.49 | 4.28 | 62.74 | 65.96 | 16.10 | 10.8 | 5.0 |
| D2 | 0.353 | -4510.0 | 1631.0 | -40.59 | 2.04 | 45.52 | 3.07 | 48.42 | 52.37 | 14.54 | 28.8 | 5.0 |

**Extended Data Table 1: Extracted Parameters of the THF model for D2/D1 devices.** Summary of the model parameters. The localization length $\lambda$ of the **f**-electrons Wannier function governs the decay of **f**-**c** hybridization at large momenta. The velocity of **c**-electrons in the absence of hybridization is denoted by $\nu_*$. The parameter $\nu'_*$ represents momentum-dependent **f**-**c** hybridization, setting the slope of the dispersion near the Rashba point (where $\nu'_* = 0$ corresponds to the chiral limit). The parameter $\gamma$ defines the **f**-**c** hybridization at the $\Gamma_M$ point, such that $2\gamma$ determines the energy difference between the Rashba points of the two remote bands. $M$ represents the splitting of the $\Gamma_1$ and $\Gamma_2$ **c**-electrons, with $2M$ setting the bandwidth of the non-interacting bands. Interaction terms include the on-site ($U_1$) and nearest-neighbor ($U_2$) Hubbard repulsion of the **f**-electrons. Finally, $W_1$ and $W_3$

describe the repulsion between **f**-electrons and the $\Gamma_3$ and $\Gamma_1$ **c**-electrons, respectively, while $J$ represents the exchange interaction between **f** and $\Gamma_1$ **c**-electrons. The atomic relaxation can be efficiently captured by the parameters $\mu_1$ and $\mu_2$, which correspond to local chemical potential shifts to the $\Gamma_3$ ($\mu_1$) and $\Gamma_1 \oplus \Gamma_2$ ($\mu_2$) **c-**electrons (see details in **Supplementary Materials**).

## Supplementary Information
## Resonant Far-Infrared Spectroscopy of Flat-Band Fermions in Magic Angle Graphene

### Appendix A: Bare Hamiltonian

In this section we provide details of the non-interacting Hamiltonian of the Topological Heavy Fermion (THF) model. As shown in the model introduced in Ref. [1], in the absence of strain and lattice relaxation the six bands closest to charge neutrality in TBG transform as the $\Gamma_1 \oplus \Gamma_2 \oplus 2\Gamma_3$ representation at the $\Gamma_M$ point. A set of localized $p_x \pm ip_y$ orbitals at the AA stacking sites accounts for one $\Gamma_3$ representation per valley and spin. To capture the nontrivial band topology of the system [2], four additional $c$-electron states are introduced, contributing the remaining $\Gamma_1 \oplus \Gamma_2 \oplus \Gamma_3$ representations per valley and spin. Projecting the Bistritzer–MacDonald continuum model onto this basis [3] yields the following non-interacting Hamiltonian:

$$\hat{H}_0 = \sum_{\substack{|\mathbf{k}|\leq\Lambda_c \\ \eta,s}} \sum_{a,a'} h^{cc,\eta}_{aa'}(\mathbf{k})\, \hat{c}^{\dagger}_{\mathbf{k},a,\eta,s}\hat{c}_{\mathbf{k},a',\eta,s} + \sum_{\substack{|\mathbf{k}|\leq\Lambda_c \\ \eta,s}} \sum_{a,\alpha} h^{cf,\eta}_{a\alpha}(\mathbf{k})\, \hat{c}^{\dagger}_{\mathbf{k},a,\eta,s}\hat{f}_{\mathbf{k},\alpha,\eta,s} + \sum_{a,\alpha} \overline{h^{cf,\eta}_{a\alpha}(\mathbf{k})}\hat{f}^{\dagger}_{\mathbf{k},\alpha,\eta,s}\hat{c}_{\mathbf{k},a,\eta,s}, \tag{A.1}$$

where $\hat{c}^{\dagger}_{\mathbf{k},a,\eta,s}$ creates a $c$-electron at momentum $\mathbf{k}$, orbital $a = 1,2,3,4$, valley $\eta$, and with spin $s$. We adopt the convention that $a = 1,2$ corresponds to the $\Gamma_3$ $c$-electrons and $a = 3,4$ corresponds to the $\Gamma_1 \oplus \Gamma_2$ $c$-electrons. In the above, $\hat{f}^{\dagger}_{\mathbf{k},\alpha,\eta,s}$ creates an $f$-electron at orbital $\alpha = 1,2$, valley $\eta$, and with spin $s$. The matrix elements in Eq. (A.1) are:

$$h^{cc,\eta}_{aa'}(\mathbf{k}) = \begin{pmatrix} 0 & v_\star\left(\eta k_x\sigma_0 + ik_y\sigma_z\right) \\ v_\star\left(\eta k_x\sigma_0 - ik_y\sigma_z\right) & M\sigma_x \end{pmatrix}_{aa'} \tag{A.2}$$

$$h^{cf,\eta}_{a\alpha}(\mathbf{k}) = \begin{pmatrix} \gamma\sigma_0 + v'_\star\left(\eta k_x\sigma_x + k_y\sigma_y\right) \\ 0 \end{pmatrix}_{a\alpha} e^{-\lambda^2|\mathbf{k}|^2/2}. \tag{A.3}$$

Here, $M$ sets the splitting of the $\Gamma_1 \oplus \Gamma_2$ $c$-electrons, $v_\star$ is the velocity of the $c$-electrons in the absence of hybridization, $\gamma$ sets the gap at the $\Gamma_M$ point between the flat and remote bands, $v'_\star$ sets the slope of the Rashba point in the remote bands, and $\lambda$ denotes the localization length of the $f$-electrons, which dictates the decay of their hybridization with the $c$-electrons at large momenta. In Sec. (I) we detail the exact parameters that we use to match the experiment; for now, we leave them undefined. The $f$-electrons have a negligible dispersion near the magic angle, and thus we neglect $h^{ff,\eta}_{\alpha\alpha'}(\mathbf{k})$. For conciseness, we denote the total number of electron flavors by $N_f = 4$, accounting for spin and valley degeneracy.

#### 1. Relaxation and strain

The experiment shows clear signatures of particle-hole symmetry breaking between hole and electron photocurrent spectra (see Fig.2 in the main text). Most notably, the peaks in the far-infrared (FIR) spectra at filling $\nu = +4$ are located at energies higher than those at $\nu = -4$. To capture this, we use the minimal relaxation model detailed in Ref.([4, 5]). Within this model, relaxation shifts the chemical potentials of the $\Gamma_1 \oplus \Gamma_2$ and $\Gamma_3$ $c-$electrons separately, such that the relaxed Hamiltonian includes the additional term

$$\hat{H}_{\text{rel}} = \sum_{\substack{|\mathbf{k}|\leq\Lambda_c \\ \eta,s}} \sum_{a,a'} h^{\text{rel},cc}_{aa'}(\mathbf{k})\, \hat{c}^{\dagger}_{\mathbf{k},a,\eta,s}\hat{c}_{\mathbf{k},a',\eta,s}, \tag{A.4}$$

where

$$h^{\text{rel},cc}_{aa'}(\mathbf{k}) = \begin{pmatrix} \mu_1\sigma_0 & 0 \\ 0 & \mu_2\sigma_0 \end{pmatrix}_{aa'}. \tag{A.5}$$

| $\kappa$ | $c$ | $c'$ | $c''$ | $M_f$ | $\gamma'$ | $M'$ | $\mu_1$ | $\mu_2$ |
|---|---|---|---|---|---|---|---|---|
| 0.6 | -8819 | 2153.58 | -3074 | 4838.3 | -4853 | -5842 | 13.8 | 5.6 |
| 0.7 | -8804 | 2152.6 | -3261 | 4572.39 | -4085 | -5242 | 14.4 | 5.0 |
| 0.8 | -8750 | 2050 | -3362 | 4380 | -3352 | -4580 | 14.1 | 4.5 |

Table I. Strain and relaxation parameters of the THF model in meV projecting the BM model including non-local hoppings and heterostrain. We list these parameters for a range of realistic values of the inter-sublattice hopping ratio $\kappa = 0.6 - 0.8$ at twist angle $\theta = 1.05°$ ( See Ref.([4, 5]) )

.

In Sec.(I) we will fix the relaxation parameters $\mu_{1,2}$; for now, we leave them unspecified. We also test the robustness of our calculations against variations in strain. To include strain in the THF model, Refs. [4, 5] introduced the strain correction:

$$\begin{aligned}\hat{H}_{\text{strain}} = &\sum_{\substack{|\mathbf{k}|\leq\Lambda_c\\ \eta,s}} \sum_{a,a'} h^{\text{str},cc,\eta}_{aa'}(\mathbf{k})\, \hat{c}^\dagger_{\mathbf{k},a,\eta,s}\hat{c}_{\mathbf{k},a',\eta,s} \\ &+ \sum_{\substack{|\mathbf{k}|\leq\Lambda_c\\ \eta,s}} \sum_{a,\alpha} h^{\text{str},cf,\eta}_{a\alpha}(\mathbf{k})\, \hat{c}^\dagger_{\mathbf{k},a,\eta,s}\hat{f}_{\mathbf{k},\alpha,\eta,s} + \sum_{a,\alpha} \overline{h^{\text{str},cf,\eta}_{a\alpha}(\mathbf{k})}\hat{f}^\dagger_{\mathbf{k},\alpha,\eta,s}\hat{c}_{\mathbf{k},a,\eta,s} \\ &+ \sum_{\substack{|\mathbf{k}|\leq\Lambda_c\\ \eta,s}} \sum_{\alpha,\alpha'} h^{\text{str},ff,\eta}_{\alpha,\alpha'}(\mathbf{k})\, \hat{f}^\dagger_{\mathbf{k},\alpha,\eta,s}\hat{f}_{\mathbf{k},\alpha',\eta,s} \end{aligned} \tag{A.6}$$

where the matrix elements are given by:

$$h^{\text{str},cc,\eta}_{aa'}(\mathbf{k}) = \begin{pmatrix} c\eta\varepsilon_-\sigma_{\text{y}} & -\eta c'\varepsilon_-\sigma_{\text{y}} \\ -\eta c'\varepsilon_-\sigma_{\text{y}} & \eta M'\varepsilon_+\sigma_{\text{y}} \end{pmatrix}_{aa'} \tag{A.7a}$$

$$h^{\text{str},cf,\eta}_{a\alpha}(\mathbf{k}) = \begin{pmatrix} i\eta\gamma'\varepsilon_+\sigma_{\text{z}} \\ -i\eta c''\varepsilon_-\sigma_{\text{z}} \end{pmatrix}_{a\alpha} \tag{A.7b}$$

$$h^{\text{str},ff,\eta}_{\alpha,\alpha'}(\mathbf{k}) = \begin{pmatrix} \eta M_f\varepsilon_-\sigma_{\text{y}} \end{pmatrix}_{\alpha\alpha'}. \tag{A.7c}$$

Typical strain and relaxation parameters for TBG near the magic angle are listed in Tab.(I). The $C_{3z}$-breaking heterostrain is parameterized by the strain tensor $\boldsymbol{\varepsilon}$ with components $\varepsilon_{ij}$. Following the same conventions of Refs.([4, 5]), we express the strain in terms of the combinations $\varepsilon_{xy}, \varepsilon_-$, and $\varepsilon_+$:

$$\varepsilon_\pm = \frac{\varepsilon_{xx} \pm \varepsilon_{yy}}{2}. \tag{A.8}$$

Here, $\varepsilon_{xy}$ corresponds to the shear component, $\varepsilon_-$ describes the anisotropic strain, and $\varepsilon_+$ represents the isotropic dilation. For uniaxial heterostrain applied along the $x$ direction, we take $\varepsilon_{xy} = 0$ and parameterize the tensor using a single strain amplitude $\varepsilon_{\text{strain}}$ together with the Poisson ratio of graphene, $\nu_G = 0.16$. In this case, the tensor components become

$$\varepsilon_{xx} = -\varepsilon_{\text{strain}}, \tag{A.9}$$

$$\varepsilon_{yy} = \nu_G\,\varepsilon_{\text{strain}}, \tag{A.10}$$

which implies

$$\varepsilon_\mp = -\frac{1 \pm \nu_G}{2}\,\varepsilon_{\text{strain}}. \tag{A.11}$$

In Sec.(J) we use the values in Tab.(I) to show how the optical conductivity maps change with varying $\varepsilon_{\text{strain}}$ as heterostrain is included phenomenologically.

## Appendix B: Current operator

In this section, we provide a brief discussion of the current operator at $B = 0$, which will later be used in Sec.(E) to obtain the finite $B$ current operator. The particle current operator of the THF model was first obtained in Ref.([6]),

and is given by:

$$\hat{J}^\eta_\mu = \sum_{\substack{|\mathbf{k}|\leq\Lambda_c \\ \eta,s}} \left[\sum_{a,a'} \left[\partial_\mu h^{cc,\eta}_{aa'}(\mathbf{k})\right] \hat{c}^\dagger_{\mathbf{k},a,\eta,s}\hat{c}_{\mathbf{k},a',\eta,s} + \sum_{a,\alpha}\left[\partial_\mu h^{cf,\eta}_{a\alpha}(\mathbf{k})\, \hat{c}^\dagger_{\mathbf{k},a,\eta,s}\hat{f}_{\mathbf{k},\alpha,\eta,s} + \text{h.c}\right]\right]. \tag{B.1}$$

In obtaining Eq. (B.1), Ref.([6]) approximates the full current operator by neglecting two types of contributions: (1) the internal structure of the $f$-electron wavefunctions, and (2) the $f$-electron hopping. The first contribution is proportional to the $f$-electron Berry connection, which is bounded by a quantity of order $\lambda^3/a_M^2$ (where $\lambda$ is the extent of the $f$-electron Wannier function and $a_M$ is the moiré length). Since $\lambda$ is roughly 20–30% of the moiré length, the Berry connection is parametrically small. In practice, this means the dipole matrix elements between $f$-electron states are at least $10^{-2}$ smaller than the $c$–$c$ and $c$–$f$ currents, as estimated in Ref.([6]). Since the hopping of the $f$-electrons vanishes near the magic angle, this contribution is also sub-leading. Evaluating the derivatives of the Hamiltonian leads to

$$\hat{J}_\mu = \sum_{\substack{|\mathbf{k}|\leq\Lambda_c \\ \eta,s}} \left[\sum_{a,a'} J^{\mu;cc,\eta}_{aa'}\hat{c}^\dagger_{\mathbf{k},a,\eta,s}\hat{c}_{\mathbf{k},a',\eta,s} + \sum_{a,\alpha} J^{\mu cf,\eta}_{a\alpha}\hat{c}^\dagger_{\mathbf{k},a,\eta,s}\hat{f}_{\mathbf{k},\alpha,\eta,s} + \text{h.c}\right] \tag{B.2}$$

where the matrix elements of the $x$ component of the current are

$$J^{\mu=\mathrm{x};cc,\eta}_{aa'} = \eta \begin{pmatrix} 0 & v_\star\sigma_0 \\ v_\star\sigma_0 & 0 \end{pmatrix}_{aa'} \quad J^{\mu=\mathrm{x};cf,\eta}_{a\alpha} = \left(\eta \begin{pmatrix} v'_\star\sigma_\mathrm{x} \\ 0 \end{pmatrix}_{a\alpha} - \lambda^2 \begin{pmatrix} \frac{\gamma}{\hbar}\sigma_0 + v'_\star(\eta k_\mathrm{x}\sigma_\mathrm{x} + k_\mathrm{y}\sigma_\mathrm{y}) \\ 0 \end{pmatrix}_{a\alpha} k_\mathrm{x}\right) e^{-\lambda^2|\mathbf{k}|^2/2} \tag{B.3}$$

and for the $y$ component of the current:

$$J^{\mu=\mathrm{y};cc,\eta}_{aa'} = \begin{pmatrix} 0 & iv_\star\sigma_\mathrm{z} \\ -iv_\star\sigma_\mathrm{z} & 0 \end{pmatrix}_{aa'}, \quad J^{\mu=\mathrm{y};cf,\eta}_{a\alpha} = \left(\begin{pmatrix} v'_\star\sigma_\mathrm{y} \\ 0 \end{pmatrix}_{a\alpha} - \lambda^2 \begin{pmatrix} \frac{\gamma}{\hbar}\sigma_0 + v'_\star(\eta k_\mathrm{x}\sigma_\mathrm{x} + k_\mathrm{y}\sigma_\mathrm{y}) \\ 0 \end{pmatrix}_{a\alpha} k_\mathrm{y}\right) e^{-\lambda^2|\mathbf{k}|^2/2}. \tag{B.4}$$

Note that only two classes of current matrix elements are nonzero: those connecting $\Gamma_1 \oplus \Gamma_2$ $c$-electrons with $\Gamma_3$ $c$-electrons, and those connecting $\Gamma_3$ $c$-electrons with $f$-electrons. The matrix element between $\Gamma_1 \oplus \Gamma_2$ and $\Gamma_3$ $c$-electrons is proportional to $v_\star$, whereas the matrix element between $\Gamma_3$ $c$- and $f$-electrons is proportional to $v'_\star$. For realistic parameters at $\theta = 1.05°$, the ratio $v'_\star/v_\star$ lies in the range 0.34–0.38 (see Ref. [1]). Because the current matrix elements enter quadratically in the Kubo formula (see Sec.(F)), the $c$–$c$ contribution to the optical conductivity is expected to be larger than the $c$–$f$ contribution by a factor of roughly $(v_\star/v'_\star)^2 \approx 7$–$9$.

We again note the absence of an $f$–$f$ current matrix element, so that $f$–$f$ excitations are "dark" even though they have a large joint density of states. Note also that in the simplified model for relaxation and strain in Eq. (A.4) and Eq.(A.6), the matrix elements of the Hamiltonian are momentum-independent and thus do not contribute to the current operator $\hat{J}^\eta_\mu$.

## Appendix C: Symmetries of the strain-less THF model

Under a spatial or time-reversal symmetry operation $g$, the creation operator transforms as:

$$\hat{g}\, c^\dagger_{\mathbf{k},a\eta s}\, \hat{g}^{-1} = \sum_{a'\eta'} c^\dagger_{g\mathbf{k},a'\eta' s} D^c_{a'\eta',a\eta}(g), \tag{C.1}$$

where $D^c(g)$ is the unitary matrix representation of the symmetry $g$ in the internal orbital-valley space. The representations for time-reversal ($T$), three-fold rotation ($C_{3z}$), two-fold rotation ($C_{2x}$), and $C_{2z}T$, are

$$D^c(T) = \sigma_0\tau_\mathrm{x} \oplus \sigma_0\tau_\mathrm{x}, \quad D^c(C_{3z}) = e^{i\frac{2\pi}{3}\sigma_z\tau_z} \oplus \sigma_0\tau_0, \quad D^c(C_{2x}) = \sigma_\mathrm{x}\tau_0 \oplus \sigma_\mathrm{x}\tau_0, \quad D^c(C_{2z}T) = \sigma_\mathrm{x}\tau_0 \oplus \sigma_\mathrm{x}\tau_0, \tag{C.2}$$

where we have introduced the valley Pauli matrices $\tau_j$. Similarly, for the $f$-electrons the transformation is

$$\hat{g} f^\dagger_{\mathbf{k},a\eta s}\hat{g}^{-1} = \sum_{a'\eta'} f^\dagger_{g\mathbf{k},a'\eta' s} D^f_{a'\eta',a\eta}(g) \tag{C.3}$$

with representations

$$D^f(T) = \sigma_0\tau_\mathrm{x}, \quad D^f(C_{3z}) = e^{i\frac{2\pi}{3}\sigma_z\tau_z}, \quad D^f(C_{2x}) = \sigma_\mathrm{x}\tau_0, \quad D^f(C_{2z}T) = \sigma_\mathrm{x}\tau_0. \tag{C.4}$$

### 1. Constraints on symmetric density matrices.

Using the spatial symmetries of the THF model, we constrain the possible density matrices at the band-insulating fillings $\nu = \pm 4$ in the absence of strain. Consider the following momentum-integrated density matrices:

$$O^{f}_{\alpha\eta s,\beta\eta' s'} = \frac{1}{N}\sum_{\mathbf{k}} \left\langle f^{\dagger}_{\mathbf{k}\alpha\eta s} f_{\mathbf{k}\beta\eta' s'} \right\rangle, \tag{C.5a}$$

$$O^{cf}_{a\eta s,\beta\eta' s'} = \frac{1}{\sqrt{N}}\sum_{\mathbf{k}} e^{-i\mathbf{k}\cdot\mathbf{R}} \left\langle c^{\dagger}_{\mathbf{k}a\eta s} f_{\mathbf{R}\beta\eta' s'} \right\rangle, \tag{C.5b}$$

$$O^{c}_{a\eta s,b\eta' s'} = \frac{1}{N}\sum_{\mathbf{k}} \left( \left\langle c^{\dagger}_{\mathbf{k}a\eta s} c_{\mathbf{k}b\eta' s'} \right\rangle - \frac{1}{2}\delta_{ab}\delta_{\eta\eta'}\delta_{ss'} \right) \tag{C.5c}$$

The condition for a state to be symmetric under a symmetry operation $\hat{g}$ is:

$$O^{f}_{\alpha\eta s,\beta\eta' s'} = \sum_{\alpha'\beta'\eta_2\eta_3} D^{f}_{\alpha\eta,\alpha'\eta_3}(g) O^{f}_{\alpha'\eta_3 s,\beta'\eta_2 s'} \left[D^{f}\right]^{-1}_{\beta'\eta_2,\beta\eta'}(g) \tag{C.6a}$$

$$O^{cf}_{a\eta s,\beta\eta' s'} = \sum_{a'\beta'\eta_2\eta_3} D^{c}_{a\eta,a'\eta_3}(g) O^{cf}_{a'\eta_3 s,\beta'\eta_2 s'} \left[D^{f}\right]^{-1}_{\beta'\eta_2,\beta\eta'}(g) \tag{C.6b}$$

$$O^{c}_{a\eta s,b\eta' s'} = \sum_{a'b'\eta_2\eta_3} D^{c}_{a\eta,a'\eta_3}(g) O^{c}_{a'\eta_3 s,b'\eta_2 s'} \left[D^{c}\right]^{-1}_{b'\eta_2,b\eta'}(g) \tag{C.6c}$$

We consider states that preserve spin rotation invariance, such that the density matrix is independent of $s, s'$:

$$O^{f}_{\alpha\eta s,\beta\eta' s'} = \rho^{f}_{\alpha\eta,\beta\eta'}\delta_{ss'},\ O^{cf}_{a\eta s,\beta\eta' s'} = \rho^{cf}_{a\eta,\beta\eta'}\delta_{ss'},\ O^{c}_{a\eta s,b\eta' s'} = \rho^{c}_{a\eta,b\eta'}\delta_{ss'} \tag{C.7}$$

We decompose the density matrix into blocks corresponding to the orbital content of each irreducible representation. Consider the $4\times 4$ blocks of the density matrix that are diagonal in each flavor $\xi \in \{c, f\}$ and transform as each irrep $\Upsilon \in \{\Gamma_3, \Gamma_1\oplus\Gamma_2\}$; we label these blocks as $\rho^{\xi,\Upsilon}$, where $\rho^{f} = \rho^{f,\Gamma_3}$, and $\rho^{c} = \rho^{c,\Gamma_3} \oplus \rho^{c,\Gamma_1\oplus\Gamma_2}$. For each $\Upsilon$, the $4\times 4$ block $\rho^{\xi,\Upsilon}$ of the density matrix (in the combined orbital$\times$valley space) can be expanded in terms of tensor products of orbital ($\sigma$) and valley ($\tau$) Pauli matrices:

$$\rho^{\xi,\Upsilon} = \sum_{i,j=0,\mathrm{x},\mathrm{y},\mathrm{z}} P^{\xi,\Upsilon}_{ij}\,\sigma_i\tau_j. \tag{C.8}$$

We impose the symmetry constraints term by term. The anti-unitary operator $C_{2z}T$ involves complex conjugation, and so the combination of $C_{2z}T$ and $C_{2x}$ restricts $\rho^{\xi,\Upsilon}$ to be a purely real matrix, forbidding any basis matrices with an odd number of $y$-Pauli matrices (e.g., $\sigma_y\tau_0$). Furthermore, commuting with $\sigma_{\mathrm{x}}\tau_0$ (from $C_{2x}$) eliminates $\sigma_z$ and $\sigma_y$. By imposing time-reversal symmetry, commuting with $\sigma_0\tau_{\mathrm{x}}$ eliminates $\sigma_0\tau_z$ and $\sigma_0\tau_y$. The terms allowed in $\rho^{\xi,\Upsilon}$ are reduced to

$$\rho^{\xi,\Upsilon} = P^{\xi,\Upsilon}_{00}\sigma_0\tau_0 + P^{\xi,\Upsilon}_{\mathrm{xx}}\sigma_{\mathrm{x}}\tau_{\mathrm{x}} + P^{\xi,\Upsilon}_{\mathrm{x}0}\sigma_{\mathrm{x}}\tau_0 + P^{\xi,\Upsilon}_{0\mathrm{x}}\sigma_0\tau_{\mathrm{x}}. \tag{C.9}$$

Finally, from $C_{3z}$ we consider the cases of the $\Gamma_3$ and $\Gamma_1\oplus\Gamma_2$ fermions separately. The $\Gamma_3$ representation transforms under $e^{i\frac{2\pi}{3}\sigma_z\tau_z}$, and terms anti-commuting with $\sigma_z\tau_z$ vanish — only $P^{\xi,\Gamma_3}_{00}$ and $P^{\xi,\Gamma_3}_{\mathrm{xx}}$ survive. However, by imposing $U(1)$ valley symmetry, the only remaining allowed term for the $\Gamma_3$ blocks is $P^{\xi,\Gamma_3}_{00}$. The $\Gamma_1\oplus\Gamma_2$ block transforms trivially under $C_{3z}$ ($\sigma_0\tau_0$), so all four terms survive, but imposing $U(1)$ valley symmetry reduces the allowed terms to $P^{\xi,\Gamma_1\oplus\Gamma_2}_{00}$ and $P^{\xi,\Gamma_1\oplus\Gamma_2}_{\mathrm{x}0}$ only.

The hybridization blocks of the density matrix $\rho^{cf}$ are also strongly constrained by symmetry. Matrix elements that connect electrons belonging to different irreducible representations vanish identically. Conversely, components connecting electrons within the same irreducible representation $\Upsilon$ transform in the same way as the diagonal density-matrix blocks $\rho^{\xi,\Upsilon}$. In the present case the only symmetry-allowed hybridization block couples the $\Gamma_3$ $c$-electrons to the $\Gamma_3$ $f$-electrons. The most general density matrix consistent with these symmetry constraints is therefore:

$$O^f_{\alpha\eta s,\beta\eta' s'} = \frac{\nu_f + N_f}{2N_f}\delta_{\alpha\beta}\delta_{\eta\eta'}\delta_{ss'}, \tag{C.10a}$$

$$O^{cf}_{a\eta s,\beta\eta' s'} = \chi_{cf}\left(\delta_{a1}\delta_{\beta 1} + \delta_{a2}\delta_{\beta 2}\right)\delta_{\eta\eta'}\delta_{ss'}, \tag{C.10b}$$

$$\begin{aligned} O^c_{a\eta s,b\eta' s'} = &\left[\frac{\nu_{c3}}{2N_f}\left(\delta_{a1}\delta_{b1} + \delta_{a2}\delta_{b2}\right) + \frac{\nu_{c1}}{2N_f}\left(\delta_{a3}\delta_{b3} + \delta_{a4}\delta_{b4}\right)\right]\delta_{\eta\eta'}\delta_{ss'} \\ &+ \Delta_c\left(\delta_{a3}\delta_{b4} + \delta_{a4}\delta_{b3}\right)\delta_{\eta\eta'}\delta_{ss'}, \end{aligned} \tag{C.10c}$$

where the $f$- and $c$-electron fillings are, respectively,

$$\nu_f = \mathrm{tr}\left[O^f\right] - 4, \qquad \nu_c = \mathrm{tr}\left[O^c\right], \tag{C.11}$$

and $\nu_{c3}$ and $\nu_{c1}$ denote the individual occupations of the $\Gamma_3$ and $\Gamma_1 \oplus \Gamma_2$ $c$-electrons, respectively. The parameters $\chi_{cf}$ and $\Delta_c$ are forced to be real by the symmetry constraints above.

### Appendix D: Interactions

In this section, we derive the interacting THF mean-field Hamiltonian used to evaluate the optical conductivity at the band-insulator fillings. The interaction terms in the THF model were obtained in Ref.([1]) by projecting the Coulomb interaction onto the THF basis. The dominant interaction terms are:

$$H_{U_1} = \frac{U_1}{2}\sum_{\mathbf{R}}\sum_{\substack{\alpha,\eta,s\\ \alpha',\eta',s'}} :\hat{f}^\dagger_{\mathbf{R},\alpha,\eta,s}\hat{f}_{\mathbf{R},\alpha,\eta,s}::\hat{f}^\dagger_{\mathbf{R},\alpha',\eta',s'}\hat{f}_{\mathbf{R},\alpha',\eta',s'}:, \tag{D.1a}$$

$$H_{U_2} = \frac{U_2}{2}\sum_{\langle\mathbf{R},\mathbf{R}'\rangle_{n.n}}\sum_{\substack{\alpha,\eta,s\\ \alpha',\eta',s'}} :\hat{f}^\dagger_{\mathbf{R},\alpha,\eta,s}\hat{f}_{\mathbf{R},\alpha,\eta,s}::\hat{f}^\dagger_{\mathbf{R}',\alpha',\eta',s'}\hat{f}_{\mathbf{R}',\alpha',\eta',s'}:, \tag{D.1b}$$

$$H_V = \frac{1}{2\Omega_0 N_0}\sum_{|\mathbf{k}_1|,|\mathbf{k}_2|\leq\Lambda_c}\ \sum_{\substack{\mathbf{q}\\ |\mathbf{k}_1+\mathbf{q}||\mathbf{k}_2-\mathbf{q}|\leq\Lambda_c}}\ \sum_{\substack{a,\eta,s\\ a',\eta',s'}} V(\mathbf{q}) :\hat{c}^\dagger_{\mathbf{k}_1+\mathbf{q},a,\eta,s}\hat{c}_{\mathbf{k}_1,a,\eta,s}::\hat{c}^\dagger_{\mathbf{k}_2-\mathbf{q},a',\eta',s'}\hat{c}_{\mathbf{k}_2,a',\eta',s'}:, \tag{D.1c}$$

$$H_W = \frac{1}{N_0}\sum_{\substack{\mathbf{k}_1\\ |\mathbf{k}_2|\leq\Lambda_c}}\ \sum_{\substack{\mathbf{q}\\ |\mathbf{k}_2-\mathbf{q}|\leq\Lambda_c}}\ \sum_{\substack{\alpha,\eta,s\\ a',\eta',s'}} W_{a'} :\hat{f}^\dagger_{\mathbf{k}_1+\mathbf{q},\alpha,\eta,s}\hat{f}_{\mathbf{k}_1,\alpha,\eta,s}::\hat{c}^\dagger_{\mathbf{k}_2-\mathbf{q},a',\eta',s'}\hat{c}_{\mathbf{k}_2,a',\eta',s'}:, \tag{D.1d}$$

$$H_J = -\frac{J}{2N_0}\sum_{\substack{\mathbf{k}_1\\ |\mathbf{k}_2|\leq\Lambda_c}}\ \sum_{\substack{\mathbf{q}\\ |\mathbf{k}_2+\mathbf{q}|\leq\Lambda_c}}\ \sum_{\substack{\alpha,\eta,s\\ \alpha',\eta',s'}}\left[\eta\eta' + (-1)^{\alpha+\alpha'}\right] :\hat{f}^\dagger_{\mathbf{k}_1+\mathbf{q},\alpha,\eta,s}\hat{f}_{\mathbf{k}_1,\alpha',\eta',s'}::\hat{c}^\dagger_{\mathbf{k}_2-\mathbf{q},\alpha'+2,\eta',s'}\hat{c}_{\mathbf{k}_2,\alpha+2,\eta,s}:, \tag{D.1e}$$

where $W_{a=1} = W_{a=2} = W_1$ and $W_{a=3} = W_{a=4} = W_3$. In the above, $H_{U_1}$ is the on-site Hubbard interaction for $f-$electrons, $H_{U_2}$ is the nearest-neighbor repulsion for the $f$ electrons, $H_V$ is the Coulomb interaction between $c$-electrons, $H_W$ is the density-density interaction between $f$- and $c$-electrons, and $H_J$ is the exchange interaction between $\Gamma_1 \oplus \Gamma_2$ $c$-electrons and $f$-electrons. We leave the interaction parameters unspecified for now; they will be determined from experiment in Sec.(I).

Next, we carry out a mean-field decomposition of the interacting Hamiltonian at the band-insulator fillings. The mean-field Hamiltonian is

$$\hat{H}^{\mathrm{MF}}_{\mathrm{int}} = \hat{H}^{\mathrm{MF}}_U + \hat{H}^{\mathrm{MF}}_W + \hat{H}^{\mathrm{MF}}_J + \hat{H}^{\mathrm{MF}}_V, \tag{D.2}$$

where each individual term is given by

$$\hat{H}_U^{\rm MF} = \sum_{\mathbf{R}}\sum_{\alpha\eta s}\{U_1(\nu_f+0.5)+6U_2\nu_f\} f^\dagger_{\mathbf{R}\alpha\eta s} f_{\mathbf{R}\alpha\eta s} - U_1\sum_{\mathbf{R}} O^f_{\alpha\eta s,\beta\eta' s'} f^\dagger_{\mathbf{R}\beta\eta' s'} f_{\mathbf{R}\alpha\eta s}, \tag{D.3a}$$

$$\hat{H}_W^{\rm MF} = \sum_{|\mathbf{k}|<\Lambda_c}\sum_{a\eta s}\nu_f W_a c^\dagger_{\mathbf{k}a\eta s}c_{\mathbf{k}a\eta s} + \sum_{\mathbf{R}}\sum_{\alpha\eta s}\nu_{c,a}W_a f^\dagger_{\mathbf{R}\alpha\eta s}f_{\mathbf{R}\alpha\eta s} - \sum_{\alpha\eta_1 s_1}\sum_{|\mathbf{k}|<\Lambda_c}\sum_{\eta_2 s_2 a} W_a\left\{O^{cf}_{a\eta_2 s_2,\alpha\eta_1 s_1} f^\dagger_{\mathbf{k}\alpha\eta_1 s_1}c_{\mathbf{k}a\eta_2 s_2} + h.c\right\}, \tag{D.3b}$$

$$\begin{aligned}\hat{H}_J^{\rm MF} =& -\frac{J}{2}\sum_{\mathbf{R}}\sum_{\alpha_1\alpha_2\eta_1\eta_2 s_1 s_2}(\eta_1\eta_2+(-1)^{\alpha_1+\alpha_2}) f^\dagger_{\mathbf{R}\alpha_1\eta_1 s_1} f_{\mathbf{R}\alpha_2\eta_2 s_2} O^c_{\alpha_2+2\eta_2 s_2,\alpha_1+2\eta_1 s_1} \\ &-\frac{J}{2}\sum_{|\mathbf{k}|<\Lambda_c}\sum_{\alpha_1\alpha_2\eta_1\eta_2 s_1 s_2}(\eta_1\eta_2+(-1)^{\alpha_1+\alpha_2}) c^\dagger_{\mathbf{k}\alpha_2+2,\eta_2 s_2} c_{\mathbf{k}\alpha_1+2,\eta_1 s_1}\left\{O^f_{\alpha_1\eta_1 s_1,\alpha_2\eta_2 s_2} - \frac{1}{2}\delta_{\alpha_1\alpha_2}\delta_{\eta_1\eta_2}\delta_{s_1 s_2}\right\} \\ &+\frac{J}{2\sqrt{N}}\sum_{\mathbf{R}}\sum_{|\mathbf{k}|<\Lambda_c}\sum_{\alpha_1\alpha_2\eta_1\eta_2 s_1 s_2} e^{i\mathbf{k}\cdot\mathbf{R}}(\eta_1\eta_2+(-1)^{\alpha_1+\alpha_2})\left\{O^{cf}_{\alpha_2+2\eta_2 s_2,\alpha_2\eta_2 s_2} f^\dagger_{\mathbf{R}\alpha_1\eta_1 s_1} c_{\mathbf{k}\alpha_1+2\eta_1 s_1} + h.c.\right\},\end{aligned} \tag{D.3c}$$

$$\hat{H}_V^{\rm MF} = V_0\nu_c\sum_{\eta s a}\sum_{|\mathbf{k}|<\Lambda_c} c^\dagger_{\mathbf{k}a\eta s}c_{\mathbf{k}a\eta s}. \tag{D.3d}$$

where $V(0)/\Omega_0 \equiv V_0$. We first focus on the case where only relaxation is present; strain will be added in Sec.(J). For the symmetric self-consistent solution at fillings $\nu = \nu_c + \nu_f = \pm 4$ in the absence of strain (see Sec.(C 1)), the density matrix takes the form of Eqs. (C.10a–C.10c). Inserting this density matrix into Eq. (D.2), we can parameterize the Hamiltonian using the energies:

$$E_{c,\Gamma_3} = V_0\nu_c + W_1\nu_f + \mu_1 - \mu \tag{D.4a}$$

$$E_{c,\Gamma_{12}} = V_0\nu_c + W_3\nu_f - \frac{J}{2N_f}\nu_f + \mu_2 - \mu \tag{D.4b}$$

$$E_f = U_1\nu_f\frac{2N_f-1}{2N_f} + 6U_2\nu_f + W_3\nu_{c1} + W_1\nu_{c3} - \frac{J}{2N_f}\nu_{c1} - \mu. \tag{D.4c}$$

In addition, we introduce the effective interaction-induced hybridization

$$\gamma_{\rm int} = -W_1\chi_{cf} \tag{D.5}$$

With these definitions, the interacting Hamiltonian of the THF model at $\nu = \pm 4$ can be written as

$$\begin{aligned}\hat{H}^{\rm MF} =& \sum_{\substack{|\mathbf{k}|\leq\Lambda_c\\ \eta,s}}\left[\sum_{a,a'} h^{{\rm BI},cc,\eta}_{aa'}(\mathbf{k})\,\hat{c}^\dagger_{\mathbf{k},a,\eta,s}\hat{c}_{\mathbf{k},a,\eta,s} + \sum_{a,\alpha} h^{{\rm BI}cf,\eta}_{a\alpha}(\mathbf{k})\,\hat{c}^\dagger_{\mathbf{k},a,\eta,s}\hat{f}_{\mathbf{k},\alpha,\eta,s} + \sum_{a,\alpha}\overline{h^{{\rm BI}cf,\eta}_{a\alpha}(\mathbf{k})}\hat{f}^\dagger_{\mathbf{k},\alpha,\eta,s}\hat{c}_{\mathbf{k},a,\eta,s}\right] \\ &+\sum_{\substack{|\mathbf{k}|\leq\Lambda_c\\ \eta,s}}\left[\sum_{\alpha,\alpha'} h^{{\rm BI}ff,\eta}_{\alpha\alpha'}(\mathbf{k})\,\hat{f}^\dagger_{\mathbf{k},\alpha,\eta,s}\hat{f}_{\mathbf{k},\alpha',\eta,s}\right]\end{aligned} \tag{D.6}$$

where the matrix elements are

$$h^{{\rm BI},cc,\eta}_{aa'}(\mathbf{k}) = \begin{pmatrix} E_{c,\Gamma_3}\sigma_0 & v_\star(\eta k_{\rm x}\sigma_0 + ik_{\rm y}\sigma_{\rm z}) \\ v_*(\eta k_{\rm x}\sigma_0 - ik_{\rm y}\sigma_{\rm z}) & M\sigma_{\rm x} + E_{c,\Gamma_{12}}\sigma_0\end{pmatrix}_{aa'} \tag{D.7}$$

$$h^{{\rm BI}cf,\eta}_{a\alpha}(\mathbf{k}) = \begin{pmatrix}\gamma\sigma_0 + v'_*(\eta k_{\rm x}\sigma_{\rm x} + k_{\rm y}\sigma_{\rm y}) \\ 0\end{pmatrix}_{a\alpha} e^{-\lambda^2|\mathbf{k}|^2/2} + \begin{pmatrix}\gamma_{\rm int}\sigma_0 \\ 0\end{pmatrix}_{a\alpha}, \tag{D.8}$$

and

$$h^{{\rm BI}ff,\eta}_{\alpha\alpha'}(\mathbf{k}) = E_f\,[\sigma_0]_{\alpha\alpha'}. \tag{D.9}$$

Note that the terms proportional to $\Delta_c$ in the density matrix can only contribute to $\hat{H}_J^{\mathrm{MF}}$. They vanish upon explicit evaluation as $\Delta_c$ is off-diagonal in the orbital indices ($a = 3 \leftrightarrow a = 4$) but diagonal in valley ($\eta_1 = \eta_2$); however, the exchange coupling $[\eta_1\eta_2 + (-1)^{\alpha_1+\alpha_2}]$ in Eq. (D.3c) contracts the $\Gamma_1 \oplus \Gamma_2$ $c$-electron indices with the $f$-electron indices as $(\alpha + 2, \eta) \leftrightarrow (\alpha' + 2, \eta')$. Therefore, the off-diagonal orbital structure of $\Delta_c$ produces a vanishing trace in the contraction.

In all our calculations, we determine $\nu_f$, $\nu_{c1}$, $\nu_{c3}$, and $\chi_{cf}$ self-consistently via Hartree-Fock. Finally, note that the interaction-induced terms are local (momentum-independent), so the current operator for the interacting Hamiltonian is identical to that of the non-interacting model derived in the previous section [6].

### 1. Density matrix in the non-interacting limit

In this subsection, we estimate the parameters of the density matrix of the non-interacting band insulator. To obtain an analytical expression we assume the chiral limit $v_\star' = 0$ and set $M = 0$. In the chiral limit, $v_\star' = 0$, the Hamiltonian decouples into two separate chiral sectors labeled by $\tau = \pm 1$:

$$\tilde{h}_{\eta,\tau}(\mathbf{k}) = \begin{pmatrix} 0 & \eta v_\star (k_x + \tau\eta k_y) & \gamma \\ \eta v_\star (k_x - \tau\eta k_y) & 0 & 0 \\ \gamma & 0 & 0 \end{pmatrix}, \tag{D.10}$$

whose eigenstates are

$$\psi_{\tau\eta}^{a,n}(\mathbf{k}) = \frac{1}{\sqrt{2}} \begin{pmatrix} -1 & 0 & 1 \\ \frac{\eta v_\star |\mathbf{k}| e^{-i\eta\theta\tau}}{\gamma\sqrt{\left(\frac{v_\star|\mathbf{k}|}{\gamma}\right)^2+1}} & -\eta\frac{\sqrt{2}\gamma}{v_\star|\mathbf{k}|}\frac{e^{-i\eta\theta\tau}}{\sqrt{\left(\frac{\gamma}{v_\star|\mathbf{k}|}\right)^2+1}} & \frac{\eta v_\star |\mathbf{k}| e^{-i\eta\theta\tau}}{\gamma\sqrt{\left(\frac{v_\star|\mathbf{k}|}{\gamma}\right)^2+1}} \\ \frac{1}{\sqrt{\left(\frac{v_\star|\mathbf{k}|}{\gamma}\right)^2+1}} & \frac{\sqrt{2}}{\sqrt{\left(\frac{\gamma}{v_\star|\mathbf{k}|}\right)^2+1}} & \frac{1}{\sqrt{\left(\frac{v_\star|\mathbf{k}|}{\gamma}\right)^2+1}} \end{pmatrix}_{a,n}. \tag{D.11}$$

We then obtain the momentum-dependent single-particle density matrix for the occupied bands as

$$\mathcal{O}_{\nu=4}^{\tau\eta as,\tau'\eta'a's'}(\mathbf{k}) = \begin{pmatrix} \frac{1}{2} & -\frac{\eta v_\star|\mathbf{k}|e^{-i\eta\theta\tau}}{2\sqrt{\gamma^2+v_\star^2|\mathbf{k}|^2}} & -\frac{\gamma}{2\sqrt{\gamma^2+v_\star^2|\mathbf{k}|^2}} \\ -\frac{\eta v_\star|\mathbf{k}|e^{i\eta\theta\tau}}{2\sqrt{\gamma^2+v_\star^2|\mathbf{k}|^2}} & \frac{1}{2}\left(1+\frac{\gamma^2}{\gamma^2+v_\star^2|\mathbf{k}|^2}\right) & -\frac{\gamma\eta v_\star|\mathbf{k}|e^{i\eta\theta\tau}}{2(\gamma^2+v_\star^2|\mathbf{k}|^2)} \\ -\frac{\gamma}{2\sqrt{\gamma^2+v_\star^2|\mathbf{k}|^2}} & -\frac{\gamma\eta v_\star|\mathbf{k}|e^{-i\eta\theta\tau}}{2(\gamma^2+v_\star^2|\mathbf{k}|^2)} & 1-\frac{\gamma^2}{2(\gamma^2+v_\star^2|\mathbf{k}|^2)} \end{pmatrix}_{aa'} \delta_{\tau\tau'}\delta_{\eta\eta'}\delta_{ss'}. \tag{D.12}$$

The filling of the $f$-electrons and the hybridization are obtained by integrating $\mathcal{O}_{\nu=4}$ over momentum. We adopt a circular momentum cutoff $\Lambda$ on the scale of the moiré Brillouin zone size, yielding

$$\frac{1}{N}\sum_{\mathbf{k}} \mathcal{O}_{\nu=4}^{\tau\eta as,\tau'\eta'a's'}(\mathbf{k}) = \frac{1}{\pi\Lambda^2}\int_0^\Lambda \mathrm{d}|\mathbf{k}| \int_0^{2\pi} d\theta\ |\mathbf{k}| \begin{pmatrix} \frac{1}{2} & -\frac{\eta v_\star|\mathbf{k}|e^{-i\eta\theta\tau}}{2\sqrt{\gamma^2+v_\star^2|\mathbf{k}|^2}} & -\frac{\gamma}{2\sqrt{\gamma^2+v_\star^2|\mathbf{k}|^2}} \\ -\frac{\eta v_\star|\mathbf{k}|e^{i\eta\theta\tau}}{2\sqrt{\gamma^2+v_\star^2|\mathbf{k}|^2}} & \frac{1}{2}\left(1+\frac{\gamma^2}{\gamma^2+v_\star^2|\mathbf{k}|^2}\right) & -\frac{\gamma\eta v_\star|\mathbf{k}|e^{i\eta\theta\tau}}{2(\gamma^2+v_\star^2|\mathbf{k}|^2)} \\ -\frac{\gamma}{2\sqrt{\gamma^2+v_\star^2|\mathbf{k}|^2}} & -\frac{\gamma\eta v_\star|\mathbf{k}|e^{-i\eta\theta\tau}}{2(\gamma^2+v_\star^2|\mathbf{k}|^2)} & 1-\frac{\gamma^2}{2(\gamma^2+v_\star^2|\mathbf{k}|^2)} \end{pmatrix}_{aa'} \delta_{\tau\tau'}\delta_{\eta\eta'}\delta_{ss'} \tag{D.13}$$

$$= \frac{2}{\Lambda^2}\int_0^\Lambda \mathrm{d}|\mathbf{k}|\ |\mathbf{k}| \begin{pmatrix} \frac{1}{2} & 0 & -\frac{\gamma}{2\sqrt{\gamma^2+v_\star^2|\mathbf{k}|^2}} \\ 0 & \frac{1}{2}\left(1+\frac{\gamma^2}{\gamma^2+v_\star^2|\mathbf{k}|^2}\right) & 0 \\ -\frac{\gamma}{2\sqrt{\gamma^2+v_\star^2|\mathbf{k}|^2}} & 0 & 1-\frac{\gamma^2}{2(\gamma^2+v_\star^2|\mathbf{k}|^2)} \end{pmatrix}_{aa'} \delta_{\tau\tau'}\delta_{\eta\eta'}\delta_{ss'} \tag{D.14}$$

$$= \begin{pmatrix} \frac{1}{2} & 0 & -\frac{\gamma\left(\gamma+\sqrt{\gamma^2+v_\star^2\Lambda^2}\right)}{v_\star^2\Lambda^2} \\ 0 & \frac{1}{2}\left(1+\frac{\gamma^2}{v_\star^2\Lambda^2}\log\left(1+\frac{v_\star^2\Lambda^2}{\gamma^2}\right)\right) & 0 \\ -\frac{\gamma\left(\gamma+\sqrt{\gamma^2+v_\star^2\Lambda^2}\right)}{v_\star^2\Lambda^2} & 0 & 1-\frac{1}{2}\frac{\gamma^2}{v_\star^2\Lambda^2}\log\left(1+\frac{v_\star^2\Lambda^2}{\gamma^2}\right) \end{pmatrix}_{aa'} \delta_{\tau\tau'}\delta_{\eta\eta'}\delta_{ss'}. \tag{D.15}$$

From the integrated density matrix, we find the following parameters:

$$\nu_{c3} = 0, \tag{D.16}$$

$$\nu_{c1} = \nu_c = \frac{4\gamma^2}{v_\star^2\Lambda^2}\log\left(1 + \frac{v_\star^2\Lambda^2}{\gamma^2}\right), \tag{D.17}$$

$$\nu_f = 4\left(1 - \frac{\gamma^2}{v_\star^2\Lambda^2}\log\left(1 + \frac{v_\star^2\Lambda^2}{\gamma^2}\right)\right), \tag{D.18}$$

$$\chi_{cf} = -\frac{\gamma\left(\gamma + \sqrt{\gamma^2 + v_\star^2\Lambda^2}\right)}{v_\star^2\Lambda^2}. \tag{D.19}$$

Note that, as expected, $\nu_c$ and $\nu_f$ have finite limits as $\gamma/v_\star\Lambda \to 0$ or $\gamma/v_\star\Lambda \to \infty$, and that the $\Gamma_3$ $c$-electrons are exactly half-filled in the chiral limit in the absence of interactions, such that $\nu_{c3} = 0$. Additionally, since $\gamma < 0$, we find that $\chi_{cf} > 0$.

Deviations from these expressions are expected in the interacting model. Numerically, however, these estimates are close to the full self-consistent calculation at $\nu = \pm 4$. This can be attributed to the fact that at the band-insulator fillings there is no symmetry breaking, and the states with and without interactions are adiabatically connected, so the density matrix does not change sharply as the interaction strength is varied.

## Appendix E: Minimal substitution

In this section, we describe our treatment of the THF model at finite magnetic field $B$. A rigorous study has been presented in Ref. [7]. We consider the THF Hamiltonian at finite $B$ obtained by minimal substitution, $k_\mathrm{x} + ik_\mathrm{y} \to -i\frac{\sqrt{2}}{\ell_B}\hat{a}$, $k_\mathrm{x} - ik_\mathrm{y} \to i\frac{\sqrt{2}}{\ell_B}\hat{a}^\dagger$, where $\hat{a}^\dagger, \hat{a}$ are the Landau-level (LL) creation and annihilation operators and $\ell_B = \sqrt{\hbar c/(eB)}$ is the magnetic length. In Ref. [7], it was shown that this substitution overcounts the number of $f$-modes; the spurious modes correspond to high angular momentum and appear as the LL cutoff $m^*$ increases. However, as we show in Sec.(G), only transitions involving the lowest LLs — which have large $c$-electron character and small angular momentum — contribute significantly to the main experimental features. The spurious high-$m$ modes therefore do not affect these transitions, and simple canonical quantization suffices to provide an appropriate physical description of the experimental observations. We perform self-consistent Hartree-Fock at $B = 0$ and use the resulting density matrix to construct the finite-$B$ Hamiltonian. Since the experiment does not directly dope carriers into the LLs, we anticipate that the effects of symmetry breaking and quantum Hall ferromagnetism can be neglected [8, 9].

In terms of the LL creation and annihilation operators, the finite-$B$ Hamiltonian in valley $\eta = +$ is:

$$\tilde{h}^{\mathrm{BI}}_{\eta=+}(B) = \begin{pmatrix} E_{c,\Gamma_3} & 0 & -i\Delta_B\hat{a} & 0 & \left(\gamma\Sigma\left(\hat{a}^\dagger\hat{a}\right) + \gamma_{\mathrm{int}}\right) & i\Delta'_B\hat{a}^\dagger\Sigma\left(\hat{a}^\dagger\hat{a}\right) \\ 0 & E_{c,\Gamma_3} & 0 & i\Delta_B\hat{a}^\dagger & -i\Delta'_B\hat{a}\Sigma\left(\hat{a}^\dagger\hat{a}\right) & \left(\gamma\Sigma\left(\hat{a}^\dagger\hat{a}\right) + \gamma_{\mathrm{int}}\right) \\ i\Delta_B\hat{a}^\dagger & 0 & E_{c,\Gamma_{12}} & M & 0 & 0 \\ 0 & -i\Delta_B\hat{a} & M & E_{c,\Gamma_{12}} & 0 & 0 \\ \left(\gamma\Sigma\left(\hat{a}^\dagger\hat{a}\right) + \gamma_{\mathrm{int}}\right) & i\Delta'_B\Sigma\left(\hat{a}^\dagger\hat{a}\right)\hat{a}^\dagger & 0 & 0 & E_f & 0 \\ -i\Delta'_B\Sigma\left(\hat{a}^\dagger\hat{a}\right)\hat{a} & \left(\gamma\Sigma\left(\hat{a}^\dagger\hat{a}\right) + \gamma_{\mathrm{int}}\right) & 0 & 0 & 0 & E_f \end{pmatrix} \tag{E.1}$$

where $\Delta_B = \sqrt{2}\hbar v_\star/\ell_B$, $\Delta'_B = \sqrt{2}\hbar v'_\star/\ell_B$, and $\Sigma(\hat{a}^\dagger\hat{a})$ is the canonically quantized Gaussian hybridization. The normal ordering of $\Sigma\left(\hat{a}^\dagger\hat{a}\right)$ is chosen to reproduce the hybridization projected onto the irreducible representations of the magnetic translation group in the limit $B \to 0$. This normal ordering implies that the action of $\Sigma\left(\hat{a}^\dagger\hat{a}\right)$ on a state $|m\rangle$ in the Landau basis with LL index $m$, yields $\Sigma\left(\hat{a}^\dagger\hat{a}\right)|m\rangle = e^{-\lambda^2(m+1/2)/\ell_B^2}|m\rangle$; see Ref. [7] for a detailed explanation. Analogously, the Hamiltonian in valley $\eta = -$ is

$$\tilde{h}^{\mathrm{BI}}_{\eta=-}(B) = \begin{pmatrix} E_{c,\Gamma_3} & 0 & -i\Delta_B\hat{a}^\dagger & 0 & \left(\gamma\Sigma\left(\hat{a}^\dagger\hat{a}\right) + \gamma_{\mathrm{int}}\right) & i\Delta'_B\hat{a}\Sigma\left(\hat{a}^\dagger\hat{a}\right) \\ 0 & E_{c,\Gamma_3} & 0 & i\Delta_B\hat{a} & -i\Delta'_B\hat{a}^\dagger\Sigma\left(\hat{a}^\dagger\hat{a}\right) & \left(\gamma\Sigma\left(\hat{a}^\dagger\hat{a}\right) + \gamma_{\mathrm{int}}\right) \\ i\Delta_B\hat{a} & 0 & E_{c,\Gamma_{12}} & M & 0 & 0 \\ 0 & -i\Delta_B\hat{a}^\dagger & M & E_{c,\Gamma_{12}} & 0 & 0 \\ \left(\gamma\Sigma\left(\hat{a}^\dagger\hat{a}\right) + \gamma_{\mathrm{int}}\right) & i\Delta'_B\Sigma\left(\hat{a}^\dagger\hat{a}\right)\hat{a} & 0 & 0 & E_f & 0 \\ -i\Delta'_B\Sigma\left(\hat{a}^\dagger\hat{a}\right)\hat{a}^\dagger & \left(\gamma\Sigma\left(\hat{a}^\dagger\hat{a}\right) + \gamma_{\mathrm{int}}\right) & 0 & 0 & 0 & E_f \end{pmatrix}. \tag{E.2}$$

Eigenstates of this Hamiltonian are expressed in the LL basis as spinors of the form

$$\Psi^{n,\eta} = \sum_m \left(\, \psi^{n,\eta}_{1m} \;\; \psi^{n,\eta}_{2m} \;\; \psi^{n,\eta}_{3m} \;\; \psi^{n,\eta}_{4m} \;\; \psi^{n,\eta}_{5m} \;\; \psi^{n,\eta}_{6m} \,\right)^T \otimes |m\rangle, \tag{E.3}$$

where $\psi_{b,m}^{n,\eta}$ are scalar coefficients, $n$ is the band index, $b = 1, \ldots, 6$ labels the orbital degrees of freedom (four $c$-electron and two $f$-electron components), and $m$ is the LL index. We determine the spectrum at finite $B$ by numerical diagonalization with a LL cutoff $m^* = \lceil (q-3)/2 \rceil$, where the total magnetic flux satisfies $\phi/\phi_0 = 1/q$, following the procedure of Ref.([7]). The $B = 0$ eigenstates of the anomalous LL sectors ($m = 0, 1, 2$; see Eqs.(G.6) and(G.14)) are analytically solvable, which provides the basis for the perturbation theory discussed in Sec.(H).

To describe the optical response at finite $B$, we now obtain the current matrix elements under canonical quantization. The $c$–$c$ current matrix elements, $J_{aa'}^{\mu;cc,\eta}$, remain unchanged as they are momentum-independent. For the $c$–$f$ hybridization current, we adopt the same normal ordering as for the Hamiltonian in Eqs.(E.1E.2). The current operators along $x$ are:

$$\sqrt{2}\frac{\hbar}{\ell_B} J_{a\alpha}^{\mu=x;cf,\eta=+} = \begin{pmatrix} \Delta_B' \sigma_x \Sigma\left(\hat{a}^\dagger \hat{a}\right) \\ 0 \end{pmatrix}_{a\alpha} - i\frac{\lambda^2}{\ell_B^2} \begin{pmatrix} \gamma\sigma_0 + i\Delta_B' \begin{pmatrix} 0 & a^\dagger \\ -a & 0 \end{pmatrix} \\ 0 \end{pmatrix}_{a\alpha} \left(a^\dagger - a\right) \Sigma\left(\hat{a}^\dagger \hat{a}\right) \tag{E.4}$$

$$\sqrt{2}\frac{\hbar}{\ell_B} J_{a\alpha}^{\mu=x;cf,\eta=-} = -\begin{pmatrix} \Delta_B' \sigma_x \Sigma\left(\hat{a}^\dagger \hat{a}\right) \\ 0 \end{pmatrix}_{a\alpha} - i\frac{\lambda^2}{\ell_B^2} \begin{pmatrix} \gamma\sigma_0 + i\Delta_B' \begin{pmatrix} 0 & a \\ -a^\dagger & 0 \end{pmatrix} \\ 0 \end{pmatrix}_{a\alpha} \left(a^\dagger - a\right) \Sigma\left(\hat{a}^\dagger \hat{a}\right) \tag{E.5}$$

and along $y$:

$$\sqrt{2}\frac{\hbar}{\ell_B} J_{a\alpha}^{\mu=y;cf,\eta=+} = \begin{pmatrix} \Delta_B' \sigma_y \\ 0 \end{pmatrix}_{a\alpha} \Sigma\left(\hat{a}^\dagger \hat{a}\right) + i\frac{\lambda^2}{\ell_B^2} \begin{pmatrix} \gamma\sigma_0 + i\Delta_B' \begin{pmatrix} 0 & a^\dagger \\ -a & 0 \end{pmatrix} \\ 0 \end{pmatrix}_{a\alpha} \left(a^\dagger + a\right) \Sigma\left(\hat{a}^\dagger \hat{a}\right) \tag{E.6}$$

$$\sqrt{2}\frac{\hbar}{\ell_B} J_{a\alpha}^{\mu=y;cf,\eta=-} = \begin{pmatrix} \Delta_B' \sigma_y \\ 0 \end{pmatrix}_{a\alpha} \Sigma\left(\hat{a}^\dagger \hat{a}\right) + i\frac{\lambda^2}{\ell_B^2} \begin{pmatrix} \gamma\sigma_0 + i\Delta_B' \begin{pmatrix} 0 & a \\ -a^\dagger & 0 \end{pmatrix} \\ 0 \end{pmatrix}_{a\alpha} \left(a^\dagger + a\right) \Sigma\left(\hat{a}^\dagger \hat{a}\right). \tag{E.7}$$

In the experiment, $B$ does not exceed 10 T such that $\ell_B > 8$nm, providing a natural hierarchy of scales between the terms in the current operators. Since $\lambda$ is typically $20 - 30\%$ of the moiré length, we find that the ratio $\frac{\lambda^2}{\ell_B^2}$ is at most $\frac{\lambda^2}{\ell_B^2} \approx 0.2 - 0.3$ for the range of the magnetic fields used in the experiment. If we consider transitions only among the lowest Landau levels, the term proportional to $\lambda^2$ is subdominant. In Sec.(G) we neglect the terms which are proportional to $\lambda^2$ to obtain approximate selection rules that simplify the interpretation of the experiment. For numerical calculations, we use current operators that include the terms proportional to $\lambda^2$ in Eq.(E.4E.6E.5E.7).

## Appendix F: Optical conductivity

In this section, we derive the form of the longitudinal optical conductivity used to calculate the optical conductivity maps shown in the manuscript. We start from the Lehmann representation of the current-current correlation function in terms of the many-body eigenstates $|i\rangle$ and many-body energies $E_i$:

$$\mathrm{Re}\sigma_{\mu\mu}(\omega) = \frac{\pi e^2}{\Omega_0} \frac{1 - e^{-\beta\omega}}{\omega} \sum_{ij} e^{-\beta E_i} \left\langle i|\hat{J}_\mu^\dagger|j\right\rangle \left\langle j|\hat{J}_{\mu'}|i\right\rangle \delta\left(\hbar\omega + E_i - E_j\right) \tag{F.1}$$

where the total particle current is defined in Eq.(B.2) for the $B = 0$, and the current matrix elements at finite $B$ are defined in Eqs.(E.4,E.5,E.6E.7). The experiment is conducted at $T \sim 500$ mK, for simplicity we then take the $\beta \to \infty$ limit:

$$\mathrm{Re}\sigma_{\mu\mu}(\omega) = \frac{\pi e^2}{\Omega_0} \frac{1}{\omega} \sum_j \left\langle \psi_{GS}|\hat{J}_\mu^\dagger|j\right\rangle \left\langle j|\hat{J}_\mu|\psi_{GS}\right\rangle \delta\left(\hbar\omega + E_0 - E_j\right) \tag{F.2}$$

where the ground state $|\psi_{\mathrm{GS}}\rangle$ is a Slater determinant with all states filled up to the Fermi level.

We now evaluate the matrix elements at finite $B$. We introduce creation (annihilation) operators $d_{n,m,\eta,s}^\dagger$ ($d_{n,m,\eta,s}$) for the eigenstates of the single-particle Hamiltonian, defined as

$$\hat{d}_{n,m,\eta,s}^\dagger = \sum_{b=1}^{4} \psi_{b,m}^{n,\eta}\, \hat{c}_{b,m,\eta,s}^\dagger + \sum_{b=5}^{6} \psi_{b,m}^{n,\eta}\, \hat{f}_{b,m,\eta,s}^\dagger. \tag{F.3}$$

Here $\hat{c}^\dagger_{b,m,\eta,s}$ and $\hat{f}^\dagger_{b,m,\eta,s}$ create $c$- and $f$-electrons in the orbital basis defined by the spinor in Eq. (E.3). The operator $\hat{d}^\dagger_{n,m,\eta,s}$ creates an electron in band $n$ with Landau-level index $m$, valley $\eta$, and spin $s$. Since the current operator creates particle–hole excitations on top of the Slater-determinant ground state, the relevant excited states are single particle–hole states obtained by promoting an electron from $(n', m', \eta, s)$ to $(n, m, \eta, s)$, $|nm\eta s; n'm'\eta s\rangle = \hat{d}^\dagger_{n,m,\eta,s}\hat{d}_{n',m',\eta,s}|\psi_{\rm GS}\rangle$ .

For interband processes, the optical conductivity must vanish as $\omega \to 0$, since photon absorption requires a finite excitation energy. In the Lehmann representation, the optical response appears as a sum over energy-conserving transitions, encoded by the factor $\delta(\hbar\omega + E_0 - E_j)$. In the absence of scattering these transitions produce sharp resonances in the optical spectrum. Using the exact identity

$$\frac{1}{\omega}\delta\left(\hbar\omega + E_0 - E_j\right) = \frac{\hbar^2\omega}{(E_j - E_0)^2}\delta\left(\hbar\omega + E_0 - E_j\right), \tag{F.4}$$

the conductivity can be written in a form that manifestly vanishes as $\omega \to 0$. This rewriting makes it explicit that the spectral weight is controlled by the finite excitation energy $E_j - E_0$, ensuring that the low-frequency limit remains well behaved even when the energy-conserving resonance acquires a small width.

Note that the combination of prefactors on the right-hand side of Eq. (F.4) arises naturally in the length-gauge formulation; using $\mathbf{J} = (i/\hbar)[H, \mathbf{P}]$, the current matrix elements generate the factor $\omega/(E_0 - E_j)^2$. For simplicity, rather than explicitly introducing the polarization operator, we express the result directly in terms of the current operator, which is straightforward to obtain in the THF model (see Sec. B).

At finite $B$, each LL has a degeneracy of

$$g_{\rm LL} = \frac{\Omega_0}{2\pi\ell_B^2}. \tag{F.5}$$

The interband optical conductivity then becomes

$$\mathrm{Re}\,\sigma_{\mu\mu}\left(\omega\right) = \frac{e^2\hbar}{2\ell_B^2}\sum_{n\neq n', mm'\eta s}\frac{\hbar\omega}{\left(\Delta\varepsilon^\eta_{nn'}\right)^2}\left\langle\psi_{\rm GS}|\hat{J}^\dagger_\mu|nm\eta s; n'm'\eta s\right\rangle\left\langle nm\eta s; n'm'\eta s|\hat{J}_\mu|\psi_{\rm GS}\right\rangle\delta\left(\hbar\omega - \Delta\varepsilon^\eta_{nn'}\right), \tag{F.6}$$

where $\varepsilon^\eta_n$ are the single-particle energies from Eqs. (E.1,E.2) and $\Delta\varepsilon^\eta_{nn'} = \varepsilon^\eta_n - \varepsilon^\eta_{n'}$.

To carry out the calculation, we approximate the $\delta$-function by a Lorentzian function of width $\zeta$. For the calculation presented in the manuscript, we use $\zeta = 3$ meV to match the experimental uncertainty in the extraction of the peak positions (see error bars in Fig. 3c). Our conclusions in the following sections are insensitive to the precise value of $\zeta$. For a discussion of how the broadening changes the optical conductivity, see Sec.(K). With this approximation, the longitudinal conductivity becomes:

$$\mathrm{Re}\sigma_{\mu\mu}\left(\omega\right) = \frac{e^2}{2\pi\hbar\ell_B^2}\hbar\omega\sum_{n\neq n', mm'\eta s}\frac{\hbar^2}{\left(\Delta\varepsilon_{nn'}\right)^2}\left\langle\psi_{GS}|\hat{J}^\dagger_\mu|nm\eta s; n'm'\eta s\right\rangle\left\langle nm\eta s; n'm'\eta s|\hat{J}_\mu|\psi_{GS}\right\rangle\frac{\zeta}{\left(\hbar\omega - \Delta\varepsilon_{nn'}\right)^2 + \zeta^2} \tag{F.7}$$

$$= \frac{e^2}{h}\hbar\omega\sum_{n\neq n', mm'\eta s}M^{nm;n'm'}_{\mu;\eta s}\frac{\zeta}{\left(\hbar\omega - \Delta\varepsilon_{nn'}\right)^2 + \zeta^2} \tag{F.8}$$

where the optical matrix elements are

$$M^{nm;n'm'}_{\mu;\eta s} = \frac{1}{2}\frac{1}{\left(\Delta\varepsilon^\eta_{nn'}\right)^2}\left|\left\langle nm\eta s|\sqrt{2}\frac{\hbar}{\ell_B}\hat{J}_\mu|n'm'\eta s\right\rangle\right|^2. \tag{F.9}$$

To evaluate the optical conductivity at finite $B$ numerically, we diagonalize the Hamiltonian in Eqs.(E.1, E.2), and evaluate the expectation value of the current matrix elements in Eqs.(E.4,E.5,E.6,E.7). Since the experiment uses unpolarized light, we sum over the components of the polarization: $\sigma(\omega) \equiv \sum_{\mu=x,y}\mathrm{Re}\,\sigma_{\mu\mu}(\omega)$. When plotting and comparing with the experimental data, we show the optical conductivity as a function of energy, $\sigma(E)$.

### Appendix G: Selection rules

To elucidate the nature of the brightest transitions observed in experiment, we derive approximate selection rules for transitions between eigenstates of Eqs. (E.1,E.2). These selection rules can be obtained in the limit $M = 0$, where the model has full $SO(2)$ rotational symmetry and the total angular momentum is conserved (see Ref.([5])). In this limit the eigenstates of the Hamiltonian can be labeled by a definite angular momentum $m$.

Since the current operator $\hat{J}^x \pm i\hat{J}^y$ carries angular momentum $\Delta m = \pm 1$, optical transitions are expected to connect states whose angular momenta differ by one unit. Consequently, the dominant transitions should satisfy the selection rule $m' = m \pm 1$, reflecting angular momentum conservation. In the following subsections we verify this expectation explicitly by evaluating the current matrix elements between the eigenstates of the Hamiltonian. We first derive the rule for the $\eta = +$ valley and then repeat the analysis for $\eta = -$, showing that the same selection rule holds in both valleys. For simplicity, throughout this section we set $\lambda = 0$.

### 1. valley $\eta = +$

Setting $M = 0$ and neglecting the Gaussian damping in the hybridization, we find:

$$\tilde{h}^{\mathrm{BI}}_{\eta=+}(B) = \begin{pmatrix} E_{c,\Gamma_3} & 0 & -i\Delta_B\hat{a} & 0 & \gamma+\gamma_{\mathrm{int}} & i\Delta'_B\hat{a}^\dagger \\ 0 & E_{c,\Gamma_3} & 0 & i\Delta_B\hat{a}^\dagger & -i\Delta'_B\hat{a} & \gamma+\gamma_{\mathrm{int}} \\ i\Delta_B\hat{a}^\dagger & 0 & E_{c,\Gamma_{12}} & 0 & 0 & 0 \\ 0 & -i\Delta_B\hat{a} & 0 & E_{c,\Gamma_{12}} & 0 & 0 \\ \gamma+\gamma_{\mathrm{int}} & i\Delta'_B\hat{a}^\dagger & 0 & 0 & E_f & 0 \\ -i\Delta'_B\hat{a} & \gamma+\gamma_{\mathrm{int}} & 0 & 0 & 0 & E_f \end{pmatrix}. \tag{G.1}$$

This Hamiltonian commutes with the total angular momentum operator $\hat{L}_{\eta=+} = \hat{a}^\dagger\hat{a}\,\mathbb{I}_{6\times6} + S^{\eta=+}_{\mathrm{ps}}$, where the pseudospin contribution, $S^{\eta=+}_{ps}$, is

$$S^{\eta=+}_{ps} = \begin{pmatrix} 1 & & & & & \\ & 2 & & & & \\ & & 0 & & & \\ & & & 3 & & \\ & & & & 1 & \\ & & & & & 2 \end{pmatrix} \tag{G.2}$$

and $\mathbb{I}_{6x6}$ is the $6\times 6$ identity matrix. To show that $\hat{L}_{\eta=+}$ commutes with the Hamiltonian, we separately calculate the commutator of the orbital part, $\mathbb{I}_{6x6}$, and the pseudospin part, $S^{\eta=+}_{\mathrm{ps}}$. The orbital part $\hat{n} = \hat{a}^\dagger\hat{a}$ acts on the operator-valued entries of the Hamiltonian. Using the relations $[\hat{n},\hat{a}] = -\hat{a}$ and $[\hat{n},\hat{a}^\dagger] = \hat{a}^\dagger$, we obtain:

$$[\hat{a}^\dagger\hat{a}\mathbb{I}_{6\times6}, \tilde{h}^{\mathrm{BI}}_{\eta=+}] = \begin{pmatrix} 0 & 0 & i\Delta_B\hat{a} & 0 & 0 & -i\Delta'_B\hat{a}^\dagger \\ 0 & 0 & 0 & -i\Delta_B\hat{a}^\dagger & i\Delta'_B\hat{a} & 0 \\ i\Delta_B\hat{a}^\dagger & 0 & 0 & 0 & 0 & 0 \\ 0 & -i\Delta_B\hat{a} & 0 & 0 & 0 & 0 \\ 0 & i\Delta'_B\hat{a}^\dagger & 0 & 0 & 0 & 0 \\ -i\Delta'_B\hat{a} & 0 & 0 & 0 & 0 & 0 \end{pmatrix} \tag{G.3}$$

The pseudospin commutator is evaluated using $[S_{ps},\tilde{h}]_{ij} = (D_i - D_j)H_{ij}$, where $D = (1,2,0,3,1,2)$. Applying this to all entries yields:

$$[S^{\eta=+}_{ps}, \tilde{h}^{\mathrm{BI}}_{\eta=+}] = \begin{pmatrix} 0 & 0 & -i\Delta_B\hat{a} & 0 & 0 & i\Delta'_B\hat{a}^\dagger \\ 0 & 0 & 0 & i\Delta_B\hat{a}^\dagger & -i\Delta'_B\hat{a} & 0 \\ -i\Delta_B\hat{a}^\dagger & 0 & 0 & 0 & 0 & 0 \\ 0 & i\Delta_B\hat{a} & 0 & 0 & 0 & 0 \\ 0 & -i\Delta'_B\hat{a}^\dagger & 0 & 0 & 0 & 0 \\ i\Delta'_B\hat{a} & 0 & 0 & 0 & 0 & 0 \end{pmatrix}. \tag{G.4}$$

The cancellation between $[S^{\eta=+}_{ps}, \tilde{h}^{\mathrm{BI}}_{\eta=+}]$ and $[\hat{a}^\dagger\hat{a}\mathbb{I}_{6\times6}, \tilde{h}^{\mathrm{BI}}_{\eta=+}]$ shows that the operator $\hat{L}_{\eta=+}$ commutes with the Hamiltonian. The eigenstates of the Hamiltonian have a definite total angular momentum $m$ and take the form

$$\Psi^{\eta=+,r=6}_{m,n} = \begin{pmatrix} \psi^{+,1}_{c3,m,n}|m-1\rangle & \psi^{+,2}_{c3,m,n}|m-2\rangle & \psi^{+,1}_{c12,m,n}|m\rangle & \psi^{+,2}_{c12,m,n}|m-3\rangle & \psi^{+,1}_{f,m,n}|m-1\rangle & \psi^{+,2}_{f,m,n}|m-2\rangle \end{pmatrix}^T. \tag{G.5}$$

Here $m \geq 3$ denotes the eigenvalue of the total angular momentum operator $\hat{L}_{\eta=+}$. The integer $r$ labels the rank of the matrix representation of the Hamiltonian within a fixed-$m$ sector (here $r = 6$ for $m \geq 3$), while $n = 1, \ldots, r$ labels the eigenstates within that sector. For $m < 3$ there are three sets of anomalous levels for which the Hamiltonian annihilates some spinor components, reducing the effective rank:

$$\Psi^{+,1}_{m=0,1} = \begin{pmatrix} 0 \\ 0 \\ |0\rangle \\ 0 \\ 0 \\ 0 \end{pmatrix}, \Psi^{+,3}_{m=1,n} = \begin{pmatrix} \psi^{+,1}_{c3,1,n}|0\rangle \\ 0 \\ \psi^{+,1}_{c12,1,n}|1\rangle \\ 0 \\ \psi^{+,1}_{f,1,n}|0\rangle \\ 0 \end{pmatrix}, \quad \text{and,} \quad \Psi^{+,5}_{m=2,n} = \begin{pmatrix} \psi^{+,1}_{c3,2,n}|1\rangle \\ \psi^{+,2}_{c3,2,n}|0\rangle \\ \psi^{+,1}_{c12,2,n}|2\rangle \\ 0 \\ \psi^{+,1}_{f,2,n}|1\rangle \\ \psi^{+,2}_{f,2,n}|0\rangle \end{pmatrix}. \tag{G.6}$$

For $r = 1$ we obtain a single state, for $r = 3$ we obtain three states, and for $r = 5$ we obtain 5 states whose energies we determine numerically in Sec.(G 3) and perturbatively in Sec.(H). The $x$ component of the simplified current ($\lambda = 0$) is

$$\sqrt{2}\frac{\hbar}{\ell_B} J^{\mu=x;\eta} = \eta \begin{pmatrix} 0 & \Delta_B\sigma_0 & \Delta'_B\sigma_x \\ \Delta_B\sigma_0 & 0 & 0 \\ \Delta'_B\sigma_x & 0 & 0 \end{pmatrix} \tag{G.7}$$

and the $y$ component is

$$\sqrt{2}\frac{\hbar}{\ell_B} J^{\mu=y;\eta} = \begin{pmatrix} 0 & i\Delta_B\sigma_z & \Delta'_B\sigma_y \\ -i\Delta_B\sigma_z & 0 & 0 \\ \Delta'_B\sigma_y & 0 & 0 \end{pmatrix} \tag{G.8}$$

We infer the selection rules by checking when the current matrix element $\Psi^{\eta,r'\dagger}_{m',n'} J^{\mu,\eta} \Psi^{\eta,r}_{m,n}$ vanishes identically. We first check the $r = r' = 6$ case. For the $x$ component of the current acting on valley $\eta = +$:

$$\sqrt{2}\frac{\hbar}{\ell_B} J^{x;+} \Psi^{+,6}_{m,n} = \begin{pmatrix} \Delta_B\psi^1_{c12,m}|m\rangle + \Delta'_B\psi^2_{f,m}|m-2\rangle \\ \Delta_B\psi^2_{c12,m}|m-3\rangle + \Delta'_B\psi^1_{f,m}|m-1\rangle \\ \Delta_B\psi^1_{c3,m}|m-1\rangle \\ \Delta_B\psi^2_{c3,m}|m-2\rangle \\ \Delta'_B\psi^2_{c3,m}|m-2\rangle \\ \Delta'_B\psi^1_{c3,m}|m-1\rangle \end{pmatrix} \tag{G.9}$$

Taking the inner product with $\Psi^{+,6}_{m',n'}$:

$$\begin{aligned} \sqrt{2}\frac{\hbar}{\ell_B} \Psi^{+,6\dagger}_{m',n'} J^{x,+} \Psi^{+,6}_{m,n} = & \\ \Delta_B \overline{\psi^{+,1}_{c3,m',n'}} \psi^{+,1}_{c12,m,n} \delta_{m'-1,m} &+ \Delta'_B \overline{\psi^{+,1}_{c3,m',n'}} \psi^{+,2}_{f,m,n} \delta_{m'-1,m-2} \\ +\Delta_B \overline{\psi^{+,2}_{c3,m',n'}} \psi^{+,2}_{c12,m,n} \delta_{m'-2,m-3} &+ \Delta'_B \overline{\psi^{+,2}_{c3,m',n'}} \psi^{+,1}_{f,m,n} \delta_{m'-2,m-1} \\ &+\Delta_B \overline{\psi^{+,1}_{c12,m',n'}} \psi^{+,1}_{c3,m,n} \delta_{m',m-1} \\ &+\Delta_B \overline{\psi^{+,2}_{c12,m',n'}} \psi^{+,2}_{c3,m,n} \delta_{m'-3,m-2} \\ &+\Delta'_B \overline{\psi^{+,1}_{f,m',n'}} \psi^{+,2}_{c3,m,n} \delta_{m'-1,m-2} \\ &+\Delta'_B \overline{\psi^{+,2}_{f,m',n'}} \psi^{+,1}_{c3,m,n} \delta_{m'-2,m-1} \end{aligned} \tag{G.10}$$

The matrix element is nonzero only if $m' = m \pm 1$. This selection rule follows from angular momentum conservation since the current operator carries total angular momentum $\Delta m = \pm 1$. Explicit calculation shows that this selection rule holds also for $\Psi^{+,r=5}_{m=2,n}$ and $\Psi^{+,r=3}_{m=1,n}$. For $\Psi^{+,r=1}_{m=0,1}$, the only other state that has a non-zero current matrix element is $\Psi^{+,r=3}_{m=1,n}$ since no spinor with $m = -1$ exists. The $y$ component of the current yields the same selection rule, with additional phases that do not alter the conclusion.

### 2. valley $\eta = -$

For valley $\eta = -$ the calculation proceeds analogously. Now the Hamiltonian is given by

$$\tilde{h}^{\mathrm{BI}}_{\eta=-}(B) = \begin{pmatrix} E_{c,\Gamma_3} & 0 & -i\Delta_B\hat{a}^\dagger & 0 & \gamma+\gamma_{\mathrm{int}} & i\Delta'_B\hat{a} \\ 0 & E_{c,\Gamma_3} & 0 & i\Delta_B\hat{a} & -i\Delta'_B\hat{a}^\dagger & \gamma+\gamma_{\mathrm{int}} \\ i\Delta_B\hat{a} & 0 & E_{c,\Gamma_{12}} & 0 & 0 & 0 \\ 0 & -i\Delta_B\hat{a}^\dagger & 0 & E_{c,\Gamma_{12}} & 0 & 0 \\ \gamma+\gamma_{\mathrm{int}} & i\Delta'_B\hat{a} & 0 & 0 & E_f & 0 \\ -i\Delta'_B\hat{a}^\dagger & \gamma+\gamma_{\mathrm{int}} & 0 & 0 & 0 & E_f \end{pmatrix} \tag{G.11}$$

The angular momentum operator in valley $\eta = -$ is now $\hat{L}_{\eta=-} = \hat{a}^\dagger\hat{a}\mathbb{I}_{6x6} + S^{\eta=-}_{ps}$, where the pseudospin contribution is

$$S^{\eta=-}_{ps} = \begin{pmatrix} 2 &&&&& \\ & 1 &&&& \\ && 3 &&& \\ &&& 0 && \\ &&&& 2 & \\ &&&&& 1 \end{pmatrix}. \tag{G.12}$$

In this case, the spinor ansatz that block-diagonalizes the eigenvalue problem is

$$\Psi^{\eta=-,r=6}_{m,n} = \begin{pmatrix} \psi^{-,1}_{c3,m,n}|m-2\rangle & \psi^{-,2}_{c3,m,n}|m-1\rangle & \psi^{-,1}_{c12,m,n}|m-3\rangle & \psi^{-,2}_{c12,m,n}|m\rangle & \psi^{-,1}_{f,m,n}|m-2\rangle & \psi^{-,2}_{f,m,n}|m-1\rangle \end{pmatrix}^T. \tag{G.13}$$

The anomalous modes in valley $\eta = -$ are

$$\Psi^{-,1}_{m=0,1} = \begin{pmatrix} 0 \\ 0 \\ 0 \\ |0\rangle \\ 0 \\ 0 \end{pmatrix}, \Psi^{-,3}_{m=1,n} = \begin{pmatrix} 0 \\ \psi^{-,2}_{c3,1,n}|0\rangle \\ 0 \\ \psi^{-,2}_{c12,1,n}|1\rangle \\ 0 \\ \psi^{-,2}_{f,1,n}|0\rangle \end{pmatrix}, \quad \text{and,} \quad \Psi^{-,5}_{m=2,n} = \begin{pmatrix} \psi^{-,1}_{c3,2,n}|0\rangle \\ \psi^{-,2}_{c3,2,n}|1\rangle \\ 0 \\ \psi^{-,2}_{c12,2,n}|2\rangle \\ \psi^{-,1}_{f,2,n}|0\rangle \\ \psi^{-,2}_{f,2,n}|1\rangle \end{pmatrix}. \tag{G.14}$$

We can again calculate the matrix elements of the current operator on these states. For the x component of the current, we have

$$\sqrt{2}\frac{\hbar}{\ell_B}J^{\mu=x;-}\Psi^{-,6}_{m,n} = -\begin{pmatrix} \Delta_B\psi^{-,1}_{c12,m,n}|m-3\rangle + \Delta'_B\psi^{-,2}_{f,m,n}|m-1\rangle \\ \Delta_B\psi^{-,2}_{c12,m,n}|m\rangle + \Delta'_B\psi^{-,1}_{f,m,n}|m-2\rangle \\ \Delta_B\psi^{-,1}_{c3,m,n}|m-2\rangle \\ \Delta_B\psi^{-,2}_{c3,m,n}|m-1\rangle \\ \Delta'_B\psi^{-,2}_{c3,m,n}|m-1\rangle \\ \Delta'_B\psi^{-,1}_{c3,m,n}|m-2\rangle \end{pmatrix} \tag{G.15}$$

Taking the inner product with $\Psi^{-,6}_{m',n'}$, we find

$$\begin{aligned} -\sqrt{2}\frac{\hbar}{\ell_B}\Psi^{-,6\dagger}_{m',n'}J^{x;-}\Psi^{-,6}_{m,n} &= \\ \Delta_B\overline{\psi^{-,1}_{c3,m',n'}}\psi^{-,1}_{c12,m,n}\delta_{m'-2,m-3} + \Delta'_B\overline{\psi^{-,1}_{c3,m',n'}}\psi^{-,2}_{f,m,n}\delta_{m'-2,m-1} & \\ +\Delta_B\overline{\psi^{-,2}_{c3,m',n'}}\psi^{-,2}_{c12,m,n}\delta_{m'-1,m} + \Delta'_B\overline{\psi^{-,2}_{c3,m',n'}}\psi^{-,1}_{f,m,n}\delta_{m'-1,m-2} & \\ +\Delta_B\overline{\psi^{-,1}_{c12,m',n'}}\psi^{-,1}_{c3,m,n}\delta_{m'-3,m-2} & \\ +\Delta_B\overline{\psi^{-,2}_{c12,m',n'}}\psi^{-,2}_{c3,m,n}\delta_{m',m-1} & \\ +\Delta'_B\overline{\psi^{-,1}_{f,m',n'}}\psi^{-,2}_{c3,m,n}\delta_{m'-2,m-1} & \\ +\Delta'_B\overline{\psi^{-,2}_{f,m',n'}}\psi^{-,1}_{c3,m,n}\delta_{m'-1,m-2}. & \end{aligned} \tag{G.16}$$

Once again, states with $|m-m'| > 1$ are not connected by the current operator. There are no inter-valley current matrix elements since the current is diagonal in $\eta$ by momentum conservation.

### 3. Spectrum in the SO(2) limit

In this section we compute the spectrum in the $SO(2)$ limit and identify the lowest-energy transitions observed in experiment. Acting with the Hamiltonian on the ansatz spinor Eq. (G.5), we obtain $\tilde{h}^{\mathrm{BI}}_{\eta=+}(B)\,\Psi^{+,6}_{m+1,n} = \tilde{h}^{\mathrm{BI},r=6}_{\eta=+,m+1}(B)\,\Psi^{+,6}_{m+1,n}$ where the Hamiltonian matrix elements are

$$\tilde{h}^{\mathrm{BI},r=6}_{\eta=+,m+1,bb'}(B) = \begin{pmatrix} E_{c,\Gamma_3} & 0 & -i\Delta_B\sqrt{m+1} & 0 & (\gamma\Sigma(m)+\gamma_{\mathrm{int}}) & i\Delta'_B\Sigma(m-1)\sqrt{m} \\ 0 & E_{c,\Gamma_3} & 0 & i\Delta_B\sqrt{m-1} & -i\Delta'_B\Sigma(m)\sqrt{m} & (\gamma\Sigma(m-1)+\gamma_{\mathrm{int}}) \\ i\Delta_B\sqrt{m+1} & 0 & E_{c,\Gamma_{12}} & 0 & 0 & 0 \\ 0 & -i\Delta_B\sqrt{m-1} & 0 & E_{c,\Gamma_{12}} & 0 & 0 \\ (\gamma\Sigma(m)+\gamma_{\mathrm{int}}) & i\Delta'_B\Sigma(m)\sqrt{m} & 0 & 0 & E_f & 0 \\ -i\Delta'_B\Sigma(m-1)\sqrt{m} & (\gamma\Sigma(m-1)+\gamma_{\mathrm{int}}) & 0 & 0 & 0 & E_f \end{pmatrix}_{bb'} \tag{G.17}$$

where $m \geq 2$ and $\Sigma(m) = e^{-\left(\frac{\lambda}{\ell_B}\right)^2\left(m+\frac{1}{2}\right)}$. The Hamiltonian annihilates a subset of components of the anomalous states defined in Eq. (G.6). The remaining nonvanishing components can therefore be collected into reduced spinors of the form

$$\tilde{\Psi}^{+,1}_{m=0,1} = \left(\,|0\rangle\,\right), \quad \tilde{\Psi}^{+,3}_{m=1,n} = \begin{pmatrix} \psi^{+,1}_{c3,1,n}|0\rangle \\ \psi^{+,1}_{c12,1,n}|1\rangle \\ \psi^{+,1}_{f,1,n}|0\rangle \end{pmatrix}, \quad \tilde{\Psi}^{+,5}_{m=2,n} = \begin{pmatrix} \psi^{+,1}_{c3,2,n}|1\rangle \\ \psi^{+,2}_{c3,2,n}|0\rangle \\ \psi^{+,1}_{c12,2,n}|2\rangle \\ \psi^{+,1}_{f,2,n}|1\rangle \\ \psi^{+,2}_{f,2,n}|0\rangle \end{pmatrix}. \tag{G.18}$$

With these definitions, the action of the Hamiltonian within the anomalous subspace can be expressed as $\tilde{h}^{\mathrm{BI}}_{\eta=+}(B)\Psi^{+,r}_{m,n} = \tilde{h}^{\mathrm{BI},r}_{\eta=+,m+1}(B)\tilde{\Psi}^{+,r}_{m,n}$, for $r = 1,3,5$. The corresponding matrix elements are

$$\tilde{h}^{\mathrm{BI},r=1}_{\eta=+,0,bb'}(B) = \left(E_{c,\Gamma_{12}}\right)_{bb'}, \tag{G.19a}$$

$$\tilde{h}^{\mathrm{BI},r=3}_{\eta=+,1,bb'}(B) = \begin{pmatrix} E_{c,\Gamma_3} & -i\Delta_B & (\gamma\Sigma(0)+\gamma_{\mathrm{int}}) \\ i\Delta_B & E_{c,\Gamma_{12}} & 0 \\ (\gamma\Sigma(0)+\gamma_{\mathrm{int}}) & 0 & E_f \end{pmatrix}_{bb'}, \tag{G.19b}$$

$$\tilde{h}^{\mathrm{BI},r=5}_{\eta=+,2,bb'}(B) = \begin{pmatrix} E_{c,\Gamma_3} & 0 & -i\Delta_B\sqrt{2} & (\gamma\Sigma(1)+\gamma_{\mathrm{int}}) & i\Delta'_B\Sigma(0) \\ 0 & E_{c,\Gamma_3} & 0 & -i\Delta'_B\Sigma(1) & (\gamma\Sigma(0)+\gamma_{\mathrm{int}}) \\ i\Delta_B\sqrt{2} & 0 & E_{c,\Gamma_{12}} & 0 & 0 \\ (\gamma\Sigma(1)+\gamma_{\mathrm{int}}) & i\Delta'_B\Sigma(1) & 0 & E_f & 0 \\ -i\Delta'_B\Sigma(0) & (\gamma\Sigma(0)+\gamma_{\mathrm{int}}) & 0 & 0 & E_f \end{pmatrix}_{bb'}. \tag{G.19c}$$

Diagonalizing the Hamiltonian within each angular-momentum sector $m$ yields the spectrum shown in Fig. 1, where we highlight the lowest-energy allowed transitions from the flat bands to the remote bands at $\nu = 4$. The dominant current matrix elements given in Eqs. (G.7,G.8) are approximately independent of $m$, which is a generic property of Dirac electrons. Since the excitation energy of the allowed transitions between the flat and remote bands increases with $m$, while the optical matrix elements in Eq. (F.9) scale inversely with the excitation energy, the overall dipole matrix elements decrease for larger $m$. Consequently, transitions involving larger $m$ occur at higher energies and are correspondingly suppressed in intensity.

In Fig. 1b we identify the transition labeled $m_3$ as arising from the anomalous $m = 0$ Landau level (LL) in the flat bands to the $m = 1$ anomalous state that emanates from the Rashba point in the remote bands. This transition is allowed due to the relative shift of the angular momentum at the Rashba point, which originates from the $\Gamma_3$ representation.

The lowest-energy transitions are labeled $m_1$. These correspond to two processes: (i) a transition between the $m = 1$ state in the flat bands and the $m = 2$ state in the remote bands, which disperses downward in energy with increasing magnetic field, and (ii) a transition between the lowest-energy $m = 3$ level in the flat bands and the same $m = 2$ level in the remote bands. A second transition from the higher-energy $m = 3$ level in the flat bands to the same $m = 2$ remote-band state is also symmetry allowed and contributes to $m_1$; however, this transition has a smaller dipole matrix element because the corresponding $m = 3$ state carries a larger $f$-electron weight, which reduces its overlap with the current operator. The intermediate-energy optical transitions between $m_1$ and $m_3$ are labeled by $m_2$. These correspond either to a transition between the $m = 2$ state in the flat bands and the lowest-energy $m = 3$ state in the remote bands, or to a transition between the lowest-energy $m = 4$ level in the flat bands and the same $m = 3$ level in the remote bands. The higher-energy $m = 4$ state in the flat bands also has an enhanced $f$-electron character, which similarly reduces the corresponding dipole matrix element.

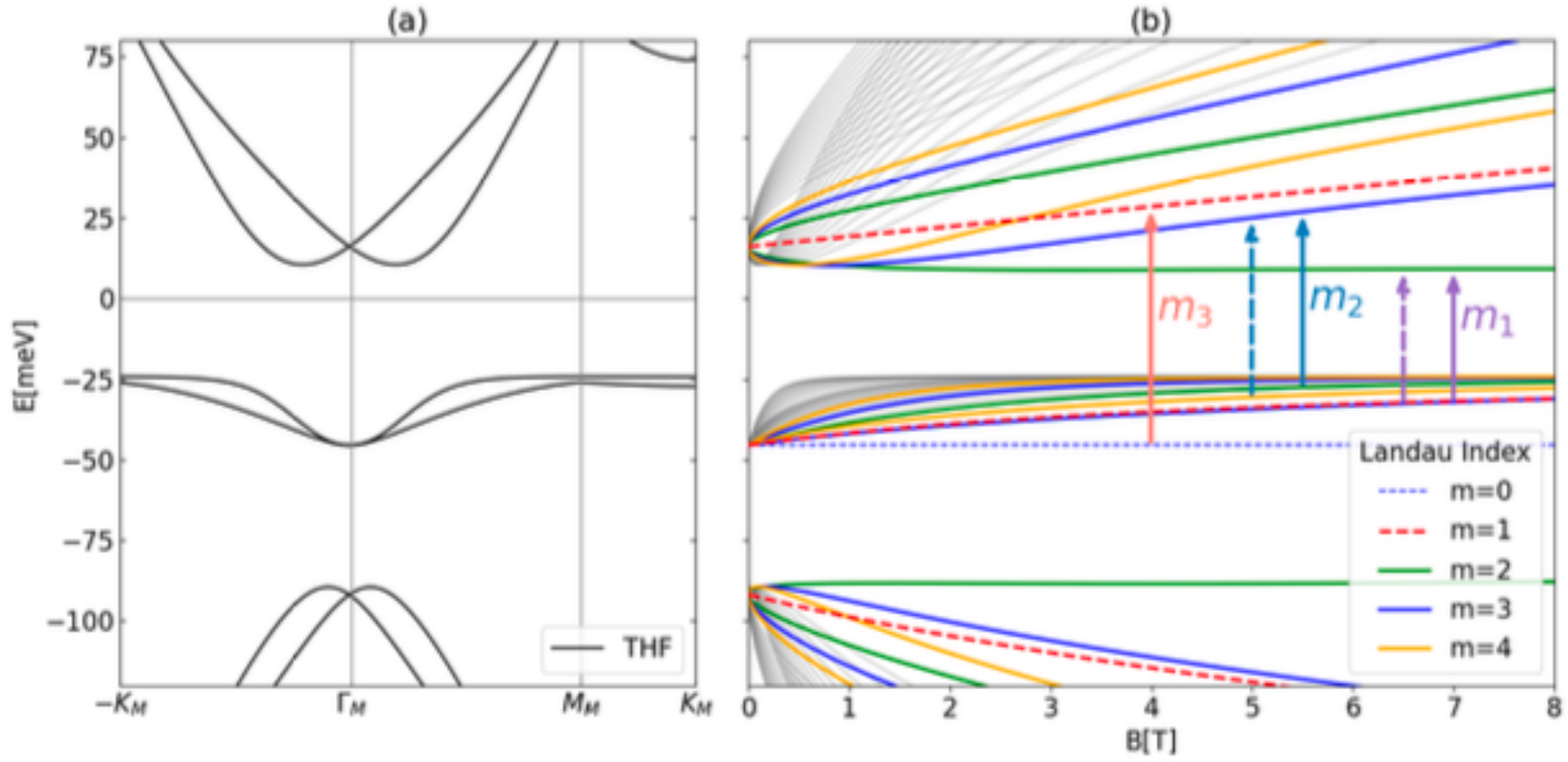

Figure 1. (a) Band structure of the THF model at $\nu = +4$ at twist angle $\theta = 1.06$ in the absence of relaxation and strain in the $M = 0$ limit using the $D2$ parameters in Tab.(III). (b) LL spectra calculated from diagonalizing Eqs.(G.19a-G.19c,G.17). We denote the lowest LL transitions that are allowed by the selection rules. Here, $m_3$ corresponds to a transition from $m = 0$ to $m = 1$. Next, $m_2$ corresponds to a transition between $m = 2$ to $m = 3$. Note that the transition between $m = 4$ to $m = 3$ has similar transition energy which we denote by a blue dashed arrow. Finally, $m_1$ corresponds to a transition between $m = 1$ to $m = 2$. Note that the transition between $m = 3$ to $m = 2$ has similar transition energy which we denote by a purple dashed arrow.

Defining the energy of the $n$th eigenstate in the $m$ angular momentum sector as $E^n_m(B)$, we denote the lowest $m = 0, 1, 2, 3, 4$ allowed transitions at $\nu = 4$ by

$$\Delta E^1_{m_1}(B) = E^{n=5}_{m=2}(B) - E^{n=2}_{m=1}(B) \tag{G.20a}$$
$$\Delta E^2_{m_1}(B) = E^{n=5}_{m=2}(B) - E^{n=4}_{m=3}(B) \tag{G.20b}$$
$$\Delta E^1_{m_2}(B) = E^{n=5}_{m=3}(B) - E^{n=4}_{m=2}(B) \tag{G.20c}$$
$$\Delta E^2_{m_2}(B) = E^{n=5}_{m=3}(B) - E^{n=4}_{m=4}(B) \tag{G.20d}$$
$$\Delta E^1_{m_3}(B) = E^{n=1}_{m=1}(B) - E^{n=1}_{m=0}(B). \tag{G.20e}$$

These transition energies are overlaid on top of the numerically calculated optical conductivity in Fig.(2) using the ($D2$) parameters in Tab.(III) (see Sec.(F) and Sec.(I)). Note that due to the size of the broadening ($\zeta$ = 3meV), different allowed transitions appear as a single peak for the $m_1$ and $m_2$ transitions. In Sec.(K) we analyze in detail the dependence of the optical map with varying $\zeta$. In Sec.(H) we obtain analytical approximations for $\Delta E^1_{m_1}(B)$ and $\Delta E^1_{m_3}(B)$ which correspond to transitions between the 'anomalous' LLs. Then in Sec.(I) we use these energies obtained numerically to obtain the best parameter match to the experiment.

### Appendix H: Perturbation theory and analytical expressions for the low $m$ states

In this section we derive analytical expressions for the Landau-level energies relevant for the lowest optical transitions. We focus on the $\eta = +$ valley and set $M = 0$. The derivation proceeds by solving the reduced Hamiltonian in each angular-momentum sector. For each sector we first determine the $B \to 0$ spectrum using perturbation theory in $\Delta_B$, then analyze the opposite $B \to \infty$ limit. Finally, we interpolate between these limits using a two-point Padé approximant explained below.

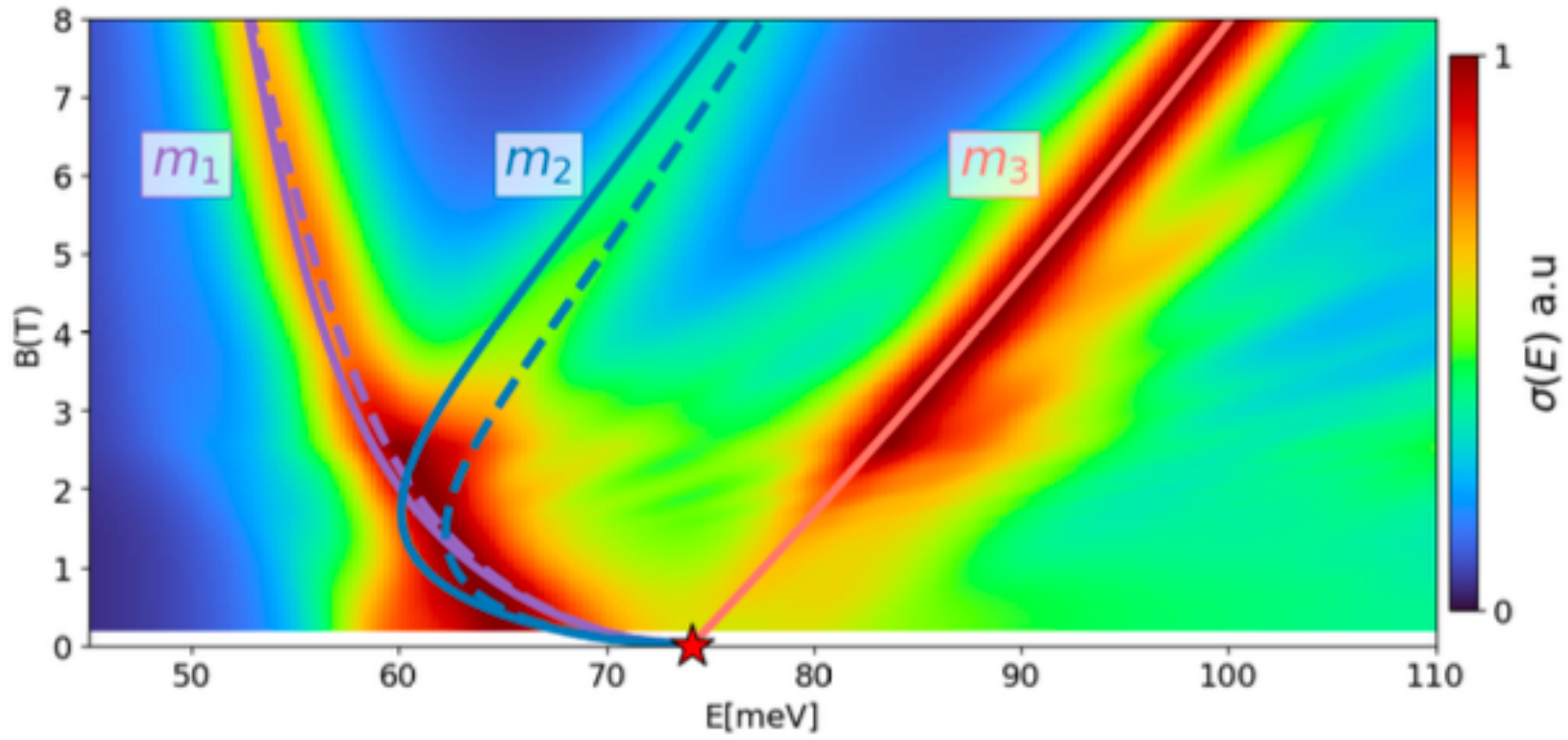

Figure 2. Optical conductivity of the THF model at $\nu = +4$ at twist angle $\theta = 1.06$ in the absence of relaxation and strain in the $M = 0$ limit using the $D2$ parameters in Tab.(III). We normalize the optical conductivity at each $B$ to match the normalization of the experimental data in the manuscript. Here, $m_3$ corresponds to a transition from $m = 0$ to $m = 1$. Next, $m_2$ corresponds to a transition between $m = 2$ to $m = 3$. Note that the transition between $m = 4$ to $m = 3$ has similar transition energy which we denote by a blue dashed line. Finally, $m_1$ corresponds to a transition between $m = 1$ to $m = 2$. Note that the transition between $m = 3$ to $m = 2$ has similar transition energy which we denote by a purple dashed line. We denote the intercept of the $m_3$ transition by a red star.

#### a. $m = 0$ *energies*

For the $m = 0$ state the Hamiltonian reduces to the flat-band sector. The corresponding Landau level has energy

$$E^{n=0}_{m=0} = E_{c,\Gamma_{12}}, \tag{H.1}$$

and its wavefunction is entirely in the $\Gamma_1 \oplus \Gamma_2$ sector, as shown in Eq. (G.6).

#### b. $m = 1$ *Energies*

For the $m = 1$ block of the Hamiltonian, we solve the reduced eigenvalue problem for the matrix in Eq.(G.19b). To obtain the spectrum at finite $B$ field we do perturbation theory on $\Delta_B$. There are two limits in which we can solve the eigenvalue problem exactly, the small and the large field limits. In the small field limit, $\Delta_B \to 0$, we can use the small field behavior of $\Sigma(0) = e^{-\frac{1}{2}\left(\frac{\lambda}{\ell_B}\right)^2} \approx 1 - \frac{1}{2}\left(\frac{\lambda}{\ell_B}\right)^2$ (see Ref.([7])). With this approximation the Hamiltonian becomes

$$\tilde{h}^{\mathrm{BI},3}_{\eta=+,1,bb'}(B) \approx \begin{pmatrix} E_{c,\Gamma_3} & -i\Delta_B & \left(\gamma\left(1-\frac{1}{2}\frac{\Delta_B^2}{\Delta_\lambda^2}\right)+\gamma_{\mathrm{int}}\right) \\ i\Delta_B & E_{c,\Gamma_{12}} & 0 \\ \left(\gamma\left(1-\frac{1}{2}\frac{\Delta_B^2}{\Delta_\lambda^2}\right)+\gamma_{\mathrm{int}}\right) & 0 & E_f \end{pmatrix}_{bb'}. \tag{H.2}$$

Here we defined $\Delta_\lambda = \sqrt{2}\frac{\hbar}{\lambda}v_\star$, and $\gamma_{\mathrm{int}}$ was defined in Eq. (D.5). For later convenience, we define

$$\Omega_0 = \sqrt{4\left(\gamma+\gamma_{\mathrm{int}}\right)^2 + \left(E_{c,\Gamma_3} - E_f\right)^2}. \tag{H.3}$$

The eigenvalues $E^n_{m=1}(B=0)$ of $\tilde{h}^{\mathrm{BI},3}_{\eta=+,1,bb'}(B=0)$ can be obtained exactly:

$$E^{n=1}_{m=1}(B=0) = \frac{1}{2}\left(E_{c,\Gamma_3} + E_f + \Omega_0\right), \tag{H.4a}$$

$$E^{n=2}_{m=1}(B=0) = E_{c,\Gamma_{12}}, \tag{H.4b}$$

$$E^{n=3}_{m=1}(B=0) = \frac{1}{2}\left(E_{c,\Gamma_3} + E_f - \Omega_0\right). \tag{H.4c}$$

The level $E_{m=1}^{n=1}(B=0)$ corresponds to the Rashba point in the remote conduction band, which we denote

$$E_{r+} = \frac{1}{2}\left(E_{c,\Gamma_3} + E_f + \Omega_0\right), \tag{H.5}$$

while $E_{m=1}^{n=3}(B=0)$ corresponds to the Rashba point in the remote valence band,

$$E_{r-} = \frac{1}{2}\left(E_{c,\Gamma_3} + E_f - \Omega_0\right). \tag{H.6}$$

We now determine the leading corrections that control the dispersion of these levels with magnetic field at small $B$. To leading order, we obtain

$$E_{m=1}^{n=1}(B\to 0) = E_{r+} + \frac{1}{2}\left(1 + \frac{E_{c,\Gamma_3} - E_f}{\Omega_0}\right)\frac{\Delta_B^2}{E_{r+} - E_{c,\Gamma_{12}}} + \gamma\sqrt{1 - \left(\frac{E_{c,\Gamma_3} - E_f}{\Omega_0}\right)^2}\frac{\Delta_B^2}{2\Delta_\lambda^2}, \tag{H.7a}$$

$$E_{m=1}^{n=2}(B\to 0) = E_{c,\Gamma_{12}} + \Delta_B^2 \frac{(E_f - E_{c,\Gamma_{12}})}{(\gamma+\gamma_{\text{int}})^2 - \left(E_{c,\Gamma_{12}} - E_{c,\Gamma_3}\right)\left(E_{c,\Gamma_{12}} - E_f\right)}, \tag{H.7b}$$

$$E_{m=1}^{n=3}(B\to 0) = E_{r-} + \frac{1}{2}\left(1 - \frac{E_{c,\Gamma_3} - E_f}{\Omega_0}\right)\frac{\Delta_B^2}{E_{r-} - E_{c,\Gamma_{12}}} - \gamma\sqrt{1 - \left(\frac{E_{c,\Gamma_3} - E_f}{\Omega_0}\right)^2}\frac{\Delta_B^2}{2\Delta_\lambda^2}. \tag{H.7c}$$

The $m=0 \to m=1$ transition energy at $\nu=4$ is therefore

$$\Delta E_{m_3}(B) \approx E_{m=1}^{n=1}(B\to 0) - E_{m=0}^{n=0}(B\to 0) \tag{H.8}$$

$$= E_{r+} - E_{c,\Gamma_{12}} + \frac{\Delta_B^2}{2\Omega_0}\frac{\Omega_0 + E_{c,\Gamma_3} - E_f}{E_{r+} - E_{c,\Gamma_{12}}} + \gamma\sqrt{1 - \left(\frac{E_{c,\Gamma_3} - E_f}{\Omega_0}\right)^2}\frac{\Delta_B^2}{2\Delta_\lambda^2}. \tag{H.9}$$

In the $B\to\infty$ limit, $\Delta_B$ becomes the dominant energy scale and we instead perform perturbation theory in the hybridization, $\gamma_{\text{int}}$. In this limit, $\Sigma(0) = e^{-\frac{1}{2}\left(\frac{\lambda}{\ell_B}\right)^2} \to 0$ as $B\to\infty$. The high-energy $2\times 2$ block then yields

$$E_{m=1}^{n=1}(B\to\infty) = \frac{E_{c,\Gamma_3} + E_{c,\Gamma_{12}}}{2} + \sqrt{\Delta_B^2 + \frac{(E_{c,\Gamma_3} - E_{c,\Gamma_{12}})^2}{4}}, \tag{H.10a}$$

$$E_{m=1}^{n=3}(B\to\infty) = \frac{E_{c,\Gamma_3} + E_{c,\Gamma_{12}}}{2} - \sqrt{\Delta_B^2 + \frac{(E_{c,\Gamma_3} - E_{c,\Gamma_{12}})^2}{4}}. \tag{H.10b}$$

The low-energy eigenvalue is obtained using second-order perturbation theory,

$$E_{m=1}^{n=2}(B\to\infty) = E_f - \begin{pmatrix}\gamma_{\text{int}} & 0\end{pmatrix}\begin{pmatrix}E_{c,\Gamma_3} & -i\Delta_B \\ i\Delta_B & E_{c,\Gamma_{12}}\end{pmatrix}^{-1}\begin{pmatrix}\gamma_{\text{int}} \\ 0\end{pmatrix} \tag{H.11}$$

$$= E_f - \frac{\gamma_{\text{int}}^2 E_{c,\Gamma_{12}}}{E_{c,\Gamma_{12}}E_{c,\Gamma_3} - \Delta_B^2}. \tag{H.12}$$

To obtain an analytical expression that interpolates between the small- and large-field limits, we use a two-point Padé approximant. A two-point Padé approximant is a rational function constructed so that it reproduces both the asymptotic value of $E_{m=1}^{n=2}(B)$ as $B\to\infty$, as well as the second order term in the Taylor expansion of $E_{m=1}^{n=2}(B)$ about $B=0$. In the present case, this yields

$$E_{m=1}^{n=2}(B) \approx E_f + \frac{E_{c,\Gamma_{12}} - E_f}{1 + \Delta_B^2 \frac{1}{(\gamma+\gamma_{\text{int}})^2 - (E_{c,\Gamma_{12}} - E_{c,\Gamma_3})(E_{c,\Gamma_{12}} - E_F)}}. \tag{H.13}$$

### c. $m = 2$ Energies

We now analyze the $m = 2$ sector of the Hamiltonian. We again begin with the $B \to 0$ limit, where the Hamiltonian takes the approximate form

$$\tilde{h}^{\mathrm{BI},5}_{\eta=+,2,bb'}(B) \tag{H.14}$$
$$\approx \begin{pmatrix} E_{c,\Gamma_3} & 0 & -i\Delta_B\sqrt{2}\left(\gamma\left(1-\frac{3}{2}\frac{\Delta_B^2}{\Delta_\lambda^2}\right)+\gamma_{\mathrm{int}}\right) & i\Delta'_B & \\ 0 & E_{c,\Gamma_3} & 0 & -i\Delta'_B & \left(\gamma\left(1-\frac{1}{2}\frac{\Delta_B^2}{\Delta_\lambda^2}\right)+\gamma_{\mathrm{int}}\right) \\ i\Delta_B\sqrt{2} & 0 & E_{c,\Gamma_{12}} & 0 & 0 \\ \left(\gamma\left(1-\frac{3}{2}\frac{\Delta_B^2}{\Delta_\lambda^2}\right)+\gamma_{\mathrm{int}}\right) & i\Delta'_B & 0 & E_f & 0 \\ -i\Delta'_B & \left(\gamma\left(1-\frac{1}{2}\frac{\Delta_B^2}{\Delta_\lambda^2}\right)+\gamma_{\mathrm{int}}\right) & 0 & 0 & E_f \end{pmatrix}_{bb'}$$

First, we consider the chiral limit where $\Delta'_B = 0$. At $B = 0$, the eigenvalues are

$$E^{n=1}_{m=2}(B=0) = E_{r+} \tag{H.15a}$$
$$E^{n=2}_{m=2}(B=0) = E_{r+} \tag{H.15b}$$
$$E^{n=3}_{m=2}(B=0) = E_{c,\Gamma_{12}} \tag{H.15c}$$
$$E^{n=4}_{m=2}(B=0) = E_{r-} \tag{H.15d}$$
$$E^{n=5}_{m=2}(B=0) = E_{r-}. \tag{H.15e}$$

We now compute the leading-order corrections produced by finite $\Delta_B$ using second-order perturbation theory in the chiral limit. To make the chiral limit explicit we indicate $\Delta'_B = 0$ in the notation $E^{n=1}_{m=2}(B, \Delta'_B = 0)$. The leading order energies are given by

$$E^{n=1}_{m=2}(B\to 0, \Delta'_B = 0) = E_{r+} + \frac{1}{2}\left(1 + \frac{E_{c,\Gamma_3} - E_f}{\Omega_0}\right)\frac{2\Delta_B^2}{E_{r+} - E_{c,\Gamma_{12}}} + \gamma\sqrt{1 - \left(\frac{E_{c,\Gamma_3} - E_f}{\Omega_0}\right)^2}\frac{3\Delta_B^2}{2\Delta_\lambda^2} \tag{H.16a}$$

$$E^{n=2}_{m=2}(B\to 0, \Delta'_B = 0) = E_{r+} + \gamma\sqrt{1 - \left(\frac{E_{c,\Gamma_3} - E_f}{\Omega_0}\right)^2}\frac{\Delta_B^2}{2\Delta_\lambda^2} \tag{H.16b}$$

$$E^{n=3}_{m=2}(B\to 0, \Delta'_B = 0) = E_{c,\Gamma_{12}} + \frac{2\Delta_B^2\left(E_{c,\Gamma_{12}} - E_f\right)}{(\gamma + \gamma_{\mathrm{int}})^2 - \left(E_{c,\Gamma_{12}} - E_{c,\Gamma_3}\right)\left(E_{c,\Gamma_{12}} - E_f\right)} \tag{H.16c}$$

$$E^{n=4}_{m=2}(B\to 0, \Delta'_B = 0) = E_{r-} - \gamma\sqrt{1 - \left(\frac{E_{c,\Gamma_3} - E_f}{\Omega_0}\right)^2}\frac{\Delta_B^2}{2\Delta_\lambda^2} \tag{H.16d}$$

$$E^{n=5}_{m=2}(B\to 0, \Delta'_B = 0) = E_{r-} + \frac{1}{2}\left(1 - \frac{E_{c,\Gamma_3} - E_f}{\Omega_0}\right)\frac{2\Delta_B^2}{E_{r-} - E_{c,\Gamma_{12}}} - \gamma\sqrt{1 - \left(\frac{E_{c,\Gamma_3} - E_f}{\Omega_0}\right)^2}\frac{3\Delta_B^2}{2\Delta_\lambda^2}. \tag{H.16e}$$

Note that the level $E^{n=2}_{m=2}(\Delta_B \to 0, \Delta'_B = 0)$, which participates in the $m = 1 \to m = 2$ transition, does not disperse with magnetic field in the strict $\lambda = 0$ limit. Since $\gamma < 0$, a finite $\lambda$ will induce a negative slope on $E^{n=2}_{m=2}(\Delta_B \to 0, \Delta'_B = 0)$. In the chiral limit, we obtain the following analytical approximation for lowest energy transition allowed by the selection rules up to angular momentum $m = 2$:

$$\Delta E^1_{m_1}(B) = E_{r+} + \gamma\sqrt{1 - \left(\frac{E_{c,\Gamma_3} - E_f}{\Omega_0}\right)^2}\frac{\Delta_B^2}{2\Delta_\lambda^2} - E_f - \frac{E_{c,\Gamma_{12}} - E_f}{1 + \Delta_B^2\frac{1}{(\gamma+\gamma_{\mathrm{int}})^2 - \left(E_{c,\Gamma_{12}} - E_{c,\Gamma_3}\right)\left(E_{c,\Gamma_{12}} - E_F\right)}} \tag{H.17}$$

Finally, we can carry out the same analysis for $E^{n=3}_{m=2}(B,, \Delta'_B = 0)$ as for $E^{n=2}_{m=1}(B)$ and find a Padé approximation

$$E^{n=2}_{m=1}(B) \approx E_f + \frac{E_{c,\Gamma_{12}} - E_f}{1 + 2\Delta_B^2\frac{1}{(\gamma+\gamma_{\mathrm{int}})^2 - \left(E_{c,\Gamma_{12}} - E_{c,\Gamma_3}\right)\left(E_{c,\Gamma_{12}} - E_F\right)}}. \tag{H.18}$$

A comparison between these analytical expressions and the numerical results is shown in Fig. (3). We find good agreement between the perturbative expressions and the numerical spectrum.

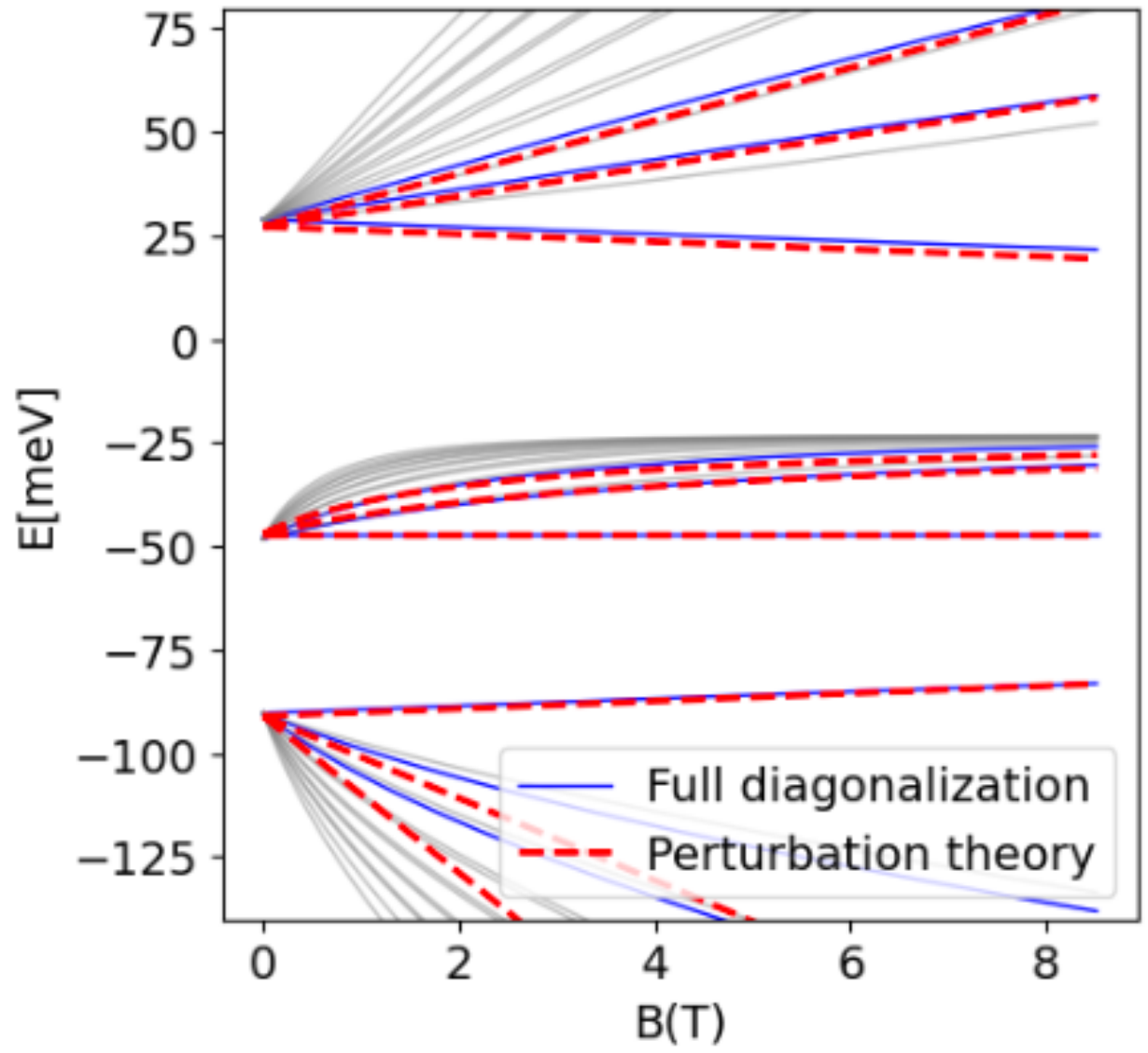

Figure 3. Comparison between the full diagonalization in the $v'_\star = 0, M = 0$ limit and the perturbation theory expressions using Eqs.(H.7a-H.7c), Eqs.(H.16a-H.16e) and Eqs.(H.13-H.18). For the flat-band LLs we show the Padé approximants instead of the perturbation theory expressions.

### 1. Simplifications in the chiral limit

In the $v'_\star = \lambda = 0$ limit, the Hartree–Fock spectrum simplifies considerably, allowing for analytic expressions for the relevant Landau-level transitions near $\nu = \pm 4$. In this section we derive approximate expressions for these transitions and show how their magnetic-field dependence can be used to extract the parameters of the THF model. The analysis proceeds in three steps: we first simplify the Hartree–Fock energies using realistic parameter estimates, then derive approximate Landau-level energies in the presence of a magnetic field, and finally relate the resulting transition energies to experimentally accessible fitting parameters.

We first express the differences between the various Hartree–Fock energies and estimate their relative magnitudes. We begin by defining $\tilde{U} = U_1 \frac{2N_f - 1}{2N_f} + 6U_2$ and $\tilde{W} = W_3 - \frac{J}{2N_f}$. We also make the simplifying assumption that $\nu_{c3} \approx 0$ which is accurate when only one of the two Rashba points per spin and valley is occupied by electrons, while the other remains empty. In the non-interacting case, we show exactly that $\nu_{c3} = 0$ (see Sec. (D 1)), which is consistent with the numerical results obtained using self-consistent Hartree-Fock, where we find $|\nu_{c1}| \gg |\nu_{c3}|$. Therefore, we approximate $\nu_{c1} \approx \nu_c$. With these simplifications, the energies that parameterize the interacting Hamiltonian at $\nu = \pm 4$ are given by

$$E_{c,\Gamma_3} = V_0 \nu_c + W_1 \nu_f + \mu_1, \tag{H.19}$$

$$E_{c,\Gamma_{12}} = V_0 \nu_c + \tilde{W} \nu_f + \mu_2, \tag{H.20}$$

$$E_f = \tilde{U} \nu_f + \tilde{W} \nu_c. \tag{H.21}$$

In practice, for realistic parameters corresponding to an inter-sublattice hopping ratio $\kappa = 0.6$–$0.7$ near the magic angle, we find $V_0 \approx \tilde{W}$. All interaction parameters, $V_0, \tilde{W}, W_3, \tilde{U}$, lie in the range of approximately 40–60 meV. We now estimate the energy differences that appear in the perturbative expressions in Sec. (H):

$$E_{c,\Gamma_3} - E_f = \left(V_0 - \tilde{W}\right) \nu_c + \left(W_1 - \tilde{U}\right) \nu_f + \mu_1, \tag{H.22}$$

$$E_f - E_{c,\Gamma_{12}} = \left(\tilde{W} - V_0\right) \nu_c + \left(\tilde{U} - \tilde{W}\right) \nu_f - \mu_2, \tag{H.23}$$

$$E_{c,\Gamma_3} - E_{c,\Gamma_{12}} = \left(W_1 - \tilde{W}\right) \nu_c + \mu_1 - \mu_2. \tag{H.24}$$

In the forthcoming discussion, we neglect relaxation for simplicity and set $\mu_1 = \mu_2 = 0$. The effective hybridization at $\nu = \pm 4$ is renormalized by interactions, we therefore introduce the notation

$$\tilde{\gamma} = \gamma + \gamma_{\text{int}}, \tag{H.25}$$
$$= \gamma - W_1 \chi_{cf}. \tag{H.26}$$

Note that if $\chi_{cf} > 0$, the magnitude of the effective hybridization, $\tilde{\gamma}$, is enhanced, $|\tilde{\gamma}| > |\gamma|$, since $\gamma < 0$. In the non-interacting limit we analytically find $\chi_{cf} > 0$ (see Sec. (D 1)), which is consistent with our self-consistent numerical results. Using realistic parameters for the inter-sublattice hopping ratio $\kappa = 0.6$–$0.7$ near the magic angle, we also find numerically that

$$|\tilde{\gamma}|^2 \gg |E_{c,\Gamma_3} - E_f|^2, \tag{H.27}$$

due to the enhancement of the hybridization by $W_1$.

Using Eq. (H.27) we then approximate

$$\Omega_0 = \sqrt{4\tilde{\gamma}^2 + (E_{c,\Gamma_3} - E_f)^2}, \tag{H.28}$$
$$\approx 2|\tilde{\gamma}|, \tag{H.29}$$

such that the Rashba-point energy becomes

$$E_{\text{r},+} = \frac{E_{c,\Gamma_3} + E_f}{2} + |\tilde{\gamma}|. \tag{H.30}$$

With these simplifications the Landau-level spectrum near the Rashba point can be obtained perturbatively in the magnetic field (see Sec.(H)). The levels relevant for the $m = 0 \to 1$ and $m = 1 \to 2$ transitions are

$$E^{n=1}_{m=0}(B) = E_{c,\Gamma_{12}}, \tag{H.31a}$$
$$E^{n=1}_{m=1}(B) = E_{\text{r},+} + \frac{1}{2}\left(1 + \frac{E_{c,\Gamma_3} - E_f}{2|\tilde{\gamma}|}\right)\frac{\Delta_B^2}{E_{\text{r},+} - E_{c,\Gamma_{12}}}, \tag{H.31b}$$
$$E^{n=2}_{m=1}(B) = E_f + \frac{E_{c,\Gamma_{12}} - E_f}{1 + \frac{\Delta_B^2}{\tilde{\gamma}^2}}, \tag{H.31c}$$
$$E^{n=2}_{m=2}(\Delta_B, \Delta'_B = 0) = E_{\text{r},+}. \tag{H.31d}$$

The relevant transition energies are

$$\Delta E^1_{m_3}(B) = E^{n=1}_{m=1}(B) - E^{n=1}_{m=0}(B), \tag{H.32}$$
$$= E_{\text{r},+} - E_{c,\Gamma_{12}} + \frac{1}{2}\left(1 + \frac{E_{c,\Gamma_3} - E_f}{2|\tilde{\gamma}|}\right)\frac{\Delta_B^2}{E_{\text{r},+} - E_{c,\Gamma_{12}}}, \tag{H.33}$$

and

$$\Delta E^1_{m_1}(B) = E^{n=2}_{m=2}(\Delta_B, \Delta'_B = 0) - E^{n=2}_{m=1}(B), \tag{H.34}$$
$$= E_{\text{r},+} - E_f + \frac{E_f - E_{c,\Gamma_{12}}}{1 + \frac{\Delta_B^2}{\tilde{\gamma}^2}}. \tag{H.35}$$

To facilitate comparison with experiment we parameterize these transitions using simple fitting forms

$$\Delta E^1_{m_3}(B) = \alpha_1 + \alpha_2 B, \tag{H.36}$$
$$\Delta E^1_{m_1}(B) = \frac{\alpha_1 + \alpha_3\alpha_4 B}{1 + \alpha_4 B}. \tag{H.37}$$

The extracted parameters correspond to

$$\alpha_1 = E_{\mathrm{r},+} - E_{c,\Gamma_{12}}, \tag{H.38a}$$

$$\alpha_2 = \frac{1}{2}\left(1 + \frac{E_{c,\Gamma_3} - E_f}{2\left|\tilde{\gamma}\right|}\right)\frac{2e\hbar v_\star^2}{E_{\mathrm{r},+} - E_{c,\Gamma_{12}}}, \tag{H.38b}$$

$$\alpha_3 = E_{\mathrm{r},+} - E_f, \tag{H.38c}$$

$$\alpha_4 = \frac{2e\hbar v_\star^2}{\tilde{\gamma}^2}. \tag{H.38d}$$

Introducing the definitions of the gap at $\Gamma$ at zero magnetic field, $\Delta_\Gamma = E_{\mathrm{r},+} - E_{c,\Gamma_{12}}$, and the bandwidth $W = E_f - E_{c,\Gamma_{12}}$, we obtain simple expressions for the parameters that determine the dispersion of each transition as a function of magnetic field:

$$\alpha_1 = \Delta_\Gamma, \tag{H.39a}$$

$$\alpha_2 = \left(1 - \frac{W}{\Delta_\Gamma}\right)\frac{e\hbar v_\star^2}{\left|\tilde{\gamma}\right|}, \tag{H.39b}$$

$$\alpha_3 = \Delta_\Gamma - W, \tag{H.39c}$$

$$\alpha_4 = \frac{2e\hbar v_\star^2}{\tilde{\gamma}^2}. \tag{H.39d}$$

These parameters have a clear physical interpretation. The intercepts of the transitions, denoted by a red star in Fig.(2), are approximately set by the renormalized gap at the $\Gamma_M$ point, $\Delta_\Gamma$. Additionally, the difference between the asymptotic values of the $m_1$ transition (see Fig.(2)) is directly related to the bandwidth of the flat bands,

$$W = \Delta E^1_{m_1}(0) - \Delta E^1_{m_1}(\infty). \tag{H.40}$$

Similarly, the slope of the $m_3$ transition (see Fig.(2)) is determined by the magnetic-field scale $B_{\star\gamma} = \left(2e\hbar\frac{v_\star^2}{\tilde{\gamma}^2}\right)^{-1}$, which is set by the renormalized Berry curvature of the flat bands at the $\Gamma_M$ point (see Refs. [1, 10]).

To relate the microscopic parameters of the THF model to the parameters $\alpha_1, \ldots \alpha_4$, we now derive simplified expressions for the energies $E_f$, $E_{c,\Gamma_{12}}$, and $E_{c,\Gamma_3}$. For the bandwidth, we obtain

$$W = \left(\tilde{W} - V_0\right)\nu_c + \left(\tilde{U} - \tilde{W}\right)\nu_f. \tag{H.41}$$

Note that at $\nu = \pm 4$ we have $|\nu_c| \ll |\nu_f|$, so that increasing $W_3 - J/2N_f$ reduces the bandwidth. For the gap at $\Gamma_M$ we obtain

$$\Delta_\Gamma = \frac{\left(\tilde{W} - V_0\right)\nu_c + \left(W_1 - \tilde{W} + \tilde{U}\right)\nu_f}{2} + \left|\tilde{\gamma}\right|. \tag{H.42}$$

Because the number of microscopic parameters exceeds the number of independent observables, the inversion problem for the $\alpha_1, \ldots, \alpha_4$ parameters is underconstrained. To simplify the analysis and allow for a crude estimate of the THF parameters, we assume $\tilde{W} \approx W_1 \approx V_0$. With this approximation, the expressions for $\Delta_\Gamma$ and $W$ simplify to

$$W = \left(\tilde{U} - \tilde{W}\right)\nu_f, \tag{H.43}$$

and the gap at the $\Gamma_M$ point becomes

$$\Delta_\Gamma = \frac{\tilde{U}\nu_f}{2} + \left|\tilde{\gamma}\right|. \tag{H.44}$$

Under these assumptions the effective THF parameters can be expressed in terms of the fitted coefficients from the $\Delta_{0\to1}$ and $\Delta_{0\to2}$ transitions as follows:

| | $\theta$ | $d$ | $\kappa$ | $\min_{\kappa,d}\chi_l^2(\kappa,d)$ |
|---|---|---|---|---|
| $D_1$ | 1.12 | 16 | 0.7 | 3.00987 |
| $D_2$ | 1.06 | 14 | 0.7 | 1.2901 |

Table II. Best-fit $d, \kappa$, and $\chi^2$ for each device obtained by selecting the parameters listed in Ref.([11]) that minimize $\chi_l^2$ in Eqn.(I.1).

| | $\lambda(a_M)$ | $\gamma$ | $M$ | $v_\star$ | $v'_\star$ | $U_1$ | $U_2$ | $J$ | $W_1$ | $W_3$ | $\mu_1$ | $\mu_2$ |
|---|---|---|---|---|---|---|---|---|---|---|---|---|
| D1 | 0.359 | -49.66 | -5.56 | -4658.0 | 1671.0 | 48.49 | 4.28 | 16.10 | 62.74 | 65.96 | 10.8 | 5.0 |
| D2 | 0.353 | -40.59 | 2.04 | -4510.0 | 1631.0 | 45.52 | 3.07 | 14.54 | 48.42 | 52.37 | 28.8 | 5.0 |

Table III. Parameters of the THF model used to match the calculated optical conductivity to the experimental FIR spectra after fixing $\epsilon$ and $\mu_1$ for each device.

$$|\gamma| = \frac{2\alpha_1\alpha_2}{\alpha_3\alpha_4} - \frac{\chi_{cf}}{\nu_F}\left(\alpha_1 + \alpha_3 - \frac{4\alpha_1\alpha_2}{\alpha_3\alpha_4}\right), \tag{H.45a}$$

$$\tilde{U} = \frac{2}{\nu_F}\left(\alpha_1 - \frac{2\alpha_1\alpha_2}{\alpha_3\alpha_4}\right), \tag{H.45b}$$

$$v_\star^2 = \frac{2\alpha_1^2\alpha_2^2}{e\hbar\alpha_3^2\alpha_4}, \tag{H.45c}$$

$$\tilde{W} = \frac{1}{\nu_F}\left(\alpha_1 + \alpha_3 - \frac{4\alpha_1\alpha_2}{\alpha_3\alpha_4}\right). \tag{H.45d}$$

where $\gamma$ is the bare hybridization.

## Appendix I: Parameter extraction

In this section we describe the procedure used to extract the THF-model parameters that best reproduce the experimental data. From the experiment, we extract the transition energies of the most prominent peaks as a function of magnetic field, $\omega_{m_i}^{D_l}(B)$, where $m_i$ $(i = 1, 2, 3)$ labels the transitions and $D_l$ $(l = 1, 2)$ labels the device. In practice, $\omega_{m_i}^{D_l}(B)$ is obtained for a discrete set of fields $B_j^{(i)}$, with $j = 1, \ldots, N_{B,i}$ for each transition $m_i$. We fix the experimentally-resolved twist angle and compute the theoretical transition energies $\Delta E_{m_i}^1\left(B_j^{(i)}\right)$ defined in Sec. G 3 [Eqs. (G.20a–G.20e)], scanning the THF parameters listed in the tables of Ref. [11] as a function of $\kappa$, the intersublattice hopping ratio, and $d$, the screening length. We then determine the parameter set that minimizes the difference between the experimentally extracted $\omega_{m_i}^{D_l}\left(B_j^{(i)}\right)$ and the calculated $\Delta E_{m_i}^1\left(B_j^{(i)}\right)$. To make the parameter dependence explicit, we write $\Delta E_{m_i}^1(B \,|\, \kappa, d)$. To quantify the goodness of fit for a given set of THF parameters, we compute

$$\chi_l^2(\kappa, d) = \frac{1}{3}\sum_{i=1}^{3}\frac{1}{N_{B,i}}\sum_{j=1}^{N_{B,i}}\frac{\left(\omega_{m_i}^{D_l}\left(B_j^{(i)}\right) - \Delta E_{m_i}^1\left(B_j^{(i)} \,|\, \kappa, d\right)\right)^2}{\zeta_{\text{exp}}^2}, \tag{I.1}$$

where $\zeta_{\text{exp}} = 3\,\text{meV}$ is the estimated uncertainty in the experimental peak location (see Fig. 3c in the main text). The optimal $\kappa$ and $d$ that minimize $\chi_l^2$ are listed in Tab. III. We find that both devices are well described by $\kappa = 0.7$, with minor differences in the screening length $d$.

After this first stage, we allow for a finite relaxation parameter $\mu_1$ (Sec. A 1) and for deviations of the dielectric constant $\varepsilon$ from the value $\varepsilon_{\text{hBN}} = 6$ assumed in Ref.([11]). We then calculate $\sigma(E)$ using Eq. (F.8), with the parameters of Ref.([11]) rescaled by $\varepsilon_{\text{hBN}}/\varepsilon$, where $\varepsilon \in \{4, 5, 6, 8, 10\}$. We determine the values of $\mu_1$ and $\epsilon$ that provide the best match to the experimental spectra, performing sweeps over these parameters and comparing $\sigma(E)$ with the experimental data while checking for consistency between $\nu = 4$ and $\nu = -4$. The results of these sweeps are shown in Figs. (4–7). The parameters that yield the best agreement are listed in Tab. III.

Most of the extracted parameters are similar between $D_1$ and $D_2$, except for $W_1$ and $W_3$, which are roughly 30% larger in $D_1$ than in $D_2$, and the relaxation parameter $\mu_1$, which is almost three times larger in $D_2$. A larger relaxation in $D_2$ is expected since its twist angle is smaller than that of $D_1$. Consequently, the adhesion energy of the layers can be minimized through a smaller local rotation that expands the AB-stacking regions without incurring a large elastic-energy penalty [12–14].

The corresponding best-fit band structures and LL spectra for $D_1$ and $D_2$ are shown in Fig. (8). The main difference between the two devices lies in the bandwidth of the flat bands: $D_1$ exhibits a narrower bandwidth than $D_2$. This trend is consistent with Eq. (H.40), where the magnetic-field dependence of the $m_1$ transition is directly related to the bandwidth. In the $M = 0$ limit the bandwidth is given by Eq. (H.41), and at $\nu = \pm 4$ we have $|\nu_c| \ll |\nu_f|$. Increasing $W_3 - J/2N_f$ therefore reduces the bandwidth, consistent with the parameters reported in Tab. III.

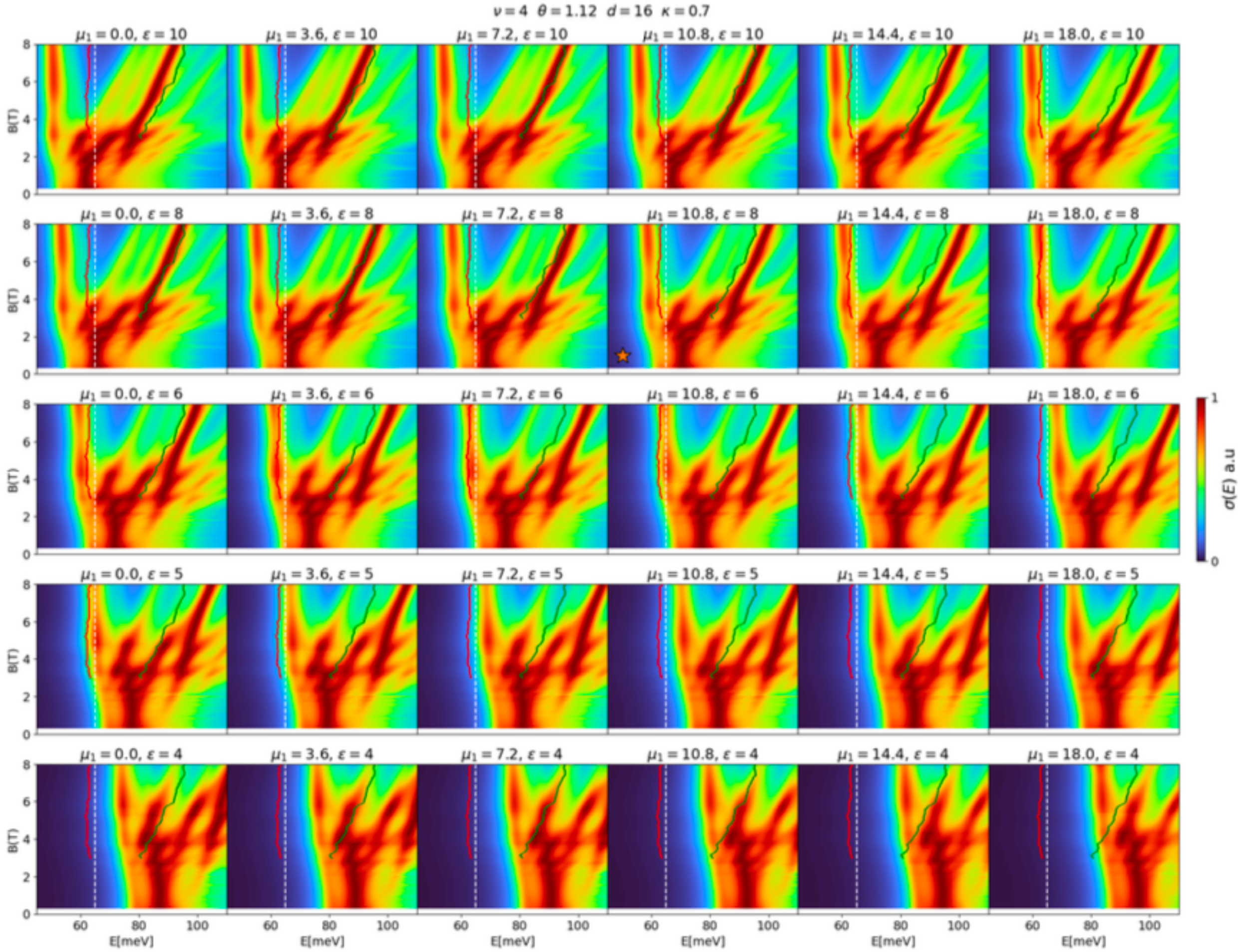

Figure 4. Theoretical optical conductivity as a function of the relaxation parameter $\mu_1$ and the dielectric constant $\varepsilon$. The grid displays $\sigma(E)$ across a systematic sweep of $\mu_1$ (0.0–18.0 meV) and $\varepsilon$ (4–10), at filling $\nu = 4$ with fixed parameters $\theta = 1.12°$, $\kappa = 0.7$, and $d = 16$ nm. The optimal parameter pair $(\mu_1, \varepsilon) = (10.8\text{meV}, 8)$, highlighted by the orange star, is determined by comparing simultaneously the calculated $\sigma(E)$ and the peak positions obtained experimentally at both $\nu = 4$ and $\nu = -4$ (see Fig. 5 for the calculated optical conductivity at $\nu = -4$). The red and green curves are the experimental peak positions extracted from the data shown in Extended Data Fig. 6(b).

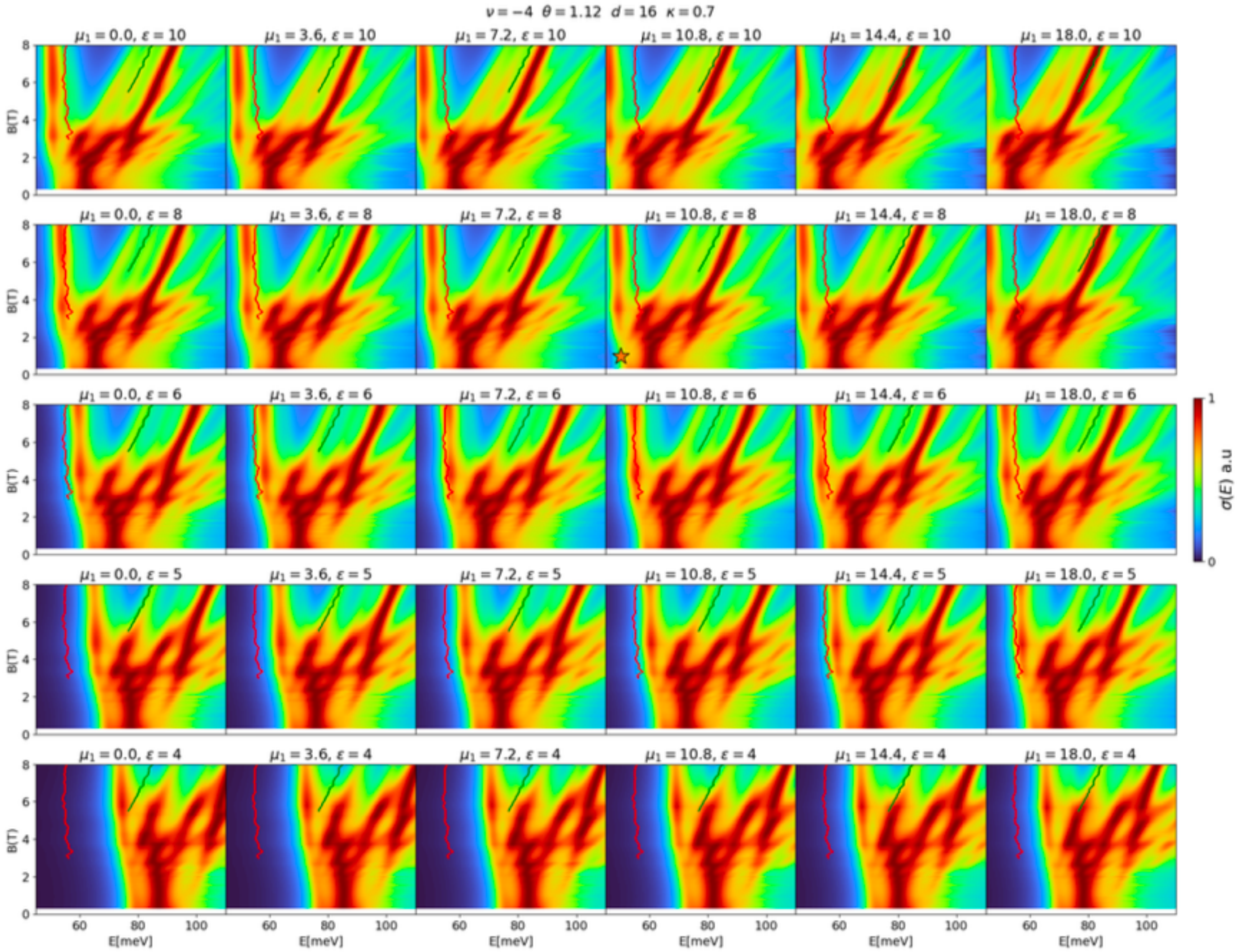

Figure 5. Theoretical optical conductivity as a function of the relaxation parameter $\mu_1$ and the dielectric constant $\varepsilon$. The grid displays $\sigma(E)$ across a systematic sweep of $\mu_1$ (0.0–18.0 meV) and $\varepsilon$ (4–10), at filling $\nu = -4$ with fixed parameters $\theta = 1.12°$, $\kappa = 0.7$, and $d = 16\,\text{nm}$. The optimal parameter pair $(\mu_1, \varepsilon) = (10.8\text{meV}, 8)$, highlighted by the orange star, is determined by comparing simultaneously the calculated $\sigma(E)$ and the peak positions obtained experimentally at both $\nu = 4$ and $\nu = -4$ (see Fig. 4 for the calculated optical conductivity at $\nu = 4$). The red and green curves are the experimental peak positions extracted from the data shown in Extended Data Fig. 6(a).

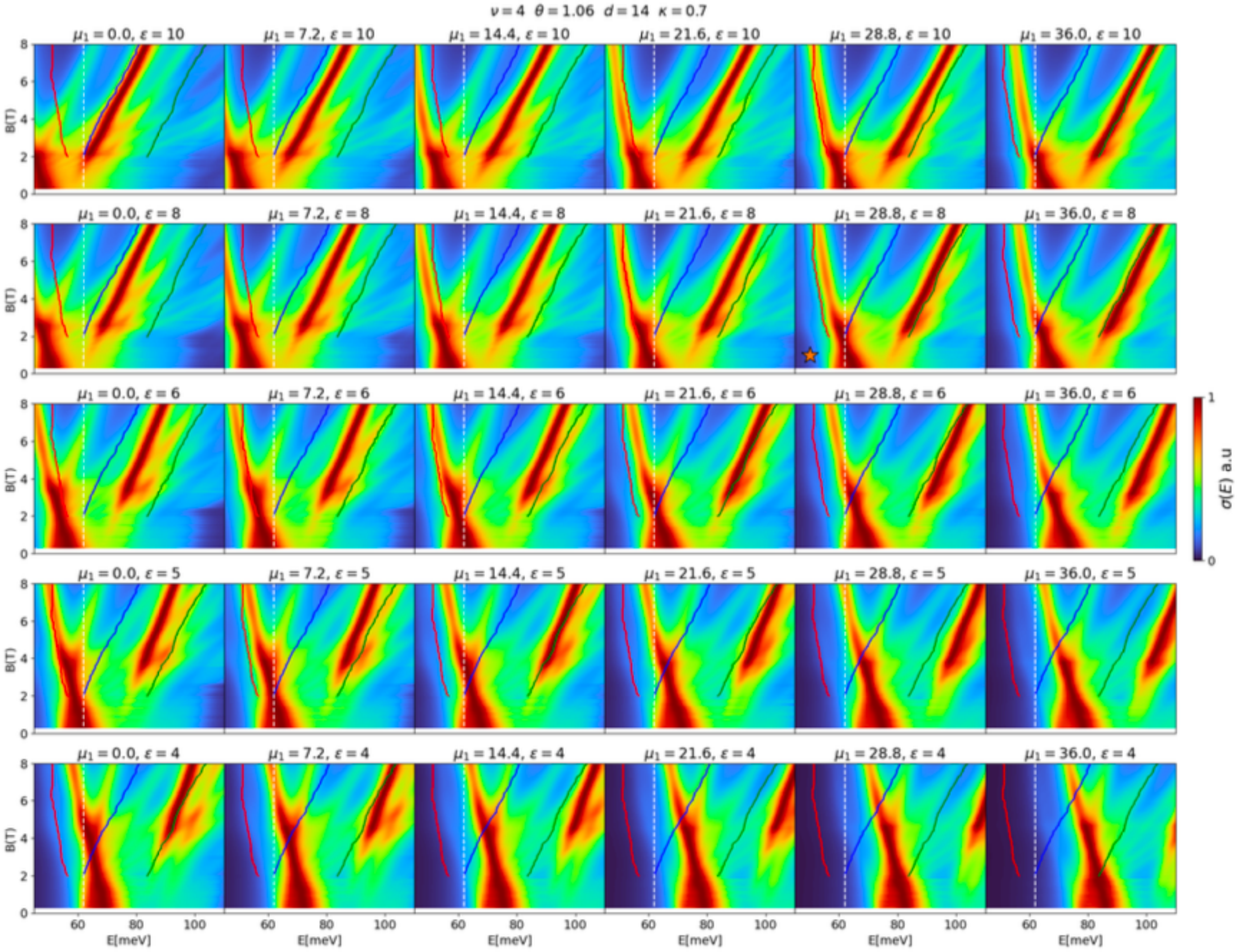

Figure 6. Theoretical optical conductivity as a function of the relaxation parameter $\mu_1$ and the dielectric constant $\varepsilon$. The grid displays $\sigma(E)$ across a systematic sweep of $\mu_1$ (0.0–36.0 meV) and $\varepsilon$ (4–10), at filling $\nu = 4$ with fixed parameters $\theta = 1.06°$, $\kappa = 0.7$, and $d = 14\,\text{nm}$. The optimal parameter pair $(\mu_1, \varepsilon) = (28.8\text{meV}, 8)$, highlighted by the orange star, is determined by comparing simultaneously the calculated $\sigma(E)$ and the peak positions obtained experimentally at both $\nu = 4$ and $\nu = -4$ (see Fig. 7 for the calculated optical conductivity at $\nu = -4$). The red, blue, and green curves are the experimental peak positions extracted from the data shown in Extended Data Fig. 3(b).

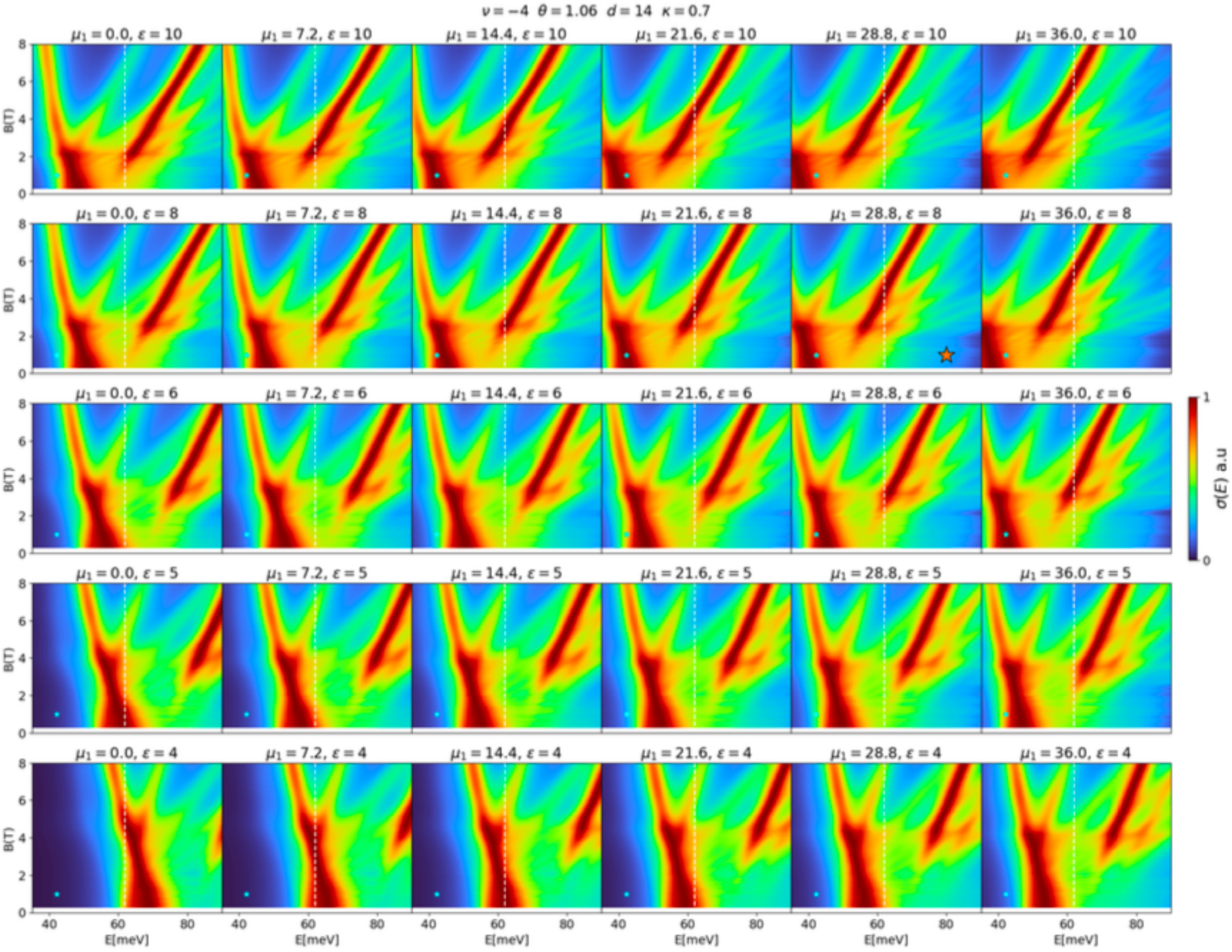

Figure 7. Theoretical optical conductivity as a function of the relaxation parameter $\mu_1$ and the dielectric constant $\varepsilon$. The grid displays $\sigma(E)$ across a systematic sweep of $\mu_1$ (0.0–36.0 meV) and $\varepsilon$ (4–10), at filling $\nu = -4$ with fixed parameters $\theta = 1.06°$, $\kappa = 0.7$, and $d = 14\,\mathrm{nm}$. The optimal parameter pair $(\mu_1, \varepsilon) = (28.8\mathrm{meV}, 8)$, highlighted by the orange star, is determined by comparing simultaneously the calculated $\sigma(E)$ and the peak positions obtained experimentally at both $\nu = 4$ and $\nu = -4$ (see Fig. 6 for the calculated optical conductivity at $\nu = 4$). To compare with the experimental data, we mark the energy of the first peak in Fig. 2(g) in the main text with a cyan star in all the subplots.

## Appendix J: Strain dependence

In extended data Fig.(9), we showed the evolution of $\sigma(E)$ as a function of heterostrain, noting that the $m_2$ transition in Fig.(2)—whose energy lies between those of $m_1$ and $m_3$—becomes diffuse as strain is added (see in Fig.(9)(i-l)). In this section we analyze the behavior of the LL spectrum at $\nu = +4$ for valley $\eta = +$ in detail to explain how the $m_1$ and $m_3$ remain sharp while the $m_2$ transition becomes diffuse.

To study the effects of heterostrain, whose main role is the breakdown of $C_{3z}$ and $C_{2x}$, we add increasing heterostrain along the +x direction. To calculate the optical conductivity, we include the additional terms in Eqs. (A.7a–A.7c), using the parameters in Tab. (I) for $\kappa = 0.7$. The results are shown in Fig. 9. As heterostrain increases, the $f$-electron states in the flat bands are split (see Fig. 9(a–h)). Note that the $m = 0$ anomalous Landau level remains robust, as it is gapped from the rest of the flat-band LLs. Similarly, $m = 1$ and $m = 3$ states remain gapped from the rest of the flat-band LLs up to $B = 10\,\mathrm{T}$.

We can track the redistribution of the optical spectral weight for transitions from the flat to the remote bands by computing the average angular momentum on each Landau level, $\langle \hat{L}_{\eta=+} \rangle$. In the remote bands, $\langle \hat{L}_{\eta=+} \rangle$ remains largely unchanged as heterostrain increases (see Fig. (9)(m–p)), with the exception of points at which Landau levels cross and therefore hybridize more strongly. In contrast, in the flat bands levels, the smaller energy splitting between LLs in the strainless limit leads to increased Landau-level mixing once heterostrain is included (see Fig. 9(q–t)).

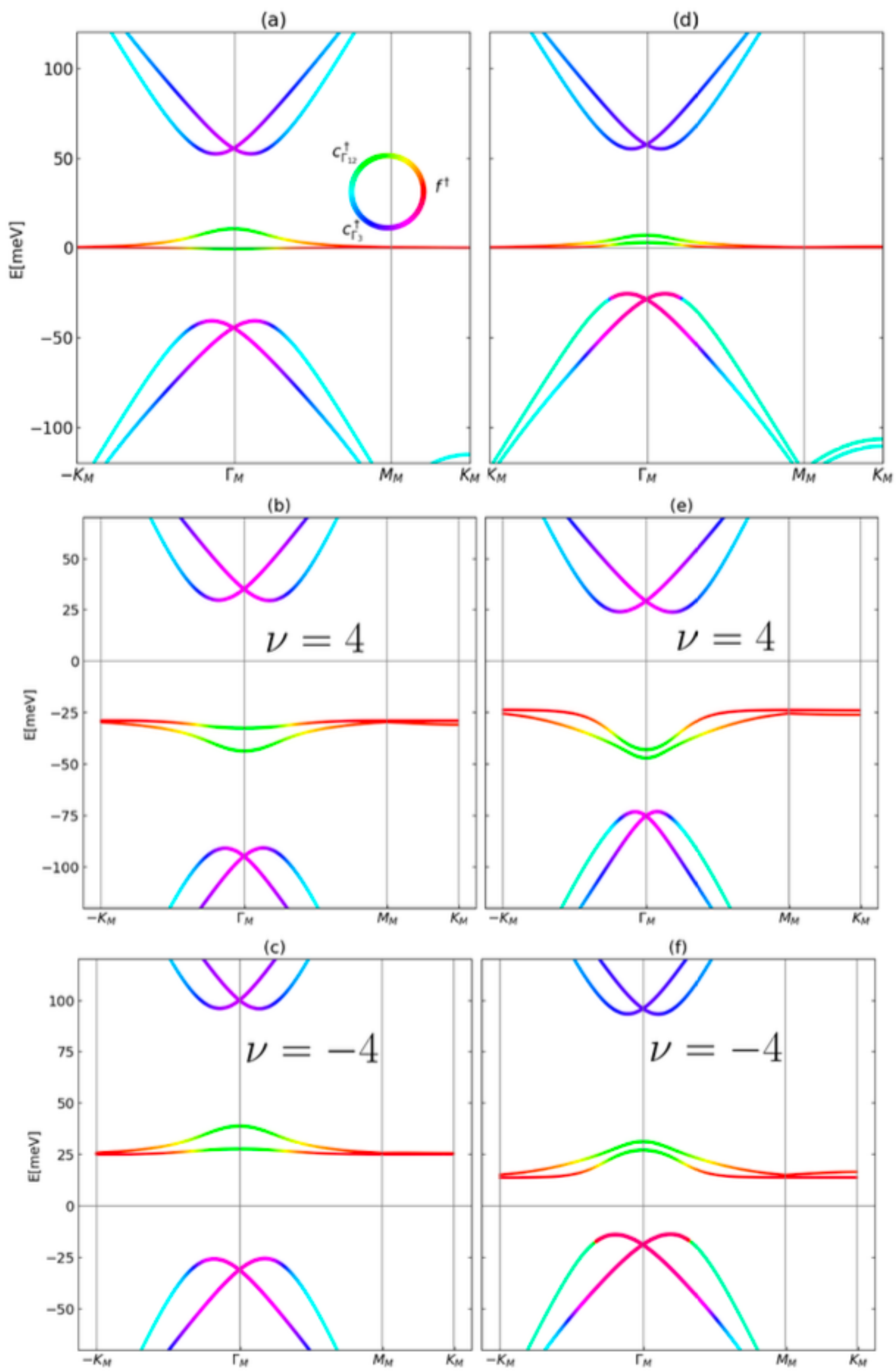

Figure 8. Band structure for the optimal THF parameters in Tab. III for $D_1$ (a–c) and $D_2$ (d–f). The color wheel denotes the orbital character of the states at each **k**-point.

We now focus on the two transitions that remain sharp with increasing heterostrain, namely $m_1$ and $m_3$ in Fig.(9)(i–l). We start by discussing the effect of heterostrain on $m_1$. Note that the $m = 1$ and $m = 3$ Landau levels in the flat bands are allowed to mix as heterostrain increases, but the two-dimensional manifold they span remains gapped from the rest of the LL spectrum (see Fig. 9(q–t)). Even though the average angular momentum in this manifold is $\langle \hat{L}_{\eta=+} \rangle \approx 2$, the states within this manifold are each a superposition of $m = 1$ and $m = 3$, so there is still a finite dipole matrix element with the lowest $\langle \hat{L}_{\eta=+} \rangle \approx 2$ Landau-level in the remote bands. The lowest-energy transition thus remains relatively sharp with increasing heterostrain (see Fig. 9(i–l)).

Next, we discuss the sharpness of the $m_3$ transition. Since the anomalous $m = 0$ level has a relatively large gap to the rest of the LLs that emanate from the flat bands, it retains $\langle \hat{L}_{\eta=+} \rangle \approx 0$. As such, it still has a strong dipole matrix element with the $\langle \hat{L}_{\eta=+} \rangle \approx 1$ remote-band LL, and the $m_3$ transition remains sharp.

In contrast to the two cases discussed above, the $m_2$ transition becomes diffuse as heterostrain increases. The Landau levels in the flat band involved in this transition ($m = 2$ and $m = 4$) lie close in energy to other Landau-level states with higher angular momentum. Mixing induced by heterostrain redistributes the optical spectral weight associated with the sharp $m_2$ transition present at zero heterostrain, causing the optical spectrum to broaden and become diffuse.

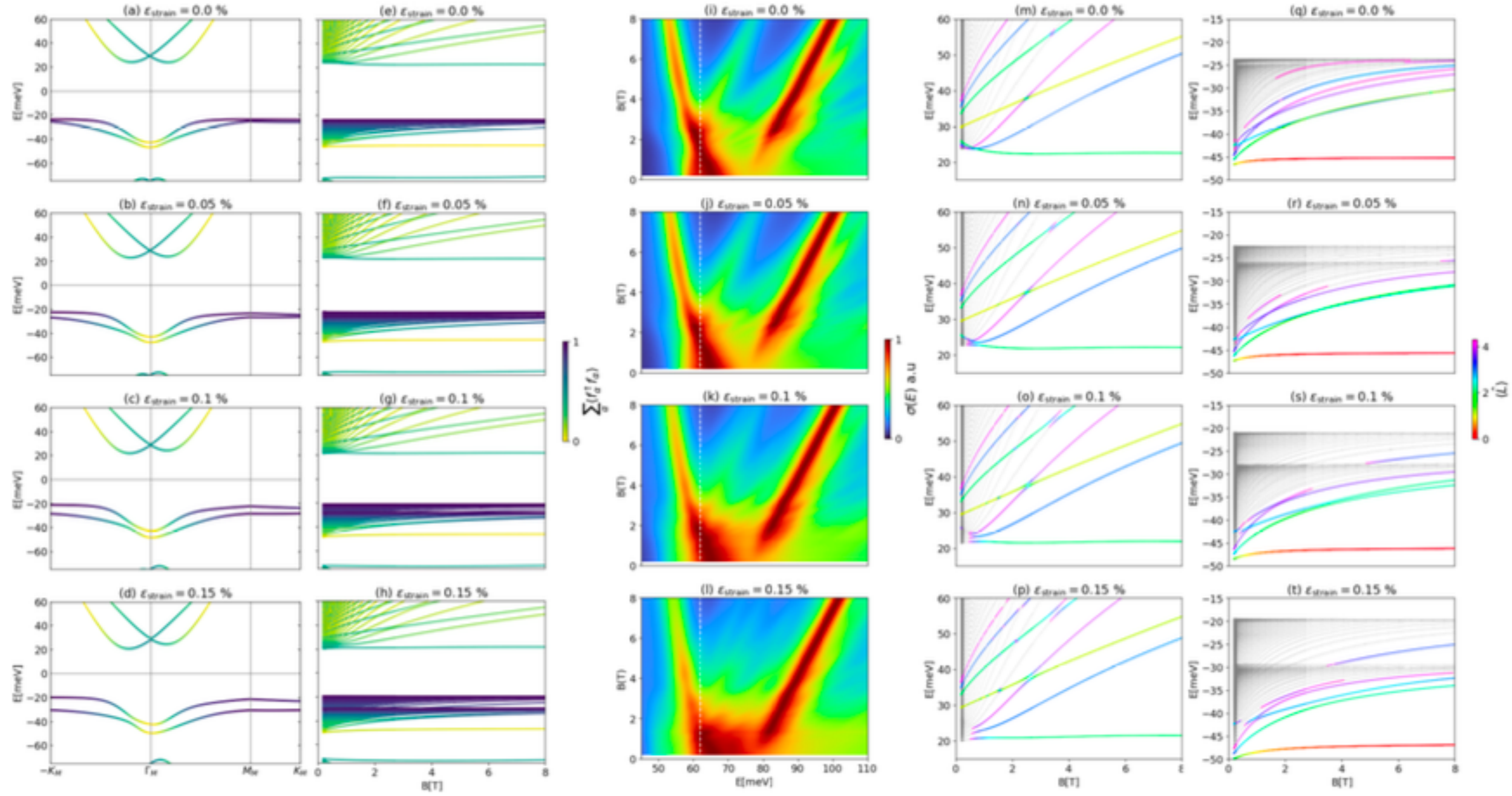

Figure 9. Evolution of the Landau-level spectra and optical conductivity of the THF model at filling $\nu = 4$ under uniaxial heterostrain, projected to valley $\eta = +$. Rows from top to bottom correspond to increasing heterostrain ($\varepsilon_{\text{strain}}$): 0%, 0.05%, 0.1%, and 0.15%. First column (a–d): zero-field band structure along the high-symmetry path. Second column (e–h): Landau-level energy spectrum versus magnetic field; the colormap shows the fractional $f$-orbital character. Third column (i–l): optical conductivity, $\sigma(E)$, as a function of $B$. Fourth column (m–p): Landau-level energy spectrum for the remote bands, colored by $\langle \hat{L}_{\eta=+} \rangle$ (shown up to 4.25 T). Fifth column (q–t): Landau-level energy spectrum focusing on the low-energy flat bands. As heterostrain increases, $\langle \hat{L}_{\eta=+} \rangle$ remains largely robust in the isolated remote bands, while the closely spaced LLs exhibit strong angular-momentum mixing, leading to the smearing of intermediate optical transitions seen in the third column.

## Appendix K: Broadening dependence

In this section we examine the effect of spectral broadening on the optical conductivity maps shown in the main text. For $D_2$, we present the optical conductivity as a function of increasing broadening $\zeta$ at $\nu = +4$ in Fig. (10)(a-e). We find that the main features discussed in the main text are largely insensitive to the choice of broadening.

Throughout this section, we use the angular-momentum labeling defined in the $M = 0$ limit and focus on the strainless case, noting that a finite $M$ only weakly mixes states with different angular momenta. The two sharpest transitions correspond to the $m = 0 \to m = 1$ and $m = 1 \to m = 2$ processes ($m_3$ and $m_1$ transition respectively). The former produces the feature that disperses linearly, while the latter has a negative slope and disperses to lower energies as $B$ increases.

In Fig.(10)(a-b), we see that reducing the broadening reveals several additional features. The most notable are: (i) the $m = 1 \to 2$ transition splits into two distinct lines corresponding to the $m = 2 \to 3$ and $m = 4 \to 3$ transitions described in Sec. (G 3); and (ii) a relatively sharp high-energy feature (appears at $E = 80$meV at $B = 0$), which we identify as a higher-energy $m = 3 \to 2$ transition.

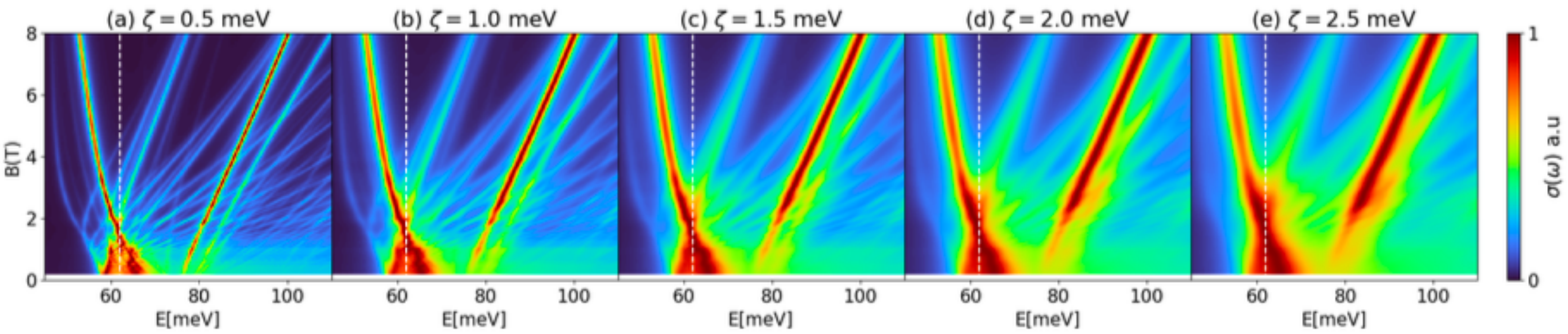

Figure 10. Dependence on the broadening, $\zeta$, of the optical conductivity $\sigma(E)$ as a function of energy, $E$, and magnetic field $B$.

---

[1] Zhi-Da Song and B Andrei Bernevig. Magic-angle twisted bilayer graphene as a topological heavy fermion problem. Physical review letters, 129(4):047601, 2022. URL https://journals.aps.org/prl/abstract/10.1103/PhysRevLett.129.047601.

[2] Zhida Song, Zhijun Wang, Wujun Shi, Gang Li, Chen Fang, and B Andrei Bernevig. All magic angles in twisted bilayer graphene are topological. Physical review letters, 123(3):036401, 2019. URL https://journals.aps.org/prl/abstract/10.1103/PhysRevLett.123.036401.

[3] Rafi Bistritzer and Allan H MacDonald. Moiré bands in twisted double-layer graphene. Proceedings of the National Academy of Sciences, 108(30):12233–12237, 2011.

[4] Jonah Herzog-Arbeitman, Jiabin Yu, Dumitru Călugăru, Haoyu Hu, Nicolas Regnault, Oskar Vafek, Jian Kang, and B Andrei Bernevig. Topological heavy fermion model as an efficient representation of atomistic strain and relaxation in twisted bilayer graphene. Physical Review B, 112(12):125128, 2025.

[5] Jonah Herzog-Arbeitman, Dumitru Călugăru, Haoyu Hu, Jiabin Yu, Nicolas Regnault, Jian Kang, B Andrei Bernevig, and Oskar Vafek. Kekulé spiral order from strained topological heavy fermions. Physical Review B, 112(12):125129, 2025.

[6] Dumitru Călugăru, Haoyu Hu, Rafael Luque Merino, Nicolas Regnault, Dmitri K. Efetov, and B. Andrei Bernevig. The thermoelectric effect and its natural heavy fermion explanation in twisted bilayer and trilayer graphene, 2024. URL https://arxiv.org/abs/2402.14057.

[7] Keshav Singh, Aaron Chew, Jonah Herzog-Arbeitman, B Andrei Bernevig, and Oskar Vafek. Topological heavy fermions in magnetic field. Nature communications, 15(1):5257, 2024.

[8] Xiaoyu Wang and Oskar Vafek. Theory of correlated chern insulators in twisted bilayer graphene. Phys. Rev. X, 14:021042, Jun 2024. doi:10.1103/PhysRevX.14.021042. URL https://link.aps.org/doi/10.1103/PhysRevX.14.021042.

[9] Xiaoyu Wang and Oskar Vafek. Narrow bands in magnetic field and strong-coupling hofstadter spectra. Phys. Rev. B, 106:L121111, Sep 2022. doi:10.1103/PhysRevB.106.L121111. URL https://link.aps.org/doi/10.1103/PhysRevB.106.L121111.

[10] Haoyu Hu, Zhi-Da Song, and B. Andrei Bernevig. Projected and solvable topological heavy fermion model of twisted bilayer graphene, 2025. URL https://arxiv.org/abs/2502.14039.

[11] Dumitru Călugăru, Maksim Borovkov, Liam L. H. Lau, Piers Coleman, Zhi-Da Song, and B. Andrei Bernevig. Twisted bilayer graphene as topological heavy fermion: II. Analytical approximations of the model parameters. Low Temperature Physics, 49(6):640–654, June 2023. ISSN 1063-777X. doi:10.1063/10.0019421. URL https://doi.org/10.1063/10.0019421. _eprint: https://pubs.aip.org/aip/ltp/article-pdf/49/6/640/18087424/640_1_10.0019421.pdf.

[12] Nguyen N. T. Nam and Mikito Koshino. Lattice relaxation and energy band modulation in twisted bilayer graphene. Phys. Rev. B, 96:075311, Aug 2017. doi:10.1103/PhysRevB.96.075311. URL https://link.aps.org/doi/10.1103/PhysRevB.96.075311.

[13] Jian Kang and Oskar Vafek. Pseudomagnetic fields, particle-hole asymmetry, and microscopic effective continuum hamiltonians of twisted bilayer graphene. Phys. Rev. B, 107:075408, Feb 2023. doi:10.1103/PhysRevB.107.075408. URL https://link.aps.org/doi/10.1103/PhysRevB.107.075408.

[14] Oskar Vafek and Jian Kang. Continuum effective hamiltonian for graphene bilayers for an arbitrary smooth lattice deformation from microscopic theories. Phys. Rev. B, 107:075123, Feb 2023. doi:10.1103/PhysRevB.107.075123. URL https://link.aps.org/doi/10.1103/PhysRevB.107.075123.